%% file: main.tex
\documentclass[twocolumn,tighten,twocolappendix]{aastex63}
\NewPageAfterKeywords
\usepackage{graphicx}
\usepackage{url}
\usepackage{color}
\usepackage{multirow}
\usepackage{ragged2e}
\usepackage{hyperref}
\usepackage{url}
\usepackage{amsmath} 
\usepackage{booktabs}
\usepackage{comment}
\usepackage{afterpage}
\usepackage{mathtools}
\usepackage{tabularx}
\usepackage{bold-extra}
\usepackage{xspace}
\usepackage{relsize}
\usepackage[utf8]{inputenc}
\usepackage{newunicodechar}
\usepackage{eht}
\usepackage{bm}
\usepackage[encapsulated]{CJK}
\usepackage{ucs}
\newcommand{\cntext}[1]{\begin{CJK}{UTF8}{gbsn}#1\end{CJK}}

\def\Gl{G$\lambda$\xspace}

\def\msun{M$_{\odot}$\xspace}
\def\model{mG-ring\xspace}
\def\uv{$(u,v)$\xspace}

\def\DiameterMeasurement{$51.8 \pm 2.3$\,\uas}
\def\MoDMeasurement{$4.8_{-0.7}^{+1.4}$\,\uas}
\def\MassMeasurement{$4.0_{-0.6}^{+1.1} \times 10^6$\,M$_{\odot}$\xspace}

\DeclareGraphicsExtensions{.pdf,.png,.jpeg}

\definecolor{ForestGreen}{rgb}{0.133,0.545,0.133}

\begin{document}

\title{
First Sagittarius~A$^*$ Event Horizon Telescope Results. IV. Variability, morphology, and black hole mass
}
\shorttitle{Variability, morphology, and black hole mass}

\input{GAL-4}
\shortauthors{The EHT Collaboration et al.}

\input{abstract}


\nocite{M87PaperI}
\nocite{M87PaperII}
\nocite{M87PaperIII}
\nocite{M87PaperIV}
\nocite{M87PaperV}
\nocite{M87PaperVI}
\nocite{M87PaperVII}
\nocite{M87PaperVIII}

\nocite{PaperI}
\nocite{PaperII}
\nocite{PaperIII}
\nocite{PaperIV}
\nocite{PaperV}
\nocite{PaperVI}

\input{intro}
\input{observations}
\input{variability}
\input{ring}
\input{image_domain}
\input{snapshot_geometric_modeling}
\input{fulltrack_geometric_modeling}
\input{results}
\input{summary}

\begin{acknowledgments}

\input{EHT-paper4-ackn}

\end{acknowledgments}

\facility{EHT}
\software{\comrade, \difmap, \dpi, \ehtim, \texttt{metronization}, \rex, \smili, Stan, \themis, \vida}

\input{appendix}

\bibliographystyle{yahapj}
\bibliography{SgrA_MCFE, EHTCPapers}

\allauthors

\end{document}

%% file: GAL-4.tex
\author[0000-0002-9475-4254]{Kazunori Akiyama}
\affiliation{Massachusetts Institute of Technology Haystack Observatory, 99 Millstone Road, Westford, MA 01886, USA}
\affiliation{National Astronomical Observatory of Japan, 2-21-1 Osawa, Mitaka, Tokyo 181-8588, Japan}
\affiliation{Black Hole Initiative at Harvard University, 20 Garden Street, Cambridge, MA 02138, USA}

\author[0000-0002-9371-1033]{Antxon Alberdi}
\affiliation{Instituto de Astrof\'{\i}sica de Andaluc\'{\i}a-CSIC, Glorieta de la Astronom\'{\i}a s/n, E-18008 Granada, Spain}

\author{Walter Alef}
\affiliation{Max-Planck-Institut f\"ur Radioastronomie, Auf dem H\"ugel 69, D-53121 Bonn, Germany}

\author[0000-0001-6993-1696]{Juan Carlos Algaba}
\affiliation{Department of Physics, Faculty of Science, Universiti Malaya, 50603 Kuala Lumpur, Malaysia}

\author[0000-0003-3457-7660]{Richard Anantua}
\affiliation{Black Hole Initiative at Harvard University, 20 Garden Street, Cambridge, MA 02138, USA}
\affiliation{Center for Astrophysics $|$ Harvard \& Smithsonian, 60 Garden Street, Cambridge, MA 02138, USA}
\affiliation{Department of Physics \& Astronomy, The University of Texas at San Antonio, One UTSA Circle, San Antonio, TX 78249, USA}

\author[0000-0001-6988-8763]{Keiichi Asada}
\affiliation{Institute of Astronomy and Astrophysics, Academia Sinica, 11F of Astronomy-Mathematics Building, AS/NTU No. 1, Sec. 4, Roosevelt Rd, Taipei 10617, Taiwan, R.O.C.}

\author[0000-0002-2200-5393]{Rebecca Azulay}
\affiliation{Departament d'Astronomia i Astrof\'{\i}sica, Universitat de Val\`encia, C. Dr. Moliner 50, E-46100 Burjassot, Val\`encia, Spain}
\affiliation{Observatori Astronòmic, Universitat de Val\`encia, C. Catedr\'atico Jos\'e Beltr\'an 2, E-46980 Paterna, Val\`encia, Spain}
\affiliation{Max-Planck-Institut f\"ur Radioastronomie, Auf dem H\"ugel 69, D-53121 Bonn, Germany}

\author[0000-0002-7722-8412]{Uwe Bach}
\affiliation{Max-Planck-Institut f\"ur Radioastronomie, Auf dem H\"ugel 69, D-53121 Bonn, Germany}

\author[0000-0003-3090-3975]{Anne-Kathrin Baczko}
\affiliation{Max-Planck-Institut f\"ur Radioastronomie, Auf dem H\"ugel 69, D-53121 Bonn, Germany}

\author{David Ball}
\affiliation{Steward Observatory and Department of Astronomy, University of Arizona, 933 N. Cherry Ave., Tucson, AZ 85721, USA}

% ...

\collaboration{0}{The Event Horizon Telescope Collaboration}

%% file: abstract.tex
\begin{abstract}

In this paper we quantify the temporal variability and image morphology of the horizon-scale emission from \sgra, as observed by the EHT in 2017 April at a wavelength of 1.3\,mm.  We find that the \sgra data exhibit variability that exceeds what can be explained by the uncertainties in the data or by the effects of interstellar scattering.  The magnitude of this variability can be a substantial fraction of the correlated flux density, reaching $\sim$100\% on some baselines.  Through an exploration of simple geometric source models, we demonstrate that ring-like morphologies provide better fits to the \sgra data than do other morphologies with comparable complexity. We develop two strategies for fitting static geometric ring models to the time-variable \sgra data; one strategy fits models to short segments of data over which the source is static and averages these independent fits, while the other fits models to the full dataset using a parametric model for the structural variability power spectrum around the average source structure.  Both geometric modeling and image-domain feature extraction techniques determine the ring diameter to be \DiameterMeasurement (68\% credible intervals), with the ring thickness constrained to have an FWHM between $\sim$30\% and 50\% of the ring diameter. To bring the diameter measurements to a common physical scale, we calibrate them using synthetic data generated from GRMHD simulations.  This calibration constrains the angular size of the gravitational radius to be \MoDMeasurement, which we combine with an independent distance measurement from maser parallaxes to determine the mass of \sgra to be \MassMeasurement.

\end{abstract}

%% file: intro.tex
\section{Introduction}\label{sec:Introduction}

Sagittarius A$^*$ (\sgra), the radio source associated with the supermassive black hole (SMBH) at the center of the Milky Way, is thought to subtend the largest angular size of all black holes in the sky.  At a distance of $D \approx 8$\,kpc and with a mass of $M \approx 4 \times 10^6$\,\msun \citep{Do_2019,Gravity_2019,Gravity_2020}, \sgra has a Schwarzschild radius of $\sim$10\,\uas.  Models of optically thin spherical accretion flows around SMBHs generically predict that they will appear to distant observers as bright rings of emission surrounding a darker central ``shadow'' \citep[e.g.,][]{Bardeen_1973,Luminet_1979,deVries_2000,Falcke_2000,
BroderickLoeb2006,BroderickNarayan2006,Broderick2011,Broderick2016,
Narayan_2019}, and a variety of more general accretion flow simulations have demonstrated that the diameter of this ring is typically $\sim$5 times larger than the Schwarzschild radius \citep[e.g.,][]{M87PaperV}.  The Event Horizon Telescope (EHT) collaboration provided observational verification of this picture, using a global very long baseline interferometry (VLBI) network of radio telescopes observing at a frequency of $\sim$230\,GHz to resolve the $\sim$40\,\uas ring of emission around the \m87 SMBH \citep[][hereafter \m87 Papers I--VIII]{M87PaperI,M87PaperII,M87PaperIII,M87PaperIV,M87PaperV,M87PaperVI,M87PaperVII,M87PaperVIII}.

The predicted ring diameter for \sgra is $\sim$50\,\uas, about 25\% larger than what the EHT observed for \m87.  However, because \sgra is more than three orders of magnitude less massive than \m87, all dynamical timescales in the system are correspondingly shorter.  In particular, the typical gravitational timescale for \sgra is $G M / c^3 \approx 20$\,s, implying that the source structure can vary substantially over the several-hour duration of a single EHT observation.  Consistent with this expectation, \sgra exhibits broadband variability on timescales of minutes to hours \citep[e.g.,][]{Genzel_2003,Ghez_2004,Fish_2011,Neilsen_2013,Goddi_2021,Wielgus2022}.  The multiwavelength properties of \sgra during the 2017 EHT observing campaign are described in \citet[][hereafter \citetalias{PaperII}]{PaperII}.

The potential for rapid structural variability complicates the analysis of EHT observations of \sgra.  A standard strategy for ameliorating the sparsity of VLBI data sets is Earth-rotation aperture synthesis, whereby Fourier coverage of the array is accumulated as the Earth rotates and baselines change their orientation with respect to the source \citep{TMS}.  This strategy is predicated on the source remaining static throughout the observing period, in which case the accumulated data measure a single image structure.  However, \sgra violates this assumption on timescales as short as minutes.  After several hours, the variable components of the image structure in \sgra are expected to be uncorrelated \citep{Georgiev_2022, Wielgus2022}.
Thus, image reconstructions from the EHT \sgra data are focused on reconstructing time-averaged source structures \citep[][hereafter \citetalias{PaperIII}]{PaperIII}.

Despite the necessity of reconstructing an average source structure, the data collected within a single multihour observation epoch are associated with many specific instances of the variable emission from \sgra, i.e., they represent an amalgam of observations of instantaneous images. The imaging strategy pursued for the EHT observations of \sgra aims to mitigate the impact of this changing source structure through the introduction of a ``variability noise budget,'' which absorbs the structural evolution into inflated uncertainties and thereby permits imaging algorithms to reconstruct a time-averaged image under the usual static source assumption.\footnote{Note that this procedure ignores the subhour correlations that are present in the \sgra data; the implications of these correlations are discussed in \autoref{sec:Variability}.} The image reconstruction procedure is described in detail in \citetalias{PaperIII}, and the results confirm that the \sgra data are consistent with being produced by a ring-like emission structure with a diameter of $\sim$50\,\uas.

For the EHT observations of \m87, morphological properties of the observed ring (e.g., diameter, thickness, orientation) were quantified using both imaging and geometrical modeling analyses \citepalias{M87PaperVI}, and the measured ring diameter was calibrated using general relativistic magnetohydrodynamic (GRMHD) simulations from \citetalias{M87PaperV} to constrain the mass of the SMBH.  The current paper applies a conceptually similar strategy to the analysis of the EHT \sgra data, though significant alterations have been made to meet the new challenges posed by \sgra and to tailor the analyses appropriately.  In this paper, we first characterize the variability seen in the \sgra data, and we develop a framework for mitigating the impact of variability when imaging or modeling the data.  We then make measurements of the ring size and other structural properties using both imaging and geometrical modeling analyses, and we derive and apply a GRMHD-based calibration to bring ring size measurements made using different techniques to a common physical scale.

This paper is organized as follows.  \autoref{sec:Observations} provides an overview of the \sgra observations and data processing.  In \autoref{sec:Variability}, we quantify the variability on different spatial scales, and we outline the strategies used to mitigate its impact during imaging and modeling.  In \autoref{sec:Ring}, we discuss salient data properties in the context of a ring-like emission structure, and we describe our procedure for using GRMHD simulations to calibrate different ring size measurement techniques to a common physical scale.  Sections \ref{sec:ImageDomain}, \ref{sec:SnapshotGeometricModeling}, and \ref{sec:FullTrackGeometricModeling} detail our three primary strategies for measuring the ring size and describe their application to the \sgra data.  Our results are presented in \autoref{sec:results}, and we summarize and conclude in \autoref{sec:Summary}.  This paper is the fourth in a series that describes the analysis of the 2017 EHT observations of \sgra.  The series is summarized in \citet[][hereafter \citetalias{PaperI}]{PaperI}.  The data processing and calibration are described in \citetalias{PaperII}, imaging is carried out in \citetalias{PaperIII}, physical simulations are described in \citet[][hereafter \citetalias{PaperV}]{PaperV}, and tests of gravity are presented in \citet[][hereafter \citetalias{PaperVI}]{PaperVI}.

%% file: observations.tex
\section{Observations and data products}\label{sec:Observations}

In this section, we briefly review the interferometric data products used for analyses in this paper (\autoref{sec:DataProducts}), and we summarize the observations (\autoref{sec:Obs}) and data processing (\autoref{sec:DataReduction}) that precede these analyses. A more comprehensive description of the \sgra data collection, correlation, and calibration can be found in \citetalias{PaperII}, \citetalias{M87PaperIII}, and references therein.

\subsection{VLBI data products} \label{sec:DataProducts}

As a radio interferometer, the EHT is natively sensitive to the Fourier transform of the sky-plane emission structure.  For a source of emission $I(\boldsymbol{x},t)$, the complex visibility $\mathcal{V}(\boldsymbol{u},t)$ is given by
\begin{equation}
\mathcal{V}(\boldsymbol{u},t) = \iint e^{- 2 \pi i \boldsymbol{u} \cdot \boldsymbol{x}} I(\boldsymbol{x},t) \text{d}^2\boldsymbol{x} ,
\end{equation}
where $t$ is time, $\boldsymbol{x} = (x,y)$ are angular coordinates on the sky, and $\boldsymbol{u} = (u,v)$ are projected baseline coordinates in units of the observing wavelength \citep[see, e.g.,][]{TMS}.

The ideal visibilities $\mathcal{V}$ are not directly observable because they are corrupted by both statistical errors and a variety of systematic effects.  For the EHT, the dominant systematics are complex station-based gain corruptions.  The relationship between an ideal visibility $\mathcal{V}_{ij}$ and the observed visibility $V_{ij}$ on a baseline connecting stations $i$ and $j$ is given by
\begin{equation}
V_{ij} = g_i g_j^* \mathcal{V}_{ij} + \sigma_{\text{th},ij} \equiv |V_{ij}| e^{i \phi_{ij}}, \label{eqn:VisibilityCorruptions}
\end{equation}
where $\sigma_{\text{th},ij}$ is the statistical (or ``thermal'') error on the baseline, $g_i$ and $g_j$ are the station gains, and we have defined the visibility amplitude $|V_{ij}|$ and phase $\phi_{ij}$.  The statistical error is well described as a zero-mean circularly symmetric complex Gaussian random variable with a variance determined (per the radiometer equation) by the station sensitivities, integration time, and frequency bandwidth \citep{TMS}.  The station gains vary in time at every site and must in general be either calibrated out or determined alongside the source structure.

The presence of station-based systematics motivates the construction and use of ``closure quantities'' that are invariant to such corruptions.  A closure phase $\psi_{ijk}$ \citep{Jennison_1958} is the sum of visibility phases around a closed triangle of baselines connecting stations $i$, $j$, and $k$,
\begin{equation}
\psi_{ijk} = \phi_{ij} + \phi_{jk} + \phi_{ki} . \label{eqn:ClosurePhase}
\end{equation}
Closure phases are invariant to station-based phase corruptions, such that the measured closure phase is equal to the ideal closure phase, up to statistical errors.  Similarly, a closure amplitude $A_{ijk\ell}$ \citep{Twiss_1960} is the ratio of pairs of visibility amplitudes on a closed quadrangle of baselines connecting stations $i$, $j$, $k$, and $\ell$,

\begin{equation}
A_{ijk\ell} = \frac{|V_{ij}| |V_{k\ell}|}{|V_{ik}| |V_{j\ell}|} .
\end{equation}

\noindent Analogous with closure phases, closure amplitudes are invariant to station-based amplitude corruptions.  Because closure quantities are constructed from nonlinear combinations of complex visibilities, they have correlated and non-Gaussian error statistics; a detailed discussion is provided in \cite{Blackburn_2020}.

\subsection{EHT observations of \texorpdfstring{\sgra}{Sgr A*}} \label{sec:Obs}

The EHT observed \sgra on 2017 April 5, 6, 7, 10, and 11 with the phased Atacama Large  Millimeter/submillimeter Array (ALMA) and the Atacama Pathfinder Experiment (APEX) on the Llano de Chajnantor in Chile, the Large Millimeter Telescope Alfonso Serrano (LMT) on Volc\'{a}n Sierra Negra in Mexico, the James Clerk Maxwell Telescope (JCMT) and phased Submillimeter Array (SMA) on Maunakea in \hawaii, the IRAM 30\,m telescope (PV) on Pico Veleta in Spain, the Submillimeter Telescope (SMT) on Mt.\ Graham in Arizona, and the South Pole Telescope (SPT) in Antarctica \citepalias{M87PaperII}.  Only the April 6, 7, and 11 observations included the highly sensitive ALMA station, and the April 11 light curve exhibits strong variability \citep{Wielgus2022} that is presumably associated with an X-ray flare that occurred shortly before the start of the track \citepalias{PaperII}.  In this paper, we thus analyze primarily the April 6 and April 7 data sets.  We note that while \citetalias{PaperIII} focuses on the April 7 data set, with the April 6 data set used for secondary validation, most of the analyses carried out in this paper instead focus on a joint data set that combines the April 6 and April 7 data.

At each site the data were recorded in two 1.875\,GHz wide frequency bands, centered around sky frequencies of 227.1\,GHz (low band; LO) and 229.1\,GHz (high band; HI), and in each of two polarization modes.  For all telescopes except ALMA and JCMT, the data were recorded in a dual circular polarization mode: right-hand circular polarization (RCP; R) and left-hand circular polarization (LCP; L).  ALMA recorded using linear feeds, and the data were later converted to a circular polarization basis during the DiFX \citep{difx} correlation \citep{Marti_2016,Matthews_2018,Goddi_2019}.  The JCMT observed only a single hand of circular polarization at a time, with the specific handedness (RCP or LCP) changing from day to day.  All other stations observed in a standard dual-polarization mode, which allows the construction of RR, RL, LR, and LL correlation products. The analyses in this paper use only the parallel-hand correlations (i.e., RR and LL), which are averaged to form Stokes $I$ data products.  Because JCMT records only a single hand at a time, we instead form ``pseudo-$I$'' data products for JCMT baselines, using whichever parallel-hand correlation is available as a stand-in for Stokes $I$.\footnote{The ``pseudo-$I$'' formation is a good approximation for Stokes $I$ when the magnitude of the Stokes $V$ contribution is small.  We expect this condition to be met for the 2017 EHT observations of \sgra \citep{Goddi_2021}, and the impact of residual Stokes $V$ is captured by the systematic error budget \citepalias{PaperII}.}

\subsection{Data reduction} \label{sec:DataReduction}

After correlation, residual phase and bandpass errors are corrected with two independent processing pipelines: EHT-HOPS \citep{Blackburn2019} producing ``HOPS'' \citep{Whitney_2004} data and rPICARD \citep{2019Janssenproc, Janssen2019} producing ``CASA'' \citep{McMullin2007} data.  Relative phase gains between RCP and LCP have been corrected based on the assumption of zero circular polarization on baselines between ALMA and other EHT stations.  Absolute flux density scales are based on a priori measurements of each station's sensitivity, resulting in a ${\sim}10$\% typical uncertainty in the amplitude gains \citepalias{M87PaperII}.  The amplitude gains of the colocated ALMA/APEX and SMA/JCMT stations have been further refined via time-variable network calibration \citepalias{M87PaperIII} using a light curve of the compact \sgra flux measured by ALMA and SMA \citep{Wielgus2022}.  For the remaining stations, gross amplitude gain errors have been corrected by a transfer of gain solutions from the J1924-2914 and NRAO\,530 calibrator sources as described in \citetalias{PaperII}.

Following the completion of the above calibration pipelines, additional preprocessing of the data has been carried out as described in \citetalias{PaperIII}, including calibration of the LMT and JCMT station gains and normalization of the visibility amplitudes by the total light curve.  The characterization of residual calibration effects (e.g., polarization leakage) into a systematic error budget, as well as a more comprehensive description of the overall EHT \sgra data reduction, is provided in \citetalias{PaperII}.

%% file: variability.tex
\section{Variability extraction and mitigation} \label{sec:Variability}

The statistical errors quoted in \citetalias{PaperII} and summarized in the preceding section do not account for three additional sources of uncertainty that can otherwise substantially bias any analysis efforts.  First, unaccounted-for nonclosing (i.e., baseline-based) systematic errors are present in the data at a level that is on the order of $\sim$1\% of the visibility amplitude, which is often larger than the formal statistical errors \citepalias[for a discussion of their magnitude and potential origins, see][]{PaperII}.  Second, significant refractive scattering in the interstellar medium produces additional substructure within the image that is not present in the intrinsic emission map \citep{Johnson_2018}.  Third, there is intraday variability in the source itself.  Source variability is theoretically expected to arise on a broad range of timescales, and it is explicitly seen in GRMHD simulations on timescales as short as minutes \citep{Georgiev_2022}.  Such variability was also observed in the light curve of \sgra during the 2017 EHT campaign on timescales from 1 minute to several hours \citep{Wielgus2022}.

In this section, we summarize the theoretical expectations for and characteristics of the variability based on GRMHD simulations, present an estimate for the degree of structural variability in \sgra directly from the visibility amplitude data, and describe the strategies pursued here and in \citetalias{PaperIII} to mitigate the impact of the three components of additional error listed above.

\subsection{Expectations from theory}
\label{sec:variability_theory}

In low-luminosity SMBH systems such as \sgra, we expect the emission to originate from the immediate vicinity of the black hole, i.e., on scales comparable to the event horizon size. Here, all characteristic speeds of the hot relativistic gas approach the speed of light. The timescales associated with these processes are therefore set by the gravitational timescale, $G M/c^3$, which is $\sim$20\,s for \sgra. This timescale is $\sim$3 orders of magnitude shorter than the nightly observations carried out by the EHT, so a single observation contains many realizations of the underlying source variability.

GRMHD simulations can model the dynamical processes in \sgra and, using ray-tracing and radiative transfer, provide a theoretical expectation for the observed emission. \citetalias{PaperV} provides a library of GRMHD simulations and associated movies, which have been scaled to the conditions during the EHT 2017 observations (e.g., the average total 230\,GHz flux is set to the EHT measurement).  We use the variability characteristics of these simulations as our expectation for the \sgra variability seen by the EHT.

GRMHD simulations are universally described by a ``red-red'' power spectrum, with the largest fluctuations in the emission occurring on the longest timescales and the largest spatial scales \citep{Georgiev_2022}. Spatially, the largest scale for variability is limited to the size of the emitting region, which for an observing frequency of 230\,GHz is typically several $G M/c^2$ and which for the EHT \sgra data is constrained to be $\lesssim$87\,\uas \citepalias{PaperII}.  Temporally, the simulations exhibit a red power spectrum that flattens on timescales ${\gtrsim}1000\ GM/c^3$. Observations of the total flux variability in \sgra corroborate this expectation, finding a red-noise spectrum extending to timescales of several hours and flattening on longer timescales \citep{Wielgus2022}.

We can, without loss of generality, express the time-variable image structure $I$ in terms of some static mean image $I_{\text{avg}}$ and a zero-mean time-variable component $\delta I$ that captures all of the variation,
\begin{equation}
I(\boldsymbol{x},t) = I_{\text{avg}}(\boldsymbol{x}) + \delta I(\boldsymbol{x},t) .
\end{equation}
The linearity of the Fourier transform ensures that an analogous decomposition holds for $\mathcal{V}$, which is thus simply the sum of an analogous $\mathcal{V}_0$ and $\delta\mathcal{V}$. The variation $\delta\mathcal{V}$ represents the component of the data we wish to mitigate.

The EHT stations ALMA and SMA are themselves interferometric arrays capable of separating out extended structure \citep[such as the Galactic center ``minispiral'';][]{Lo_1983,Goddi_2021,Wielgus2022} from the \sgra light curve,
\begin{equation}
    L(t)=\mathcal{V}(\boldsymbol{0},t)=\iint I(\boldsymbol{x},t) d^2 \boldsymbol{x} ,
\end{equation}
on the largest spatial scales, predicted by GRMHD simulations to be the most variable.
Using this motivation, the light-curve-normalized image is defined to be 
\begin{equation}
    \hat{I}(\boldsymbol{x},t) \equiv \frac{I(\boldsymbol{x},t)}{L(t)}
\end{equation}
with $\hat{I}_{\text{avg}}$ and $\delta \hat{I}$ similarly defined; here, the ``hat'' diacritic denotes light-curve normalization.
From GRMHD simulations, the expected noise is well approximated by a broken power law,
\begin{equation}
\sigma_{\text{var}}^2 
\equiv 
\<\delta \hat{\mathcal{V}}^2\>
\approx 
\frac{a^2 \left( |\boldsymbol{u}| / u_0 \right)^c}{1 + \left( |\boldsymbol{u}| / u_0 \right)^{b+c}} ,
\label{eq:noise-model}
\end{equation}
along any radial direction \citep{Georgiev_2022}. This broken power law is described by four parameters: a break at $u_0$, an amplitude $a$ representing the amount of noise at the break location, and long- and short-baseline power-law indices $b$ and $c$, respectively. 
Typically, we expect that $c\gtrsim2$, due to the compact nature of the source.

\begin{figure}
  \begin{center}
    \includegraphics[width=\columnwidth]{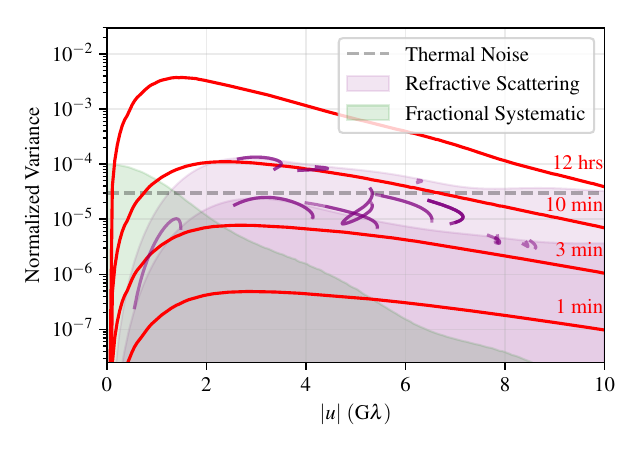}
  \end{center}
\caption{The amount of excess variance in the visibility amplitudes expected for a GRMHD simulation (red lines, $\sigma_\text{var}^2$), after subtraction of the average image and normalization of the flux density by the light curve. Shown also are the typical thermal noise (black dashed line) and a 1\% fractional systematic noise (green band) proportional to the mean image visibility amplitudes. The expected degree of refractive scattering  is shown by the purple bands, with purple lines evaluated for a Gaussian source at the projected location of EHT data \citepalias[see][]{PaperIII}. The variability is shown about a mean image constructed on different observationally relevant timescales. The fractional systematic and variability noise have been averaged over azimuth and over the position angle of the diffractive screen.}
\label{fig:GRMHD_PSD}
\end{figure}

In \autoref{fig:GRMHD_PSD}, red lines show $\sigma_{\text{var}}^2$ measured for an example GRMHD simulation about average images that have been constructed on observationally relevant timescales. The variability has been averaged in azimuth and across different black hole spin orientations.  As the timescale over which the average image is constructed increases, the location of the break $u_0$ decreases and the amount of power at the break 
increases.\footnote{This behavior is generic across a large number of simulations of \sgra and explored in detail in \citet{Georgiev_2022}.} This behavior can intuitively be understood as the GRMHD simulations changing less for short timescales.  For comparison, we show the thermal, systematic, and refractive scattering noise. For timescales longer than $\sim$10\,min, the variability noise dominates on EHT VLBI baselines.

\subsection{Intraday variability in the \texorpdfstring{\sgra}{Sgr A*} data}  \label{sec:CrossFollow}

\begin{figure*}
  \begin{center}
    \includegraphics[width=\textwidth]{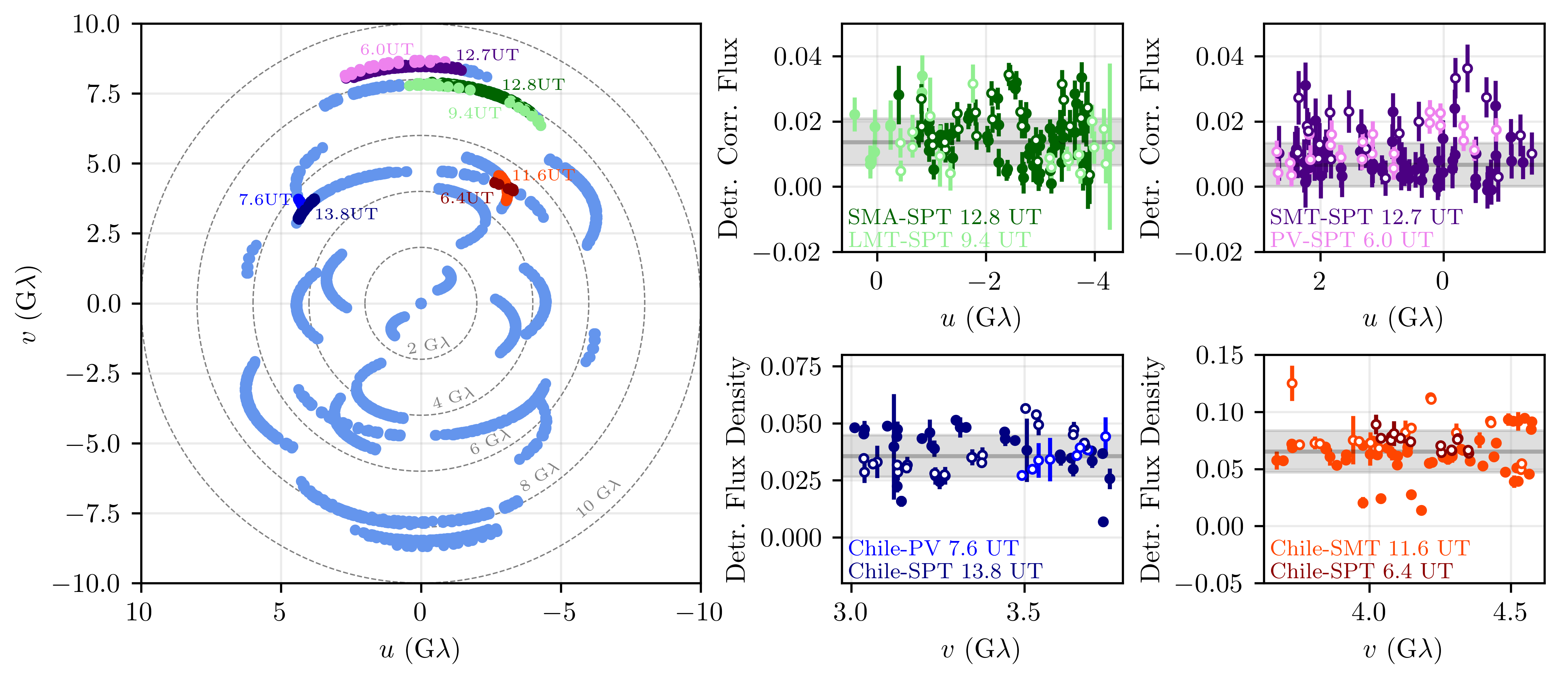}
  \end{center}
\caption{Location and detrended visibility amplitudes for crossing and following tracks from the \sgra observations on April 5, 6, 7, and 10.  \textit{Left}: the locations of all data points in the $(u,v)$-plane.  Individual baselines correspond to the curving tracks seen in this plot, and we have highlighted the crossing and following tracks using colored points.  For reference, circles of constant baseline length are shown by the dashed gray lines.  The central time stamps for the highlighted tracks are labeled in the corresponding colors. \textit{Right}: linearly detrended and light-curve-normalized visibility amplitudes for the crossing and following baselines highlighted in the left panel, shown in corresponding point colors.  For crossing tracks (bottom panels), the points within $1~{\rm G}\lambda$ of the crossing point are shown. Error bars show the thermal errors. For following tracks (top panels) all points for which the baselines overlap are shown.  All highlighted points are employed in the linear detrending for each panel.  The associated estimates of the mean and standard deviation are shown by the gray dashed line and horizontal band in each panel.  The April 7 visibility amplitudes are indicated by open circles.}
\label{fig:crossing_tracks}
\end{figure*}

The intraday variability expected from theoretical considerations can be observed directly in the \sgra data.
\autoref{fig:crossing_tracks} shows the combined baseline coverage for the EHT's 2017 \sgra campaign, including the observations on April 5, 6, 7, and 10.  
The upper limit on the source size of 87~\uas \citepalias[see the second-moment analysis in][]{PaperII} implies that the complex visibilities will be correlated in regions of the \uv-plane smaller than ${\sim}2$\,\Gl.
In practice, the visibility amplitudes exhibit variations on scales smaller than this and otherwise appear strongly correlated on scales of $1$\,\Gl (see \citetalias{PaperIII}, Figure 3).
Therefore, among the baseline tracks in \autoref{fig:crossing_tracks} there are four regions where the \uv-coverage is redundant, i.e., multiple baselines pass within 1\,\Gl of the same \uv-position.
We separate the redundant baseline combinations into ``crossing tracks,'' in which two baseline tracks intersect at a single \uv-point, and ``following tracks,'' in which two baselines follow a nearly identical extended track in the \uv-plane.  Both sets of redundant baselines provide an opportunity to directly probe the degree of intraday variability in the visibilities at specific locations in the \uv-plane.

Prior to making comparisons, we apply the data preprocessing steps outlined in \autoref{sec:DataReduction} to mitigate unphysical sources of variability.  To avoid addressing the unknown atmospheric phase delays, we focus exclusively on visibility amplitudes.
Because source structure will produce additional variations in the visibility amplitudes that are hard to visualize in projection and obscure the relative degree of variability, we detrend the visibility amplitudes with a linear model.
The crossing and following tracks discussed below are shown in the top and bottom subpanels of \autoref{fig:crossing_tracks}, respectively.

{\em Chile--PV vs.\ Chile--SPT}: The first crossing track we consider contains baselines between the Chile stations (ALMA, APEX) and PV and SPT, which both cross near $(u,v) = ($4\,\Gl, 3.5\,\Gl) at times separated by 6.2\,hr.  The concurrent ALMA and APEX baselines are consistent within the reported statistical errors, and thus there is no evidence for unaddressed baseline-specific dominant systematic errors. The normalized visibility amplitudes for the Chile--PV and Chile--SPT baselines individually vary smoothly with time.  Nevertheless, they differ significantly at the crossing point, and this difference is consistent in magnitude with the variation found across days (indicated by the gray band in the relevant panel of \autoref{fig:crossing_tracks}).

{\em Chile--SMT vs.\ Chile--SPT}: The second crossing track we consider contains baselines between the Chile stations (ALMA, APEX) and SMT and SPT, which both cross near $(u,v) = ($3\,\Gl, 4.5\,\Gl) at times separated by 5.2\,hr. Again, we find excellent agreement between ALMA and APEX baselines, individually smooth variations on the Chile--SMT and Chile--SPT baselines, and significant differences in the visibility amplitudes between those baselines.

{\em SMA--SPT vs.\ LMT--SPT}: The first following track we consider contains baselines between the SPT, which is located at the South Pole, and SMA and LMT, which have similar latitudes.  Because the baseline tracks are coincident across a large range of locations in the \uv-plane, this following track permits many direct comparisons at a baseline length of 8\,\Gl at times separated by 3.4\,hr.  As with both crossing tracks, significant differences exist between the two sets of baselines, consistent with the range across multiple days.

{\em SMT--SPT vs.\ PV--SPT}: The final following track we consider again involves the SPT, and now the SMT and PV, which also have similar latitudes.  This is the longest set of baselines that we consider, with a length of roughly 8.5\,\Gl and covering similar regions in the \uv-plane at times separated by 6.7\,hr.  Again, significant variations are exhibited, consistent with those across days.

In summary, intraday variability is observed on multiple baselines with lengths ranging from 5 to 8.5\,\Gl and on timescales as short as 3.4\,hr.  In all cases, this variability is broadly consistent with that observed on interday timescales.
Furthermore, the variability behavior is consistent with theoretical expectations from GRMHD simulations and empirical expectations from the \sgra light curve, both of which imply that the variable elements of the \sgra emission should be uncorrelated beyond a timescale of a few hours \citep{Georgiev_2022,Wielgus2022}.  Any average image of \sgra reconstructed from data spanning a time range longer than several hours captures the long-timescale asymptotic source structure; the intrinsic image averaged over a single day or multiple days is thus expected to exhibit similar structure.

\subsection{Model-agnostic variability quantification} \label{sec:pre-modeling}

To quantify the variability observed in the EHT \sgra data, we make use of the procedure described in \citet{NoiseModeling}.  This procedure provides an estimate of the excess variability -- i.e., the visibility amplitude variance in excess of that caused by known sources, such as average source structure, statistical and systematic uncertainties, and scattering -- as a function of baseline length.  We apply the same data preparation steps summarized in \autoref{sec:CrossFollow} and described in \citet{NoiseModeling}, combining the visibility amplitudes measured on April 5, 6, 7, and 10 in both observing bands.  All data points are weighted equally.

\begin{figure}
  \begin{center}
    \includegraphics[width=\columnwidth]{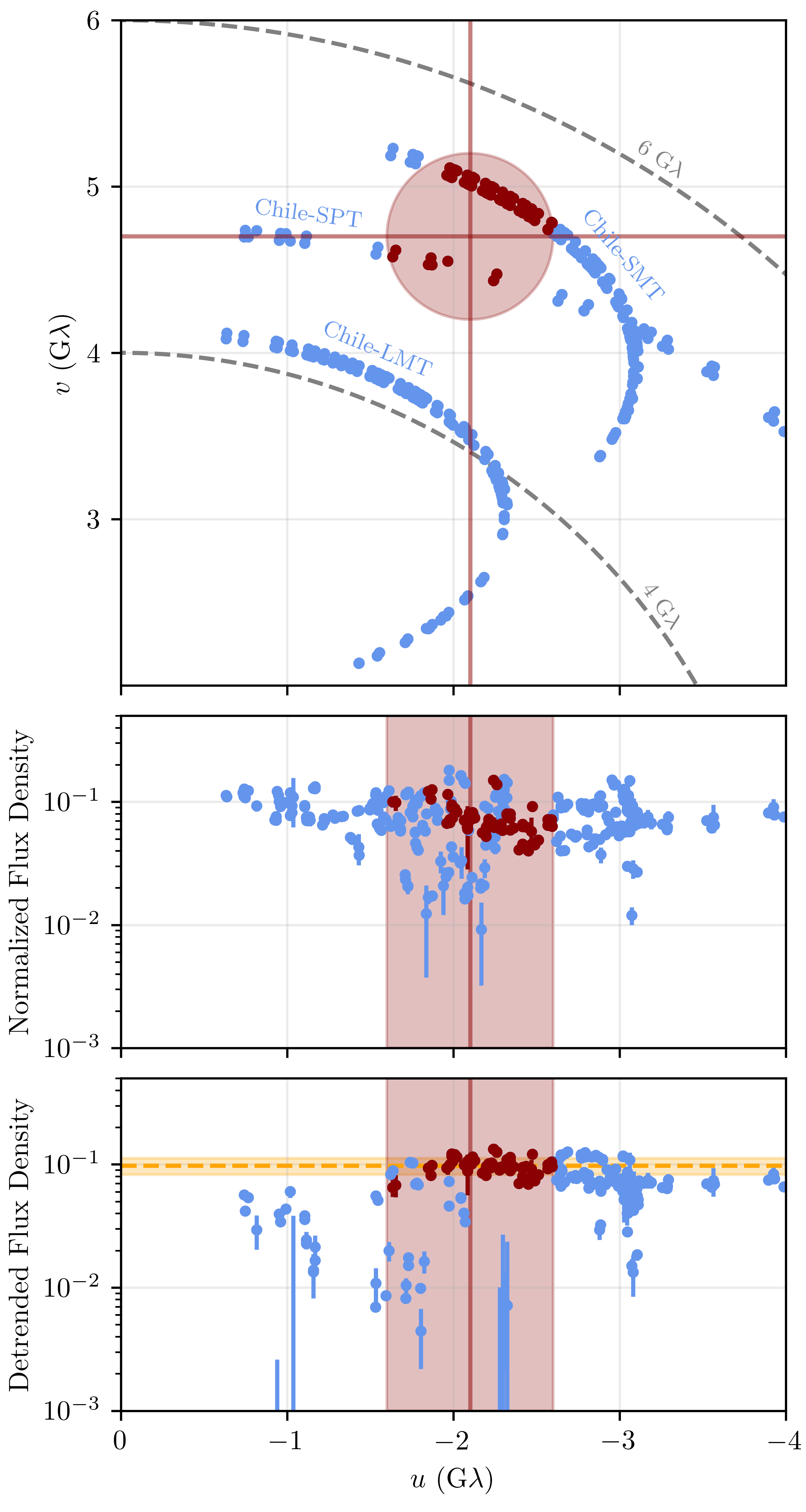}
  \end{center}
\caption{Illustration of detrended visibility amplitudes and associated variance estimate.
\textit{Top}: scan-averaged tracks in \uv-coordinates with a circular region of diameter 1\,\Gl superposed (red disk), centered at $(-2.1$\,\Gl, 4.7\,\Gl).  Scans within the region are dark red, while those outside are blue.
\textit{Middle}: light-curve-normalized visibility amplitudes as a function of $u$, projected in $v$ (limited to points within the top panel).
\textit{Bottom}: light-curve-normalized visibility amplitudes after detrending with a linear model defined by the scans within the 1\,\Gl circular region.  The estimated mean and standard deviation are shown by the orange dashed line and horizontal band.
}
\label{fig:premod_illustration}
\end{figure}

\begin{figure}
  \begin{center}
    \includegraphics[width=\columnwidth]{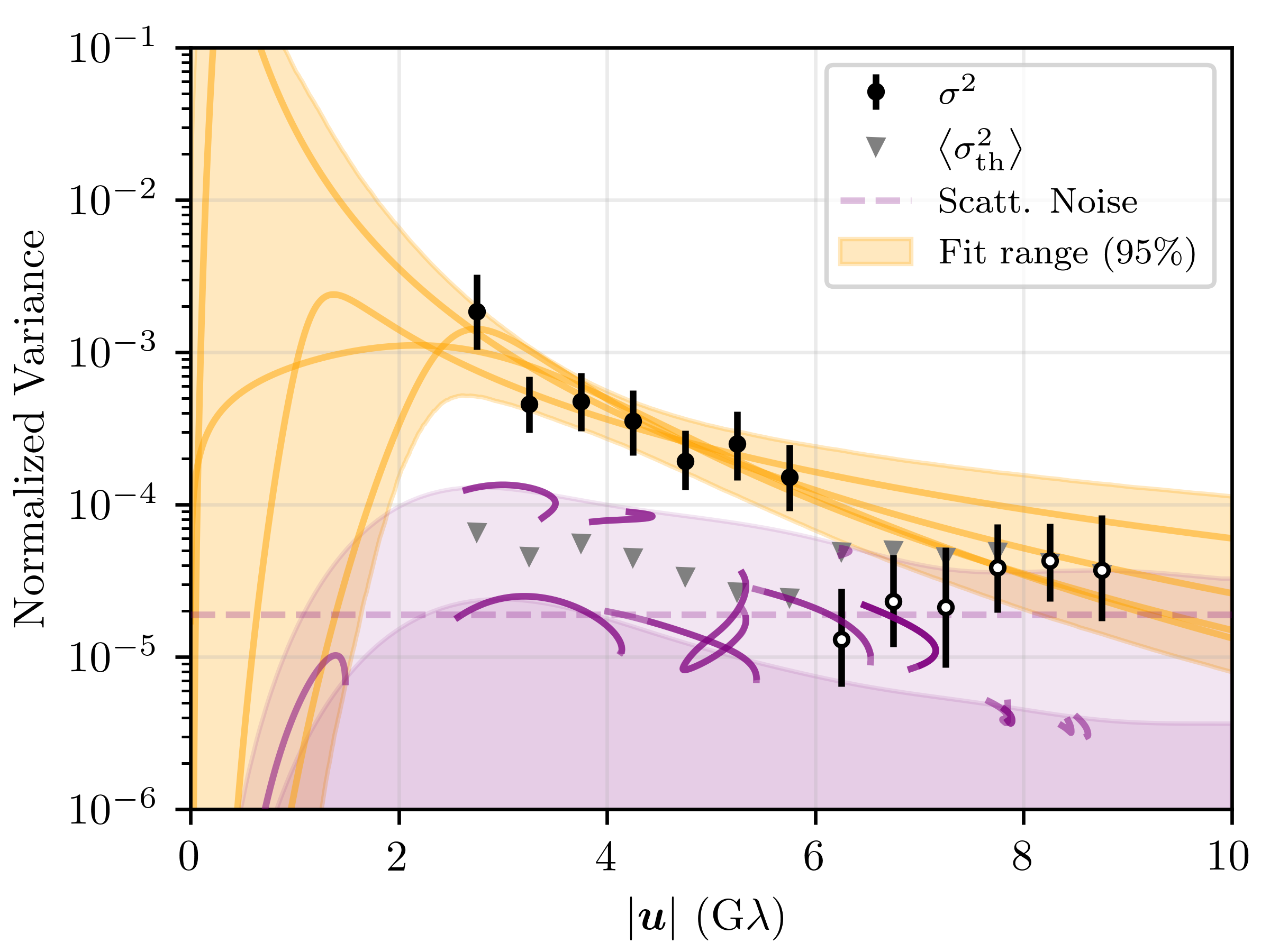}
  \end{center}
\caption{Model-agnostic estimate of the azimuthally averaged excess variance of the visibility amplitudes, after subtracting the variance from the reported statistical errors, as a function of baseline length.  Nonparametric estimates (filled and open black circles) are obtained across April 5, 6, 7, and 10 and using both high- and low-band data.  The filled black circles indicate significant detections of source variability, while the open black circles indicate variance measurements that are dominated by the other sources of uncertainty; only the former are used in the parametric fitting.  Uncertainties associated with the thermal errors, uncertain station gains, and polarization leakage are indicated by the error bars.  Azimuthally averaged thermal errors are shown by the gray triangles and provide an approximate lower limit on the range of accurate variance estimates. For comparison, the magnitudes of the variance induced by refractive scattering are shown in purple along the minor (top) and major (bottom) axes of the diffractive scattering kernel \citepalias[see Section 4 of][]{PaperIII}; the variance along individual tracks on April 7, as well as a $\sim$10\,mJy floor (assuming a fixed 2.5\,Jy total flux), is shown by the solid and dashed purple lines, respectively.  The orange band indicates the 95th-percentile range of fits to the filled variance estimates shown by filled points by the broken power law of the form in \autoref{eq:noise-model}, with a handful of specific examples shown explicitly.}
\label{fig:sevv1}
\end{figure}

The procedure is illustrated in \autoref{fig:premod_illustration}.  We again make use of the strong correlations induced by the finite source size, and for every location in the \uv-plane we consider only those data points falling within a circular region of diameter 1\,\Gl centered at that point (red circular region in the top panel of \autoref{fig:premod_illustration}).  Within each such region containing at least three data points, we linearly detrend the light-curve-normalized visibility amplitudes with respect to $u$ and $v$ to remove variations due to physical structure (bottom panel of \autoref{fig:premod_illustration}), and we compute the variance of the residuals.  This variance is then debiased to remove the contributions from the reported statistical errors, as described in \cite{NoiseModeling}.  Finally, the variances from all regions having a common baseline length are averaged to produce an azimuthally averaged set of variances.  The uncertainty in the variance estimates is obtained via Monte Carlo sampling of the unknown gains, leakage terms, and statistical errors.

\autoref{fig:sevv1} shows the results of applying this procedure to the \sgra data, with the normalized visibility amplitude variance measurements given by the black points.  For baselines shorter than $2.5$\,\Gl, the LMT calibration procedure precludes an accurate estimate of the variance, and thus these baselines have been excluded.  For baselines between ${\sim}2.5$ and 6\,\Gl in length, our empirical estimates of the noise exceed the typical contributions from statistical errors and refractive scattering, indicating the presence of an additional source of structural variability.  The degree of inferred variability is consistent with that seen in prior millimeter-VLBI data sets, which is discussed further in \autoref{app:premodeling_2013}.  For baselines longer than 6\,\Gl, our measurements are consistent with the degree of variability expected from the statistical uncertainties in the data; we thus do not directly constrain the source variability on these long baselines.

To characterize the variability behavior within the \uv-plane, we fit a broken power law of the form in \autoref{eq:noise-model} to the normalized variance measurements.  As indicated by the filled black circles in \autoref{fig:sevv1}, significant measurements exist only in the range of baselines with lengths $\sim$2.5--6\,\Gl; on baselines longer than 6\,\Gl, we are unable to distinguish the variability from its associated measurement uncertainties.  We thus perform the broken power-law fit only to the $\sim$2.5--6\,\Gl range of baselines, where we have significant measurements, and we find no evidence for a break in the power law in this region.  As a result, only an upper limit can be placed on $u_0$, and we are not able to constrain the short-baseline power-law index, $c$.
The range of permitted broken power-law fits is illustrated in \autoref{fig:sevv1} by the orange shaded region, with several samples from the posterior distribution explicitly plotted as orange lines.

\begin{figure}
  \begin{center}
    \includegraphics[width=\columnwidth]{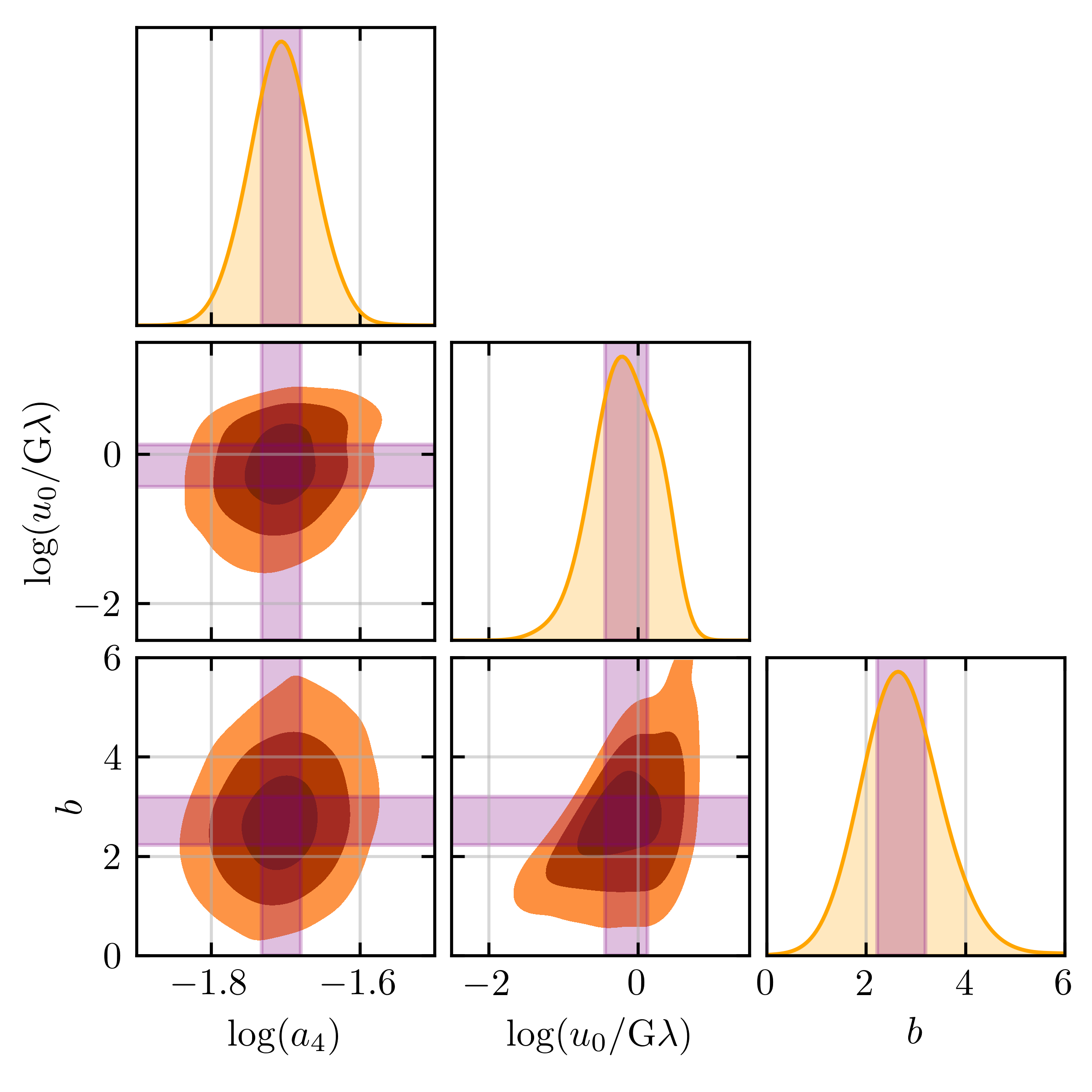}
  \end{center}
\caption{Joint posteriors of the constrained parameters after fitting a broken power law to the model-agnostic normalized variances estimates.  Because the amplitude is well constrained within the range of baseline lengths for which good estimates of the variability exist, we set the normalization at $|\boldsymbol{u}|=4~{\rm G}\lambda$, denoted as $a_4$. Contours show the enclosed 50th, 90th and 99th percentiles.  The purple bands indicate the ranges used as priors during the full-track modeling, associated with the interquartile ranges.}
\label{fig:bpl_tri}
\end{figure}

Because the location of the broken power-law break is poorly constrained, the parameters $u_0$ and $a$ (describing the location of the break and the amplitude of the power law at the break, respectively) are strongly correlated and highly uncertain.  However, it is clear from the orange shaded region in \autoref{fig:sevv1} that there is only a narrow range of variances permitted over the 2.5--6\,\Gl range of baselines over which the data are constraining.  We thus choose to characterize the amplitude of the excess variability noise at $|\boldsymbol{u}|=4\,{\rm G}\lambda$, which we denote as $a_4$. Joint posteriors for $a_4$, the break location $u_0$, and the long-baseline power-law index $b$ are shown in \autoref{fig:bpl_tri}.  These constraints are used to inform the prior distributions for the full-track geometric modeling described in \autoref{sec:FullTrackGeometricModeling}; the associated prior ranges on each parameter are indicated by the purple shaded regions in \autoref{fig:bpl_tri}.

\subsection{Description of variability mitigation approaches}
\label{sec:mitigation}

Having established the existence of structural variability and quantified its magnitude in the \sgra data, 
we now turn to strategies for mitigating its impact on downstream analyses.
We employ the light-curve-normalized visibility data, which eliminates large-scale variations and correlations by construction.
In principle, there are four methods that we might pursue to address the remaining structural variability:
\begin{enumerate}
\item Analyze time-averaged data products. 
\item Employ explicitly time-variable models.
\item Analyze short time segments of the data and combine the results afterward to characterize the average source structure.
\item Simultaneously reconstruct the average source structure and a statistical characterization of the structural variability.
\end{enumerate}
The first of these options is complicated substantially by the uncertain visibility phases, which limit our ability to coherently average the data on timescales longer than several minutes.  The second option can be employed either when a descriptive low-dimensional model for the source structure can be constructed \citep[e.g.,][]{Miller-Jones_2019,3c279_EHT}, or when there is sufficient \uv-coverage for nonparametric dynamical imaging algorithms to be successful \citep[e.g.,][]{Johnson_2017,Bouman_2018,Arras_2022}.  The latter approach is explored in the dynamical imaging analyses described in \citetalias{PaperIII}, ultimately demonstrating that the \sgra \uv-coverage is insufficient to permit unambiguous reconstructions of the variable source structure.

We dub the third option ``snapshot'' modeling, whereby a simple geometric model of the source structure is fit to segments of the data that are short enough in duration (${\lesssim}3$ minutes) for the impact of structural variability to be subdominant to other sources of visibility uncertainty (e.g., refractive scattering; see \autoref{fig:GRMHD_PSD}).  Though the data sparsity is exacerbated by restricting the reconstructions to only a single snapshot at a time, the model itself is also correspondingly restricted in its parameterization of the source structure.  The results of the fits to each individual snapshot are then combined across the entire data set, effectively averaging over the source variability.  Details of our snapshot modeling analyses as applied to \sgra are presented in \autoref{sec:SnapshotGeometricModeling}.

The fourth option we refer to as ``full-track'' modeling, which aims to simultaneously reconstruct both the average source structure and a set of parameters describing the contribution of the structural variability to the visibility data variances \citep{NoiseModeling}.  In contrast to the snapshot modeling, full-track modeling considers the entire data set at once and uses a parameterized ``variability noise'' model to appropriately modify the data uncertainties as part of the fitting procedure.  In this way, the full-track modeling retains access to sufficient \uv-coverage to permit fitting a nonparametric image model to the data \citepalias[see][]{PaperIII}, though in \autoref{sec:FullTrackGeometricModeling} we also pursue full-track geometric modeling to provide a cross-comparison with the results from the snapshot geometric modeling.  Our parameterization of the variability noise follows \autoref{eq:noise-model}, with the amplitude specified at a baseline length of 4\,\Gl as described in \autoref{sec:pre-modeling}.  A detailed description of our full-track modeling approach as applied to \sgra is presented in \autoref{sec:FullTrackGeometricModeling}.

Both the snapshot and full-track modeling approaches focus on describing the average source structure and treating the structural variability in a statistical manner.  This goal is formally mismatched with what the EHT data measure for a single day, which is instead a collection of complex visibilities that sample different instantaneous realizations of the intrinsic \sgra source structure.  The nature of this mismatch impacts the full-track analyses significantly.

The variability mitigation scheme employed by full-track modeling presumes that the variability may be modeled as excess uncorrelated fluctuations in the complex visibility data.  This assumption is well justified on timescales exceeding a few hours, but significant correlations between visibilities exist on shorter timescales.  Within a single day, subhour correlations that are localized in the \uv-plane can induce significant biases in the source structure reconstructed from the sparse EHT \uv-coverage of \sgra. The noise model is thus fundamentally misspecified for EHT data, with the level of misspecification increasing as shorter-in-time segments of data are analyzed; \autoref{app:SingleDayBias} describes pathological behavior that can arise when analyzing EHT \sgra data from only a single day.  While prominent artifacts associated with these subhour correlations are present in the April 6 reconstructions shown in \autoref{app:SingleDayBias} and \autoref{app:SingleDayFits}, we note that the underlying origin of these artifacts is no less present on April 7.

The impact of unmodeled correlations on the reconstructed source structure can be ameliorated by combining multiple days, which provides visibility samples associated with independent realizations of the source structure.  This additional sampling rapidly brings the statistical properties of the data into better agreement with the assumptions underpinning the full-track analyses; even the combination of just 2 days is often sufficient to mitigate the subhour correlations in analysis experiments that make use of GRMHD simulation data.  For this reason, we combine both the April 6 and April 7 data sets during the analysis of the \sgra data.  For comparison, \autoref{app:SingleDayFits} presents the results of equivalent analyses applied to the April 6 and April 7 data sets individually.

%% file: ring.tex
\section{Ring characterization and calibration}\label{sec:Ring}

We have a strong prior expectation -- from both prior millimeter-VLBI observations of a different black hole \citepalias[i.e., the EHT images of \m87; see][]{M87PaperIV} and theoretical simulations of the accretion flow around \sgra itself \citepalias[see][]{PaperV} -- that \sgra ought to contain a ring of emission, and we thus aim to determine the characteristics of the ring-like image structure that best describes the \sgra data.  In this section, we first review the evidence from the \sgra data for a ring-like image structure, and we then present a geometric model for fitting parameters of interest and describe our procedure for bringing ring size measurements made using different techniques to a common physical scale.

\subsection{Evidence for a ring}

In reconstructing images of \sgra, \citetalias{PaperIII} explores a large space of imaging algorithms and associated assumptions. The resulting ``top sets'' of images
contain primarily ``ring-like'' image structures, though a small fraction of the images are morphologically ambiguous.  These ``nonring'' images still nominally provide a reasonable fit to the \sgra data and so are not ruled out from the \citetalias{PaperIII} results.

\begin{figure*}[t]
  \begin{center}
    \includegraphics[width=\textwidth]{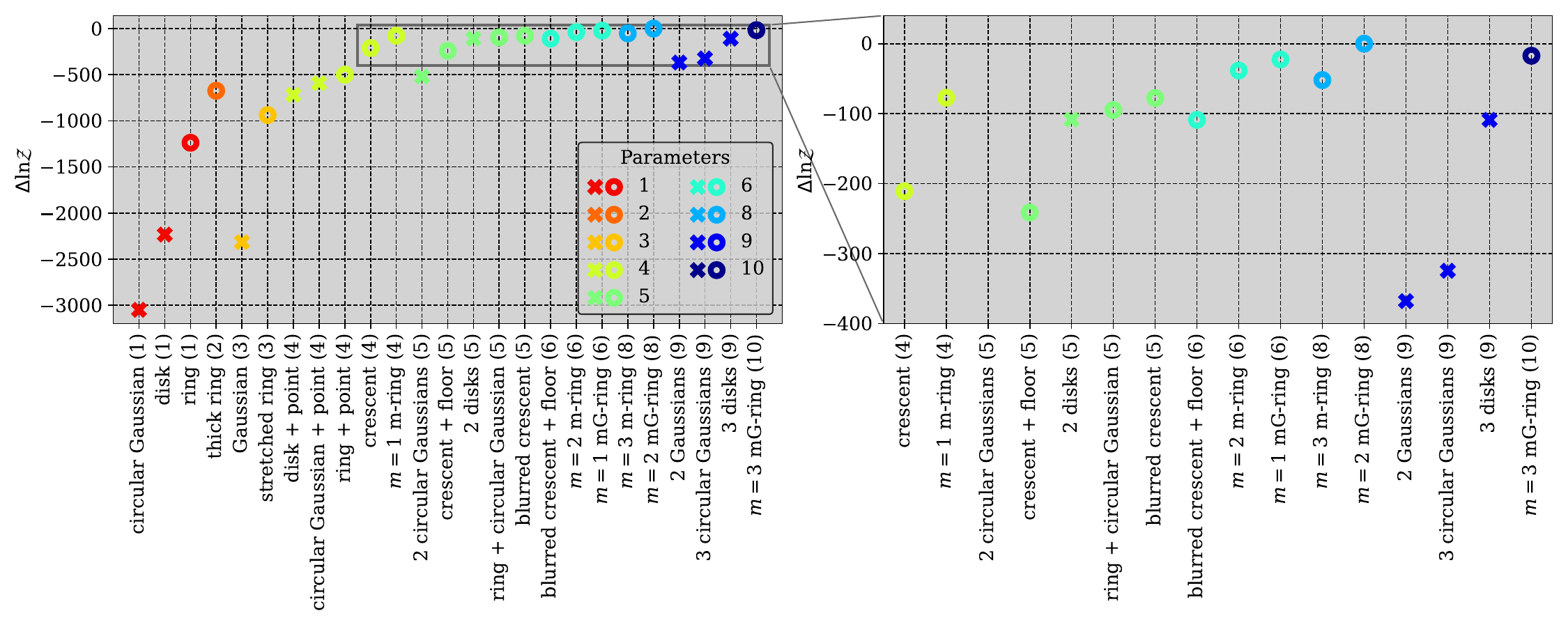}
  \end{center}
\caption{Comparison of the relative Bayesian evidence, $\Delta \ln \mathcal{Z}$, for a series of increasingly complex geometric models fitted using closure amplitudes and closure phases within the snapshot modeling formalism described in detail in \autoref{sec:SnapshotGeometricModeling}.  The fits have been carried out using \ehtim on the the HOPS April 7 \sgra data, and each point in the figure is colored according to the number of free parameters in the model; the number of free parameters in each model is also indicated in the horizontal axis labels.  The panel on the right shows a zoom-in to the highest-evidence region of the left panel.  Ring-like models are indicated with circles, and nonring models are indicated with crosses.  All Bayesian evidence values are quoted relative to the highest value attained across all models. The parameter counts reflect the fact that all models are normalized to have unit total flux density and are centered at the image origin. The crescent model consists of a smaller disk subtracted from an offset larger disk. In the crescent+floor model, the smaller disk may have a nonzero flux density. The m-ring and mG-ring models are defined in \autoref{sec:mring}.  The maximum value of $\ln\mathcal{Z}$ among the models explored in this figure is obtained for an $m=2$ \model model, in agreement with the \texttt{DPI} analysis described in \autoref{sec:SnapshotGeometricModeling}.}
\label{fig:model_selection_snapshot}
\end{figure*}

We can quantify the preference for a ring-like image structure by fitting the data with a set of simple geometrical models. Employing the snapshot geometric modeling technique detailed in \autoref{sec:SnapshotGeometricModeling}, we compare the Bayesian evidence, $\mathcal{Z}$, between these different geometric models. 
The value of $\mathcal{Z}$ serves as a model comparison metric that naturally balances improvements in fit quality against increases in model complexity, with larger values of $\mathcal{Z}$ indicating preferred models \citep[see, e.g.,][]{Trotta_2008}. \autoref{fig:model_selection_snapshot} shows the results of a survey over simple geometric models with varying complexity, captured here by the number of parameters required to specify the model.  At all levels of complexity, ring-like models outperform the other tested models.  This disparity is most stark for the simplest models but continues to hold as the models increase in complexity.

The remainder of this paper proceeds with analyses that presuppose a ring-like emission structure for \sgra.

\subsection{Salient features in the context of a ring model} \label{sec:SalientFeatures}

\begin{figure*}[t]
  \centering
  \includegraphics[width=\textwidth]{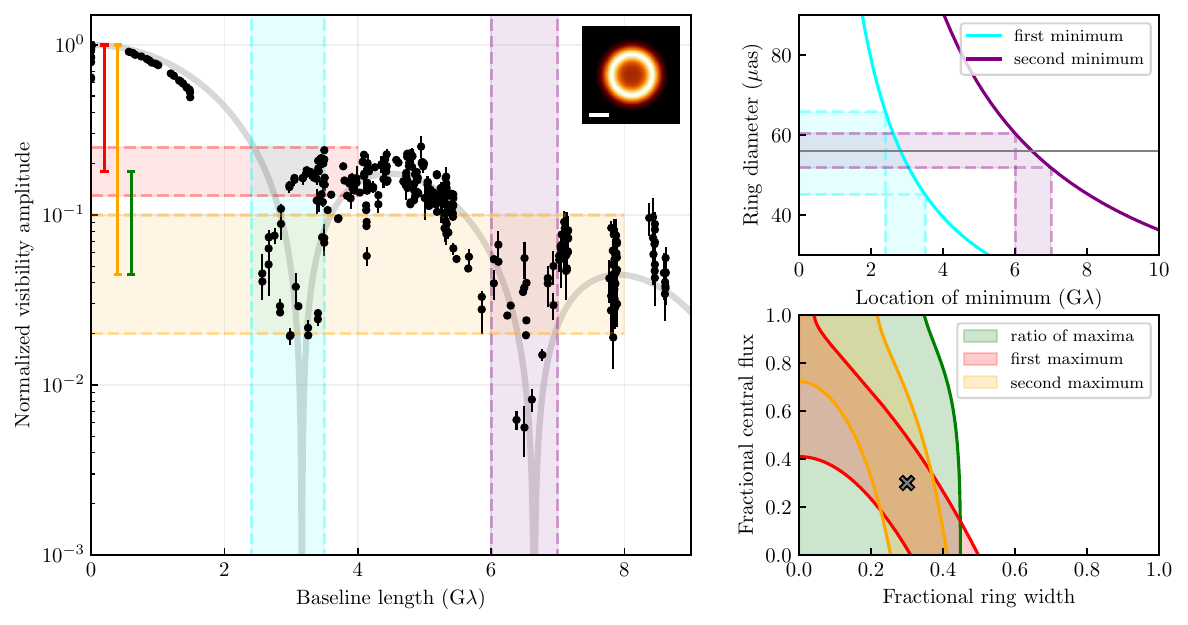}
  \caption{An illustration of how the observed visibilities constrain the source structure, within the context of a simple geometric model (see \autoref{sec:SalientFeatures}).  The left panel shows the debiased visibility amplitudes from April 7 (black points) after normalizing by the light curve, averaging in time across scans, and deconvolving the diffractive scattering kernel.  Only data points with a signal-to-noise ratio greater than 3 are shown.  The ranges of baseline lengths corresponding to the locations of the first and second visibility minima are highlighted in cyan and purple, respectively.  The locations of these minima constrain the diameter of an emitting ring, as shown in the top right panel.  Similarly, the ranges of normalized amplitudes corresponding to the first and second visibility maxima are highlighted in red and orange, respectively.  Both the absolute fractional amplitudes and their relative values (shown in green) constrain a combination of the fractional ring width and the fractional central flux.  The ``best-fit'' model visibility amplitudes are shown by the gray curve in the left panel, with the corresponding parameters marked by the gray line in the top right panel and the gray cross in the bottom right panel.  The image structure corresponding to this model is shown in the upper right corner of the left panel, with a 20\,\uas scale bar shown in white.}
  \label{fig:salient_features}
\end{figure*}

The overall structure of the \sgra visibility amplitudes (see the left panel of \autoref{fig:salient_features}) exhibits at least three distinct regions:

\begin{enumerate}
  \item A ``short-baseline'' region containing baselines shorter than $\sim$2\,\Gl.  The effects of data calibration and pre-processing -- particularly the light-curve normalization and LMT calibration procedures \citepalias{PaperIII} -- are evident in the unit total flux density and the Gaussian structure of the visibility amplitudes in this region.
  \item An ``intermediate-baseline'' region containing baselines between $\sim$2 and 6\,\Gl.  The visibility amplitudes in this region exhibit a general rise and then fall with increasing baseline length, peaking at a flux density approximately $\sim$20\% of the total at a baseline length of $\sim$4\,\Gl.
  \item A ``long-baseline'' region containing baselines with lengths in excess of $\sim$6\,\Gl.  The visibility amplitudes in this region generally rise with increasing baseline length from a deep minimum near $\sim$6.5\,\Gl, approximately flattening out at longer baselines to a level that is $\sim$3\%--10\% of the total flux density.
\end{enumerate}

The visibility amplitudes exhibit indications of asymmetric source structure, particularly on baselines with lengths of $\sim$3\,\Gl that fall near the first minimum.  Here, the baselines between the SMT and \hawaii stations (oriented approximately in the east--west direction) have systematically higher correlated flux densities than the similar-length baselines between the LMT and Chile stations (oriented approximately in the north--south direction).  The implication for the source morphology is that we would expect to see more symmetric structure in the north--south than in the east--west direction.  Detailed geometric modeling analyses that are able to capture this asymmetry are described in \autoref{sec:SnapshotGeometricModeling} and \autoref{sec:FullTrackGeometricModeling}; here, we consider only a simple azimuthally symmetric toy model that captures some salient features of interest.

We attempt to understand the visibility behavior in light of expectations for a ring-like emitting structure.  Specifically, we consider a geometric construction whereby an infinitesimally thin circular ring bordering an inner disk of emission is convolved with a Gaussian blurring kernel.  The visibility function $V$ produced by such an emission structure is given by
\begin{equation}
\begin{aligned}
    V & = F_0 V_{\text{Gauss}} \left[ f_d V_{\text{disk}} + (1 - f_d) V_{\text{ring}} \right] \\
    & = F_0 \exp\left( - \frac{w^2 \xi^2}{4 \ln(2)} \right) \left[ \frac{2 f_d}{\xi} J_1(\xi) + (1 - f_d) J_0(\xi) \right] ,
\end{aligned}
\end{equation}
where $F_0$ is the total flux in the image, $f_d$ is the fraction of that flux that is contained in the disk component, $w = W / d$ is a fractional ring width, $W$ is the FWHM of the Gaussian convolving kernel, $d$ is the diameter of the ring and disk components, $\xi \equiv \pi |\boldsymbol{u}| d$ is a normalized radial visibility-domain coordinate, and $J_n(\xi)$ is a Bessel function of the first kind of order $n$.

The three regions of \sgra data identified above are separated by apparent minima in the visibility amplitudes, and they can be approximately characterized by the baseline locations of those minima and the peak flux density levels achieved at the visibility maxima between them. \autoref{fig:salient_features} illustrates how this characterization manifests as constraints on the defining parameters of the geometric toy model.  The cyan and purple shaded regions in the left panel indicate the approximate ranges of baseline lengths corresponding to the locations of the first and second visibility minima, respectively.  The locations of these minima constrain the diameter of the emitting structure, as shown in the top right panel of \autoref{fig:salient_features}.  To be consistent with both a first minimum falling between $\sim$2.5 and 3.5\,\Gl and a second minimum falling between $\sim$6 and 7\,\Gl, the emitting region must be between $\sim$50 and 60\,\uas across.  The amplitudes of two visibility maxima -- one falling between the first and second visibility minima, and the second following the second minimum -- constrain a combination of the fractional disk flux $f_d$ and the fractional ring width $w$.  The bottom right panel of \autoref{fig:salient_features} shows the constraints from the first and second visibility maxima in red and orange, respectively, and from the ratio of the two in green.

Taken together, even these few, simple, and only modestly constrained visibility features result in a rather narrow permitted range of model parameter values for $d$, $w$, and $f_d$; an example of a ``best-fit'' model from within the permitted range is shown by the gray curve in the left panel of \autoref{fig:salient_features}.  However, we stress that the above constraints only strictly hold within the context of the specific toy model used to derive them.  More general and robust constraints on the emission structure require a model that can accommodate more than just the gross features; such models are produced as part of the imaging (\citetalias{PaperIII} and \autoref{sec:ImageDomain}) and geometric modeling analyses (see \autoref{sec:SnapshotGeometricModeling} and \autoref{sec:FullTrackGeometricModeling}) carried out in this paper series.

\subsection{Geometric ring model specification}\label{sec:mring}

The ring-like images reconstructed in \citetalias{PaperIII} are not azimuthally symmetric, but instead show pronounced azimuthal brightness variations that we would like to capture in our geometric modeling analyses.  In this section, we specify the ``\model'' model that we use in \autoref{sec:SnapshotGeometricModeling} and \autoref{sec:FullTrackGeometricModeling} to quantify the morphological properties of the observed \sgra emission.

\subsubsection{Image-domain representation of \model model}

Adopting the construction developed by \cite{Johnson_2020}, we can model an infinitesimally thin circular ring with azimuthal brightness variations using a sum over angular Fourier modes indexed by integer $k$,

\begin{align}
    I_{\text{ring}}(r,\phi)=\frac{F_{\rm ring}}{\pi d} \, \delta\!\left(r-\frac{d}{2}\right) \sum_{k=-m}^m\beta_k e^{i k \phi}. \label{eqn:RingImage}
\end{align}

\noindent Here $r$ is the image radial coordinate, $\phi$ is the azimuthal coordinate (east of north), $d$ is the ring diameter, $\{ \beta_k \}$ are the set of (dimensionless) complex azimuthal mode coefficients, and $m$ sets the order of the expansion.  Because the image is real, $\beta_{-k} = \beta_k^\ast$; we enforce $\beta_0 \equiv 1$ so that $F_{\rm ring}$ sets the total flux density of the ring.  Given that the images from \citetalias{PaperIII} show a ring of radius ${\sim}25$\,\uas and the diffraction-limited EHT resolution is ${\sim}20$\,\uas, we expect the data to primarily constrain ring modes with $m \lsim \pi \left(25/20\right) \approx 4$.  We refer to this asymmetric ring as an ``m-ring'' of order $m$.

\begin{figure}
  \begin{center}
    \includegraphics[width=\columnwidth]{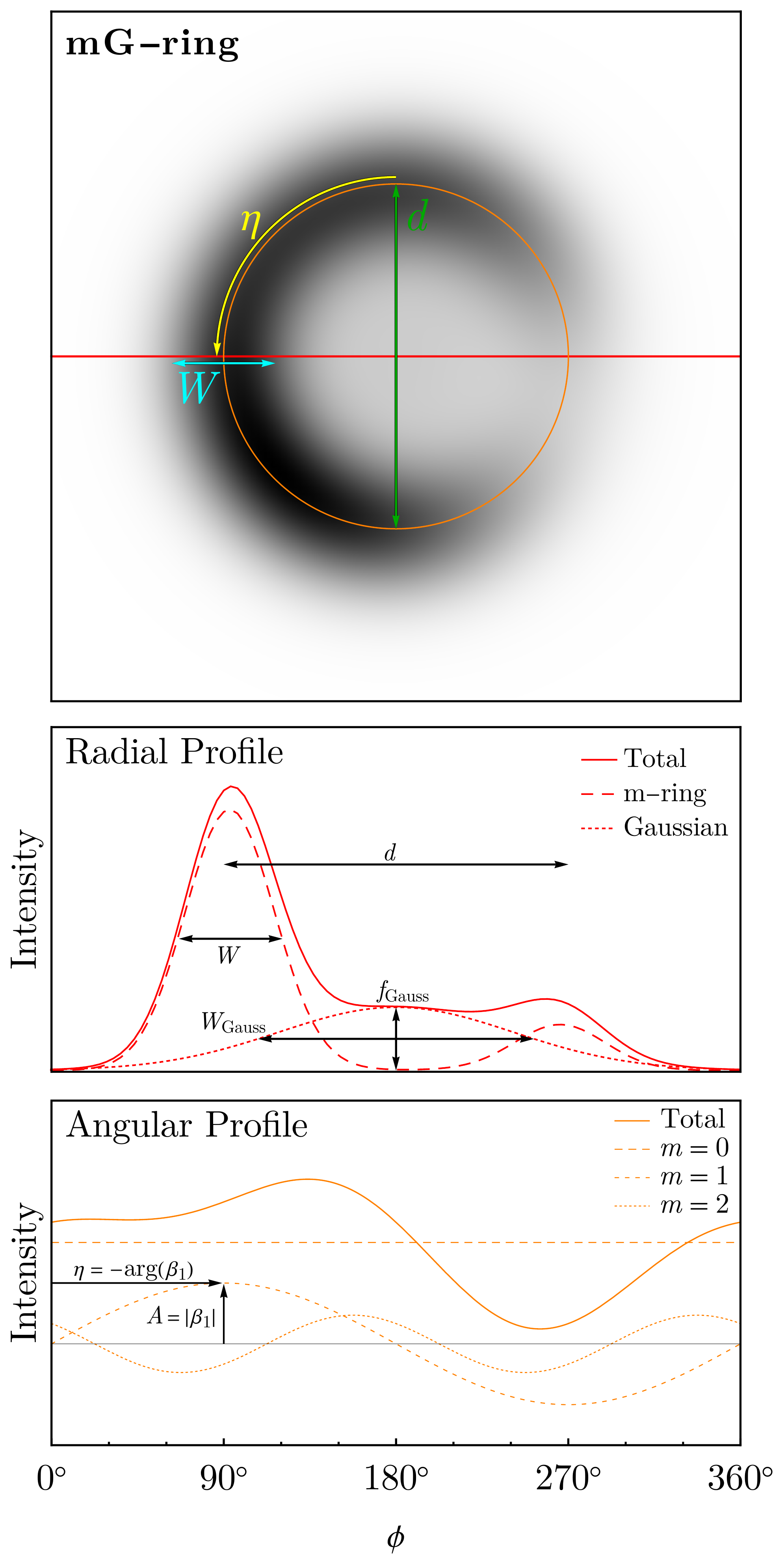}
  \end{center}
\caption{Example \model model. The model consists of an m-ring of diameter $d$, with azimuthal variations determined by Fourier coefficients $\beta_m$ and with thickness determined by convolution with a circular Gaussian of FWHM $W$. The \model model also includes a circular Gaussian of FWHM $W_{\text{Gauss}}$ and fractional relative flux density $f_{\rm Gauss}$ (\autoref{eqn:FractionalGaussFlux}). The position angle $\eta$ of the ring is determined by the phase of the \mbox{$m=1$} mode (\autoref{eqn:PositionAngle}), while the magnitude of its asymmetry $A$ is determined by the amplitude of the \mbox{$m=1$} mode (\autoref{eqn:Asymmetry}).  The red curve in the middle panel shows a radial profile in the horizontal direction; the orange curve in the bottom panel shows the azimuthal profile and its decomposition into its three modes.  The plotted model has $w = 0.3$, $f_{\rm Gauss} = 0.2$, $W_{\text{Gauss}}/d=0.8$, $\beta_1 = -0.3i$, and $\beta_2 = 0.1+0.1i$.}
\label{fig:mring_cartoon}
\end{figure}

For the purposes of constraining additional image structures, we augment this m-ring in two ways.  First, we convolve the m-ring with a circular Gaussian kernel of FWHM $W$,
\begin{equation}
I_{\text{ring}}(r,\phi;W) = I_{\text{ring}}(r,\phi) \ast \left[\frac{4 \ln(2)}{\pi W^2} \exp\!\left( - \frac{4 \ln(2) r^2}{W^2} \right)\right]. \label{eqn:ModelImage}
\end{equation}
Second, we add a circular Gaussian component that is concentric with the ring, which serves to provide a nonzero brightness floor interior to the ring. The Gaussian component has a total flux density of $F_{\text{Gauss}}$ and an FWHM of $W_{\text{Gauss}}$, 
\begin{align}
I_{\text{Gauss}}(r,\phi) = \frac{4 \ln(2) F_{\text{Gauss}}}{\pi W_{\text{Gauss}}^2} \exp\!\left( - \frac{4 \ln(2) r^2}{W_{\text{Gauss}}^2} \right). \label{eqn:GaussImage}
\end{align}
We refer to the resulting composite model $I(r,\phi)$, where
\begin{equation}
    I(r,\phi) = I_{\text{ring}}(r,\phi; W) + I_{\text{Gauss}}(r,\phi) ,
\end{equation}
\noindent as an ``\model.''  An example \model is shown in \autoref{fig:mring_cartoon}.

An \model of order $m$ has $5 + 2 m$ model parameters: the flux density in the ring ($F_{\text{ring}}$), the diameter of the ring ($d$), the flux density in the central Gaussian ($F_{\text{Gauss}}$), the FWHM of the central Gaussian ($W_{\text{Gauss}}$), the FWHM of the ring convolving kernel ($W$), and two parameters for each complex Fourier coefficient $\beta_k$ with $1\leq k \leq m$.

\subsubsection{Visibility-domain representation of \model model}

To aid in efficient parameter space exploration, the \model model is intentionally constructed using components and transformations that permit analytic Fourier transformations. The Fourier transform of the m-ring image (\autoref{eqn:RingImage}) is given by
\begin{equation}
V_{\text{ring}}(|\boldsymbol{u}|,\phi_u) = F_{\text{ring}} \sum_{k=-m}^m \beta_k J_k(\pi |\boldsymbol{u}| d) e^{ik (\phi_u - \pi/2)} ,
\end{equation}
where $(|\boldsymbol{u}|,\phi_u)$ are polar coordinates in the Fourier domain.
The convolution with a circular Gaussian in the image plane corresponds to multiplication of this function by the Fourier transform of the convolving kernel,
\begin{equation}
    V_{\text{ring}}(|\boldsymbol{u}|,\phi_u;W) = \exp\left( -\frac{\pi^2 W^2 |\boldsymbol{u}|^2}{4 \ln(2)} \right) V_{\text{ring}}(|\boldsymbol{u}|,\phi_u) .
\end{equation}
\noindent The Fourier transform of the Gaussian image (\autoref{eqn:GaussImage}) is given by
\begin{equation}
V_{\text{Gauss}}(|\boldsymbol{u}|,\phi_u) = F_{\text{Gauss}} \exp\left( - \frac{\pi^2 W_{\text{Gauss}}^2 |\boldsymbol{u}|^2}{4 \ln(2)} \right) .
\end{equation}
\noindent By the linearity of the Fourier transform, the visibility-domain representation of the \model model is then simply the sum of these two components,
\begin{equation}
    V(|\boldsymbol{u}|,\phi_u) = V_{\text{ring}}(|\boldsymbol{u}|,\phi_u;W) + V_{\text{Gauss}}(|\boldsymbol{u}|,\phi_u) .
\end{equation}

When interpreting model-fitting results in subsequent sections, we are interested in a number of derivative quantities.  We will typically work with the fractional thickness of the ring, $w$, defined to be
\begin{equation}
w \equiv \frac{W}{d} .
\end{equation}
\noindent Similarly, we are typically interested in fractional representations of flux densities. We define
\begin{equation*}
F_0 \equiv F_{\text{ring}} + F_{\text{Gauss}}
\end{equation*}
\noindent to be the total flux density, and then
\begin{equation}
f_{\text{ring}} \equiv \frac{F_{\text{ring}}}{F_0}
\end{equation}
\noindent and
\begin{equation}
f_{\text{Gauss}} \equiv \frac{F_{\text{Gauss}}}{F_0} \label{eqn:FractionalGaussFlux}
\end{equation}
\noindent are the fraction of the total flux density that is contained in the ring and in the Gaussian components, respectively.  Note that $F_0$ is typically close to or fixed to unity as a consequence of normalizing the data by the light curve.  We also define a fractional central flux as
\begin{equation}
f_{\text{c}} \equiv \frac{F_{\text{Gauss}}(r<d/2)}{F_{\text{ring}} + F_{\text{Gauss}}(r<d/2)} , \label{eqn:FractionalCentralFlux}
\end{equation}
\noindent where $F_{\text{Gauss}}(r<d/2)$ is the integrated flux density of the central Gaussian component interior to the ring radius, given by
\begin{equation}
F_{\text{Gauss}}(r< d/2) = F_{\text{Gauss}}  \left[ 1 - \exp\left( - \frac{d^2 \ln(2)}{W_{\text{Gauss}}^2} \right) \right] .
\end{equation}
\noindent Following \citetalias{M87PaperIV}, the m-ring position angle $\eta$ and degree of azimuthal asymmetry $A$ are both determined by the coefficient of the $m=1$ mode,
\begin{align}
    \eta &\equiv {\rm arg}\left( \int_0^{2\pi} I(\phi) e^{i \phi} d\phi \right) \nonumber \\
    &= -{\rm arg}\left( \beta_1 \right) , \label{eqn:PositionAngle}\\
    A &\equiv \frac{ \left| \int_0^{2\pi} I(\phi) e^{i \phi} d\phi \right| }{\int_0^{2\pi} I(\phi) e^{i \phi} d\phi} \nonumber \\
    &= |\beta_1| . \label{eqn:Asymmetry}
\end{align}
\noindent A number of these derivative quantities are illustrated in the example \model shown in \autoref{fig:mring_cartoon}.

\subsection{Calibrating ring size measurements to a common physical scale} \label{sec:Calibration}

The parameters returned by the geometric modeling and feature extraction analyses used in this paper to describe the \sgra emission structure do not correspond directly to physical quantities.  Instead, the relationship between measured and physical quantities must be calibrated using data for which we know the correct underlying physical system's defining parameters.  For ring size measurements, the associated physical quantity of interest is related to the angular size of the gravitational radius,
\begin{equation}
\theta_g = \frac{G M}{c^2 D} . \label{eqn:ThetaG}
\end{equation}
which sets the absolute scale of the system.

Under the assumption that the emission near the black hole originates from some ``typical'' radius, a measurement of the angular diameter $d$ of the emitting region will be related to $\theta_g$ by a scaling factor $\alpha$,
\begin{equation}
d = \alpha \theta_g . \label{eqn:Alpha}
\end{equation}
If the observations were directly sensitive to the critical curve bounding the black hole shadow, then $\alpha$ could be determined analytically and would take on a value ranging from $\sim$9.6 to 10.4 depending on the black hole spin and inclination \citep{Bardeen_1973,Takahashi_2004}.  For more realistic emission structures and measurement strategies, the value of $\alpha$ cannot be determined from first principles and must instead be calibrated.

Our $\alpha$ calibration strategy generally follows the procedure developed in \citetalias{M87PaperVI}.  Using the library of GRMHD simulations described in \citetalias{PaperV}, we generate a suite of 100 synthetic data sets that emulate the cadence and sensitivity of the 2017 EHT observations and that contain a realistic character and magnitude of data corruption; \autoref{sec:MoDSyntheticData} describes the generation of these synthetic data sets.  In the analyses described in Sections \ref{sec:ImageDomain}, \ref{sec:SnapshotGeometricModeling}, and \ref{sec:FullTrackGeometricModeling}, 90 of these 100 synthetic data sets are used to derive the $\alpha$ calibration for each analysis pathway, while the remaining 10 data sets are used to validate the calibration.

After carrying out ring size measurements on each of the data sets in the suite, we determine $\alpha$ (for each specific combination of data set and measurement technique) by dividing the measured ring diameter by the known value of $\theta_g$ (per \autoref{eqn:Alpha}).  For a given measurement technique, the distribution of $\alpha$ values that results from applying this procedure to the entire suite of synthetic data sets then provides a measure of $\alpha$ and its theoretical uncertainty.  The $\alpha$ value associated with each measurement technique can then be used to translate \sgra ring size measurements into their corresponding $\theta_g$ constraints.  We note that this calibration strategy assumes that the images contained in the GRMHD library provide a reliable representation of the emission structure in the vicinity of \sgra; a separate calibration strategy that relaxes this GRMHD assumption is presented in \citetalias{PaperVI}.

\autoref{app:Validation} describes elements of the calibration and validation strategy that are specific to each of the analysis pathways detailed in Sections \ref{sec:ImageDomain}, \ref{sec:SnapshotGeometricModeling}, and \ref{sec:FullTrackGeometricModeling}.

%% file: image_domain.tex
\section{Image-domain feature extraction}
\label{sec:ImageDomain}

The imaging carried out in \citetalias{PaperIII} permits very flexible emission structures to be reconstructed from the \sgra data, but the majority of these images exhibit a ring-like morphology whose properties we seek to characterize.  In this section, we describe our image-domain feature extraction (IDFE) procedure, which uses a topological classification scheme to identify the presence of a ring-like structure in an image and quantifies the parameters that best describe this ring using two different algorithms.  We apply this IDFE procedure to the \sgra image reconstructions from \citetalias{PaperIII}.

\subsection{Imaging methods and products}
\label{subsec:ImageDomain_ImagingMethods}

\begin{figure*}[t]
    \centering
    \includegraphics[width=\columnwidth]{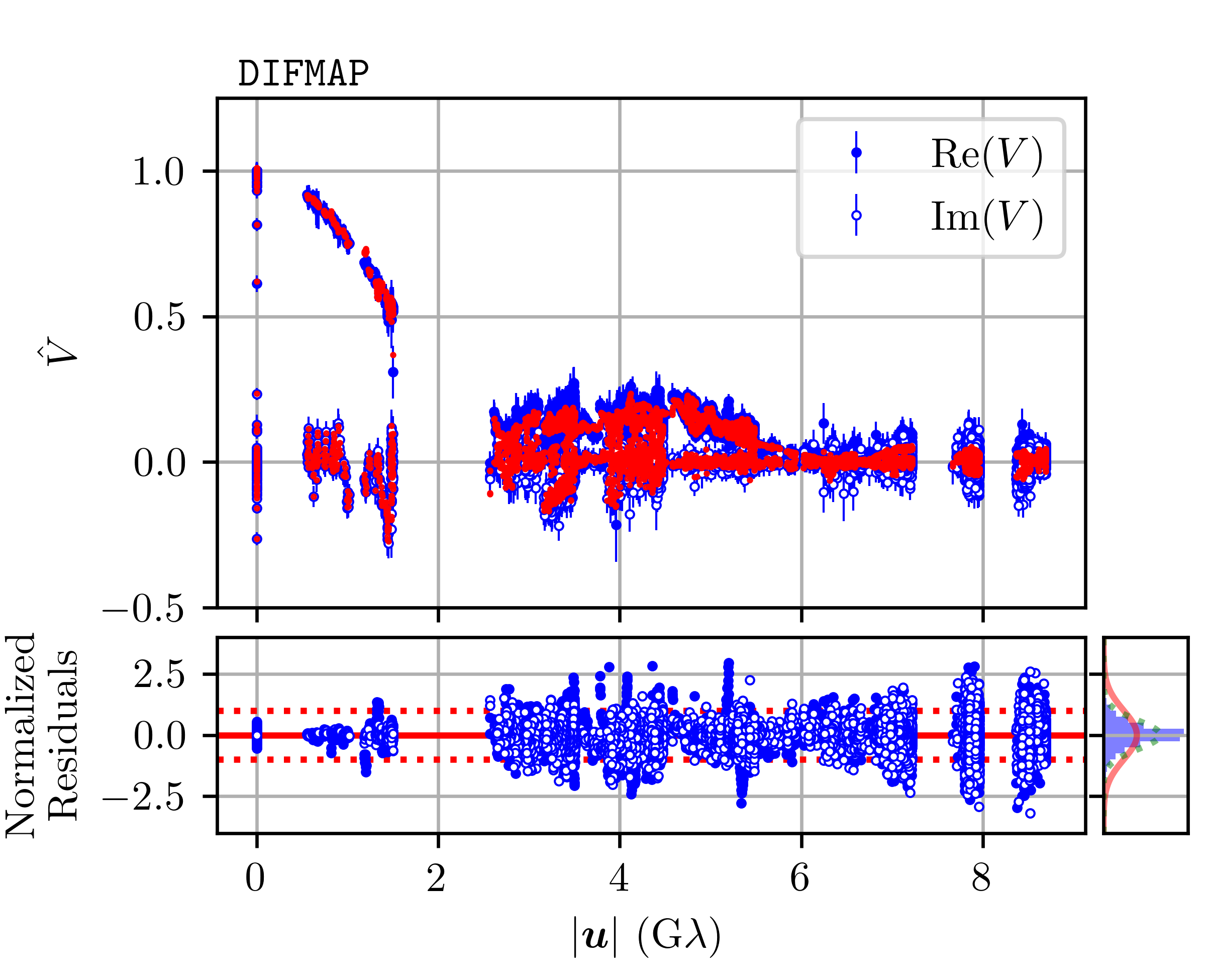}
    \includegraphics[width=\columnwidth]{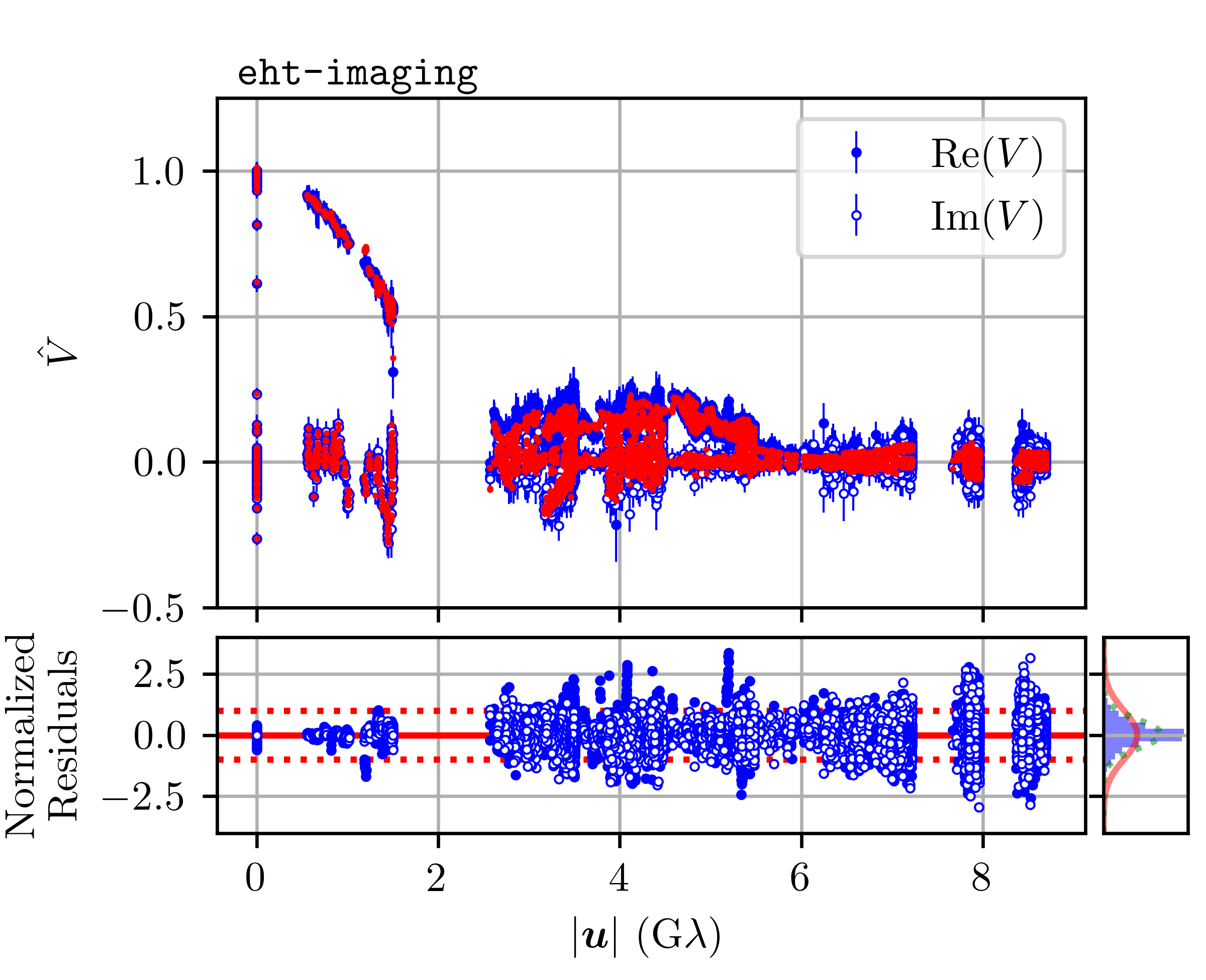} \\
    \includegraphics[width=\columnwidth]{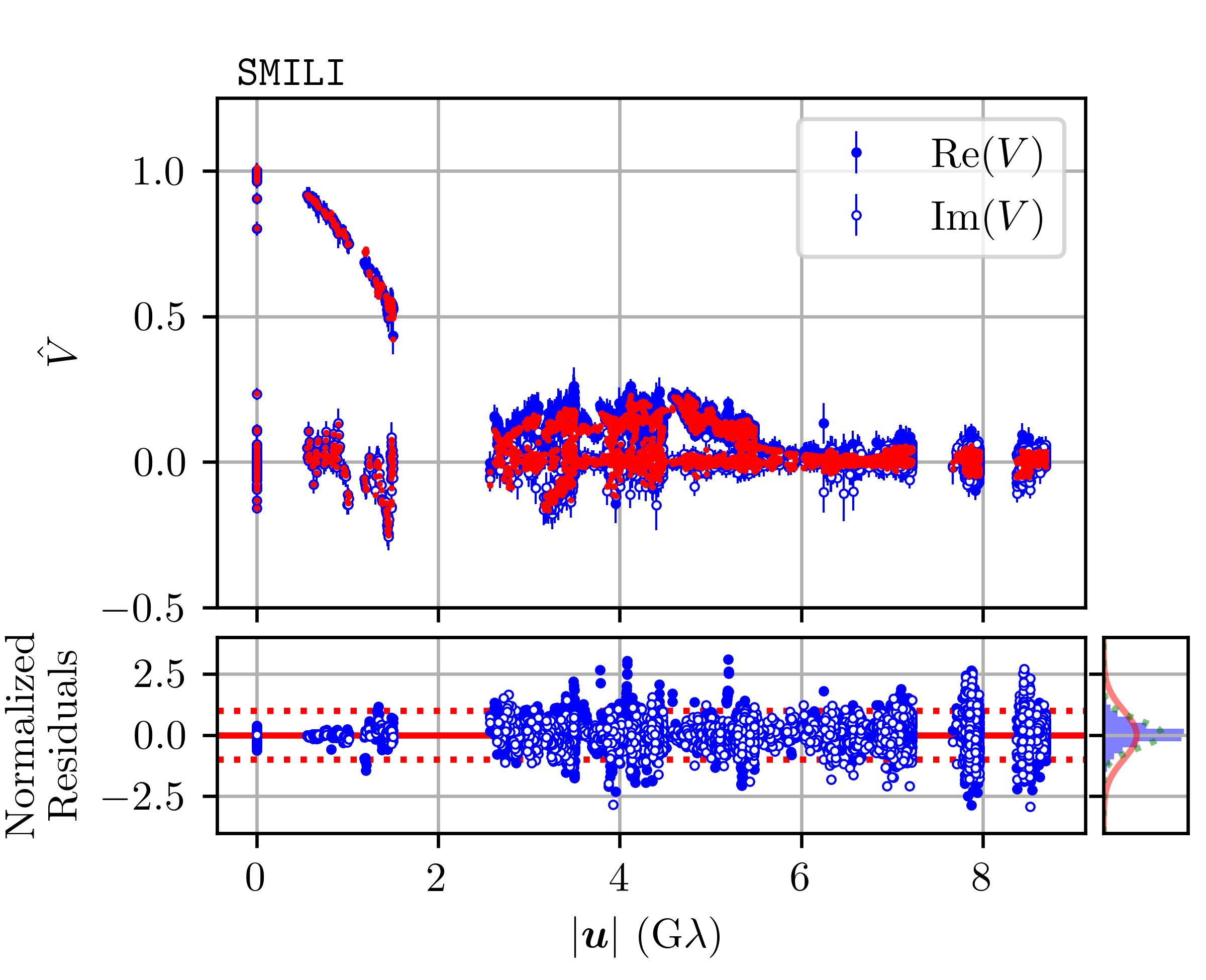}
    \includegraphics[width=\columnwidth]{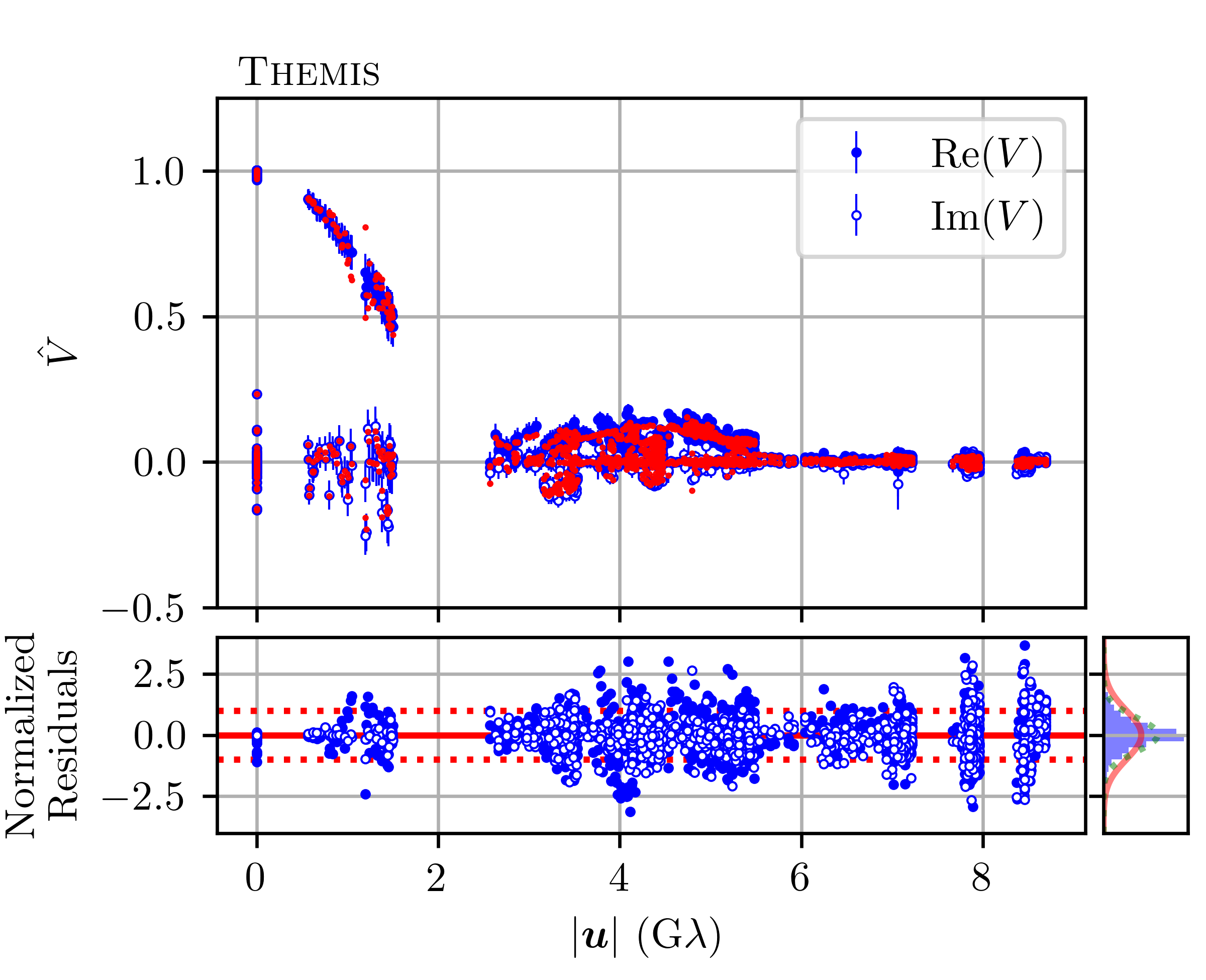}
    \caption{Representative examples of imaging results for each of the four imaging pipelines used in \citetalias{PaperIII}; \difmap is shown in the top left panel, \ehtim in the top right panel, \smili in the bottom left panel, and \themis in the bottom right panel.  The top section of each panel shows the light-curve-normalized complex visibility data (in blue) as a function of baseline length; the light-curve-normalized visibilities are denoted as $\hat{V}$.  The real parts of the complex visibilities are plotted as filled markers, and the imaginary parts of the complex visibilities are plotted as open markers; the corresponding model visibilities are overplotted as red points.  The plotted data have been through the pre-analysis and pre-imaging calibration procedures described in \citetalias{M87PaperIII}, \citetalias{PaperII}, and \citetalias{PaperIII}.  The bottom section of each panel shows the normalized residuals -- i.e., the difference between the model and data visibilities, normalized by the data uncertainties -- as a function of baseline length.  The solid red horizontal line marks zero residual, and the two dotted horizontal red lines mark $\pm$ one standard deviation.  The blue histogram on the right side of each bottom panel shows the distribution of normalized residuals, with the solid red curve showing a unit-variance normal distribution and the dotted green curve showing a normal distribution with variance equal to that of the normalized residuals.  We note that the visibilities for the \difmap, \ehtim, and \smili pipelines have been ``descattered'' and so have somewhat larger typical amplitudes than the visibilities for the \themis pipeline (for which the scattering is incorporated as part of the forward model; see \autoref{subsec:ImageDomain_ImagingMethods} and \citetalias{PaperIII}).  We also note that the different imaging pipelines make different choices about data averaging: \difmap and \ehtim average the data over 60\,s intervals, \smili averages over 120\,s intervals, and \themis averages over scans.  Detailed descriptions of each of the imaging methods are provided in \citetalias{PaperIII}.}
    \label{fig:imaging_residuals}
\end{figure*}

The imaging analyses carried out in \citetalias{PaperIII} use four different algorithms classified into three categories: one sampling-based posterior exploration algorithm (\themis, \citealt{Broderick_2020a,Broderick_2020b}), one CLEAN-based deconvolution algorithm (\difmap, \citealt{Shepherd_1997}), and two ``regularized maximum likelihood'' (RML) algorithms (\ehtim, \citealt{Chael_2016,Chael_2018}; and \smili, \citealt{Akiyama_2017a,Akiyama_2017b}). 
All methods produce image reconstructions using band-combined data (i.e., both low band and high band) and the latter three are run on two versions of the \sgra data: a ``descattered'' version that attempts to deconvolve the effects of the diffractive scattering kernel from the data, and a ``scattered'' version that applies no such deconvolution.  The posterior exploration imaging method \themis instead applies the effects of diffractive scattering as part of its internal forward model, rather than deconvolving the data; the analogous ``scattered'' and ``descattered'' versions of the \themis images thus correspond simply to those for which the scattering kernel has been applied or not, respectively.  The posterior exploration imaging jointly reconstructs the combined April 6 and April 7 data sets (see \autoref{app:SingleDayBias}), while the CLEAN and RML imaging reconstructs each day individually, focusing primarily on the April 7 data and using the April 6 data for cross-validation.  Example fits and residuals for each of the imaging pipelines are shown in \autoref{fig:imaging_residuals}\footnote{Note that the preprocessing and data products used during imaging are not the same across imaging pipelines; \difmap and \themis fit to complex visibilities, while \ehtim and \smili fit iteratively to different combinations of data products that include visibility amplitudes, closure phases, and log closure amplitudes \citepalias[see][for details]{PaperIII}.  For clarity in \autoref{fig:imaging_residuals}, we simply show residual complex visibilities for each imaging pipeline using a representative image from that pipeline's top set or posterior.}, and $\chi^2$ statistics for each image are provided in \autoref{app:Chisq}; detailed descriptions of the data preprocessing and imaging procedures for each imaging algorithm are provided in \citetalias{PaperIII}.

For the CLEAN and RML imaging methods, there are a number of tunable hyperparameters associated with each algorithm whose values are determined through extensive ``parameter surveys'' carried out on synthetic data sets. During a parameter survey, images of each synthetic data set are reconstructed using a broad range of possible values for each hyperparameter. Settings that produce high-fidelity image reconstructions across all synthetic data sets are collected into a ``top set'' of hyperparameters, and these settings are then applied for imaging the \sgra data. The resulting top sets of \sgra images capture emission structures that are consistent with the data, and we use these top-set images for the feature extraction analyses in this paper.

The \themis imaging algorithm explores a posterior distribution over the image structure, and there are no hyperparameters that require synthetic data surveys to determine.  Rather than producing a top set of images, \themis instead produces a sample of images drawn from the posterior determined from the \sgra data. We use these posterior image samples for the feature extraction analyses in this paper.

\begin{figure*}
    \centering
    \includegraphics[width=\linewidth]{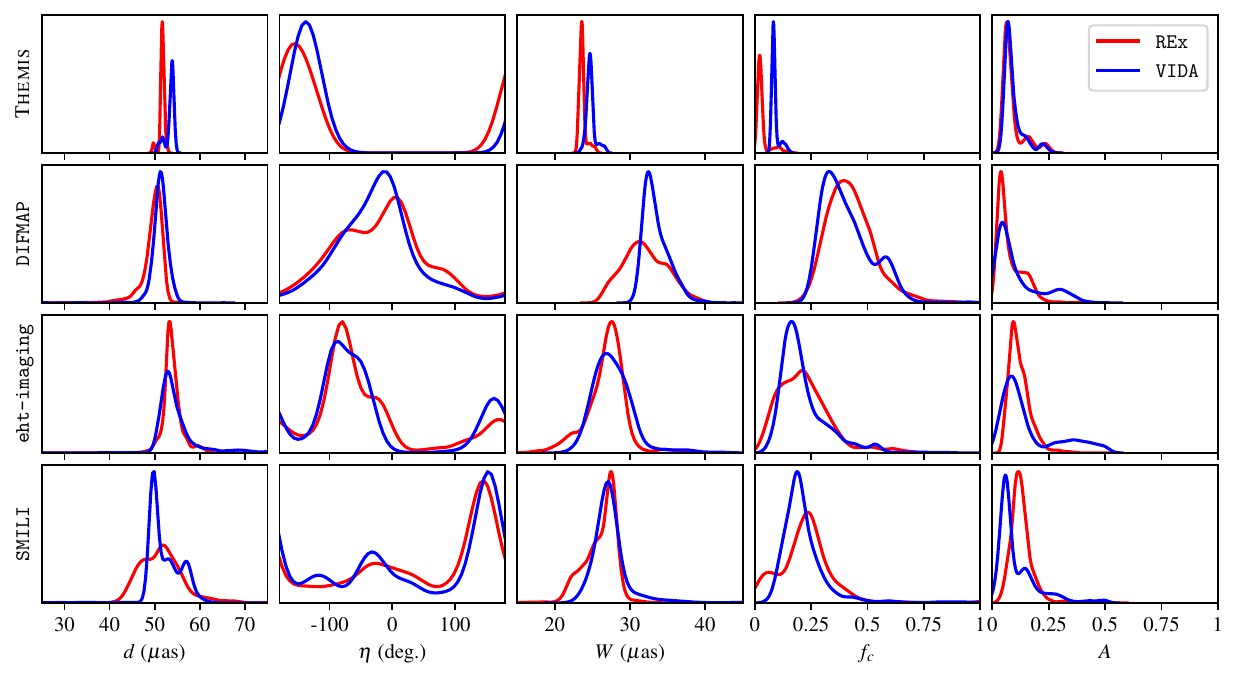}
    \caption{Morphological parameter distributions from IDFE analyses, applied to the descattered \sgra top-set and posterior images corresponding to the combined LO+HI band data from the HOPS calibration pipeline. The distributions shown correspond to combined April 6+7 results for posterior imaging and April 7 data for top-set imaging. No \texttt{metronization}-based filtering has been applied.}
    \label{fig:idfe-sgra-deblur-hops-1nometron}
\end{figure*}

\subsection{Image-domain feature extraction methods}
\label{subsec:ImageDomain_FEM}

Given the top-set and posterior images from \citetalias{PaperIII}, we carry out IDFE analyses using two separate tools: \rex and \vida.  An independent cross-validation of both IDFE tools has been carried out in \cite{Tiede2020}.  In this section, we provide a brief overview of each method and specify the details relevant for the analyses presented in this paper.

\subsubsection{\rex}
\label{ssec:rex}

The Ring Extractor (\rex) is an IDFE tool for quantifying the morphological properties of ring-like images. It is available as part of the \ehtim software library and is described in detail in \citet{ChaelThesis}. \rex was the main tool used in \citetalias{M87PaperIV} to extract ring properties from the \m87 images, and detailed definitions of the various \rex parameters are provided in that paper.

For the majority of the \rex-derived ring parameters, we retain the same definitions as used in \citetalias{M87PaperIV}.  \rex first defines a ring center $(x_0,y_0)$, which is determined to be the point in the image from which radial intensity profiles have a minimum dispersion in their peak intensity radii.  The ring radius, $r_0$, is then taken to be the average of these peak intensity radii over all angles, and the ring thickness $w$ is taken to be the angular average of the FWHM about the peak measured along each radial intensity profile.  To avoid biases associated with a nonzero floor to the image brightness outside of the ring, we subtract out the quantity

\begin{equation}
I_{\text{floor}} = \frac{1}{2\pi} \int_0^{2\pi} I(r_{\mathrm{max}}=60\,\uas,\phi) \text{d}\phi
\end{equation}
\noindent when computing the FWHM, i.e., we compute the average FWHM of $I(r,\phi) - I_{\text{floor}}$.
For all other ring parameters, the definitions remain the same as those used in \citetalias{M87PaperIV}.

\rex defines the ring position angle $\eta$ and asymmetry $A$ as the argument and amplitude, respectively, of the first circular mode,

\begin{equation}
\beta_1 = \left\langle \frac{\int_0^{2\pi} I(\phi) \cos(\phi) \text{d}\phi}{\int_0^{2\pi} I(\phi) \text{d}\phi} \right\rangle_r ,
\end{equation}

\noindent where the angled brackets denote a radial average between $r_0 - w/2$ and $r_0 + w/2$.\footnote{We note that this definition for the position angle $\eta$ does not necessarily return the azimuthal location of the intensity peak; rather, it tracks the circular mean of the azimuthal intensity profile.}  These definitions are analogous to those used to define the corresponding position angle and asymmetry of the \model model (\autoref{eqn:PositionAngle} and \autoref{eqn:Asymmetry}, respectively).  The fractional central brightness $f_c$ is defined to be the ratio of the mean brightness within 5\,\uas of the center to the azimuthally averaged brightness along the ring (i.e., along $r = r_0$).

As in \citetalias{PaperIII}, we replace the negative pixels in \themis images with zero values before performing \rex analyses.

\begin{figure*}
    \centering
    \includegraphics[width=\linewidth]{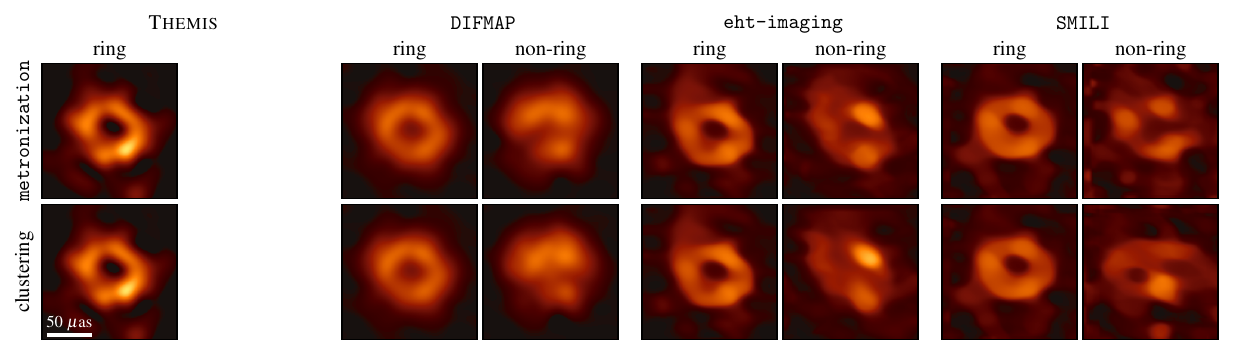}
    \caption{Comparison of two ring classification procedures.  Each panel shows a mean \sgra ring and nonring image for a single imaging pipeline, with the top row showing how the images are classified by \texttt{metronization} in the ``permissive'' mode and the bottom row showing the classification determined by the clustering analysis from \citetalias{PaperIII}.  All of the images have been produced using descattered \sgra data from the HOPS calibration pipeline.  The results correspond to combined April 6 and 7 data for posterior imaging and April 7 data for top-set imaging.  All of the images share a common brightness color scale; the absolute brightness scale is arbitrary because each image has been normalized to have unit total flux density.}
    \label{fig:idfe-sgra-mean-metron-vs-clustering}
\end{figure*}

\subsubsection{\vida}
\label{ssec:vida}

Variational Image-Domain Analysis \citep[\vida,][]{Tiede2020} is an IDFE tool for quantifying the parameters describing a specifiable image morphology; it is written in Julia \citep{bezanson2017julia} and contained in the package \texttt{VIDA.jl}\footnote{\url{https://github.com/ptiede/VIDA.jl}}.  \vida employs a template-matching approach for image analysis, using parameterized templates to approximate an image and adjusting the parameters of the templates
until a specified cost function is minimized.  Within \vida, the cost function takes the form of a probability divergence, which provides a distance metric between the image and template; the template parameters that minimize this divergence are taken to provide the best description of the image.  The \VIDA optimization strategy and additional details are provided in \citet{Tiede2020}.

For the IDFE analyses in this paper, we use \vida's \texttt{SymCosineRingwFloor} template and the least-squares divergence (for details, see Section 8 of \citetalias{PaperIII}).  This template describes an image structure that is similar to the \model model (\autoref{sec:mring}), and it is characterized by a ring center ($x_0$, $y_0$), a ring diameter $d=2r_0$, an FHWM fractional ring thickness $w$, and a cosine expansion describing the azimuthal brightness distribution $S(\phi)$,
\begin{align}
    S(\phi) = 1 - 2\sum_{k=1}^m A_k \cos\left[k (\phi - \eta_k)\right] .
\end{align}
\noindent To maintain consistency with the geometric modeling analyses (see \autoref{sec:SnapshotGeometricModeling} and \autoref{sec:FullTrackGeometricModeling}), we use $m=4$.  We also restrict the value of the $A_1$ parameter to be $<0.5$ to avoid negative flux in the template.  As with the \model model, the orientation $\eta$ is equal to the first-order phase $\eta_1$, and the asymmetry $A$ is equal to the first-order coefficient $A_1$.

To permit the presence of a central brightness floor, the \texttt{SymCosineRingwFloor} template contains an additional component in the form of a circular disk whose center point is fixed to coincide with that of the ring. The disk radius is fixed to be $r_0$. A Gaussian falloff is stitched to the outer edge of the disk, such that for radii larger than $r_0$ the intensity profile becomes a Gaussian with mean $r_0$ and an FWHM that matches the ring thickness. The flux of this disk component is a free parameter in the template.  We then retain the same definition of the fractional central brightness $f_c$ as used by \rex.

\subsection{Identifying rings via topological classification}
\label{subsec:ImageDomain_TDA}

The output of the IDFE analysis is a set of distributions for the ring parameters from each imaging method; \autoref{fig:idfe-sgra-deblur-hops-1nometron} shows an example set of results from applying both IDFE software packages to the descattered \sgra posterior and top-set images.  However, both \rex and \vida implicitly assume that the images fed into them contain a ring-like emission structure. If the input image does not contain a ring, then the output measurements may not be meaningful.  For each input image, we thus wish to determine both whether the image contains a ring-like structure and how sensitive the IDFE results are to the specific manner in which ``a ring-like structure'' is defined.

To determine whether the images we are analyzing with \rex and \vida contain ring-like structures, we use \texttt{metronization}\footnote{\url{https://github.com/focisrc/metronization}}, a software that preprocesses the images into a form suitable for topological analysis and extracts topologically relevant features with the help of the open-source computational topology code \texttt{Dionysus 2}\footnote{\url{https://mrzv.org/software/dionysus2}} \citep{Dionysus2}. A detailed description of \texttt{metronization} can be found in \cite{metronization_paper}.

The \texttt{metronization} preprocessing procedure  consists of the following steps:

\begin{enumerate}
    \item First, the image undergoes a ``robust'' thresholding step, in which the pixels are sorted by brightness in a cumulative sequence, and all pixels below a certain threshold in this sequence have their values set to zero and the rest are set to a value of one.
    \item Next, in a process called ``skeletonization,'' the Boolean image produced in the first step is reduced to its topological skeleton that preserves the topological characteristics of the original shape. This step thins large contiguous areas of flagged pixels and enlarges the ``holes.''
    \item The topological skeleton is rebinned and downsampled.  Holes smaller than the rebinning resolution are preserved by the skeletonization in the previous step.
    \item The downsampled image undergoes skeletonization once more.
\end{enumerate}

\noindent The resulting output is a low-resolution image that preserves the topologically relevant information from the original image, speeding up the application of computationally expensive topological algorithms that follow. A technique known as persistent homology is then used to convert this low-resolution image into a topological space that preserves features that are topologically invariant. It computes a quantity known as the first Betti number that provides a metric for measuring the number of holes present in the image.

The \texttt{metronization} software contains a number of tunable parameters that determine how closely the emission structure in the input image must resemble that of a topological ring, and for how many cumulative threshold levels it must persist, for it to be classified as a ring.  We identify three modes for these parameters -- a ``permissive'' mode, a ``moderate'' mode, and a ``strict'' mode -- and explore the impact on the \rex and \vida measured parameter distributions when the input top sets and posterior images are restricted only to those that are classified as containing rings.  We compare these results with those of a fourth, default setting in which the top sets and posterior images are not filtered by the classification prescribed by \texttt{metronization}.

We note that \texttt{metronization} differs from the ring identification methods presented in \citetalias{PaperIII} in that it searches for the presence of a topological ring in the input image. \autoref{fig:idfe-sgra-mean-metron-vs-clustering} compares the mean ring and nonring descattered images for each imaging pipeline as classified by \texttt{metronization} in the ``permissive'' mode and the clustering analysis from \citetalias{PaperIII}. Both methods classify all the posterior imaging samples as rings, while the top-set imaging samples contain both ring and nonring images.  We find that the mean ring and nonring images for each imaging pipeline are broadly consistent between the two classification methods.

\begin{figure}
    \centering
    \includegraphics[width=\linewidth]{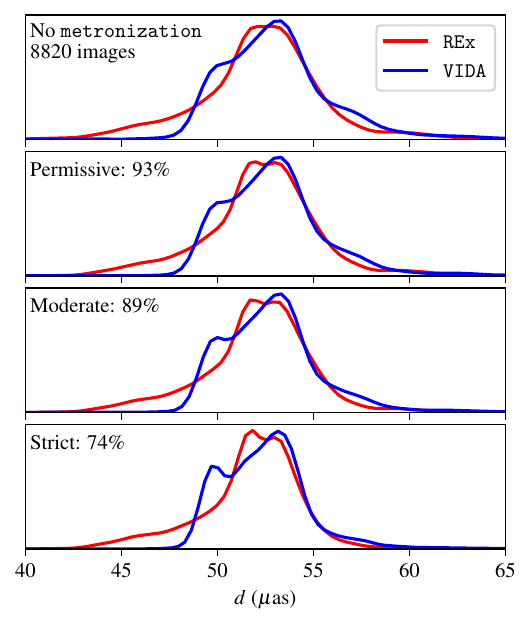}
    \caption{Diameter distributions determined by \rex and \VIDA for all descattered \sgra images from the HOPS pipeline, organized by \texttt{metronization} mode. Each panel shows the fraction of images found to possess topological ring structure. For posterior imaging we use the combined April 6+7 results, and for top-set imaging we use the April 7 results.
    }
    \label{fig:idfe-sgr-dia-by-metronmode-hops-deblur-3599}
\end{figure}

The definition of what constitutes a ring is subjective, and there will always be images that are ambiguous to the human eye. Different automated methods will classify these images differently. Hence, it is important to verify that the ring parameters measured by \rex and \vida are robust against the specifics of the ring identification scheme used. \autoref{fig:idfe-sgr-dia-by-metronmode-hops-deblur-3599} shows the resulting diameter distributions from ring fitting to the descattered \sgra images from all imaging pipelines, split out by \texttt{metronization} setting. As we move from the most to the least permissive classification scheme, the tails in the distributions are diminished while the primary peaks are sharpened, but the mean and general shape of the distribution remain largely unchanged. This trend indicates that while \texttt{metronization} penalizes images with emission structures deviating from a topological ring, the distributions of the \rex and \vida measurements are robust against the choice of \texttt{metronization} mode employed.

%% file: snapshot_geometric_modeling.tex
\section{Snapshot geometric modeling}\label{sec:SnapshotGeometricModeling}

Because the \sgra data are observed to be time-variable (see \autoref{sec:Variability}), a static model cannot reproduce the observed data.  As described in \autoref{sec:mitigation}, one method for mitigating the effects of this variability on the reconstructed source structure is through the use of an inflated variability noise budget, as pursued during the imaging (\citetalias{PaperIII}; \autoref{sec:ImageDomain}) and full-track geometric modeling (\autoref{sec:FullTrackGeometricModeling}) analyses.  In this section we instead pursue ``snapshot'' geometric modeling, whereby we fit a geometric model -- the \model model described in \autoref{sec:mring} -- for which the parameters are allowed to vary as a piecewise constant function of time. To this end, we divide the \sgra data into many independent and short ``snapshots'' over which the source is assumed to be static. In this section, we detail our formalism for fitting the \model model to snapshots of data and for combining the fits from across snapshots into a global posterior distribution.

\subsection{Data preparation}\label{sec:snapshot:dataprep}

Prior to fitting the \model model to real or synthetic data, we process the data using the pre-imaging pipeline described in \citetalias{PaperIII}. This preprocessing procedure entails light-curve normalization and an inflation of the error budget to account for residual calibration uncertainties and the effects of refractive scattering in the interstellar medium toward \sgra.  Specifically, the total error budget $\sigma_{sb}$ for a visibility measured on the baseline $b$ during snapshot $s$ is given by
\begin{equation}
    \sigma_{sb}^2 = \sigma_{\text{th},sb}^2 + f^2 |V|_{sb}^2 + \sigma_{\text{ref}, sb}^2 . \label{eqn:SnapshotNoise}
\end{equation}
Here the first term corresponds to the baseline-specific thermal noise (see \autoref{eqn:VisibilityCorruptions}), the second term is a component
that is multiplicative in the visibility amplitude and is intended to capture residual (nongain) calibration errors (e.g. residual polarization leakage), and the third term is the \texttt{J18model1} refractive scattering noise from \citetalias{PaperIII}. 
For the snapshot modeling, we fix $f=0.02$ per the analyses carried out in \citetalias{PaperII}. The pre-imaging pipeline also mitigates the impact of diffractive scattering by ``deblurring'' the data using the \citet{Johnson_2018} model.

Following the application of the pre-imaging pipeline, we split the data into 120\,s segments, or ``snapshots,'' and coherently average the visibilities in each snapshot over the 120\,s window. Finally, we flag snapshots that contain fewer than four unique stations, so as to retain snapshots during which closure amplitudes can be formed.

\subsection{Snapshot fitting procedure}\label{sec:snapshot:likelihood}

The first step of our snapshot modeling procedure is to determine the posterior distribution for the \model model parameters on each snapshot of data. The observation is divided up into $N_s$ independent snapshots, which we label using a snapshot index $s$. Within each snapshot we fit the \model model described in \autoref{sec:mring}, whose parameter vector we denote as $\boldsymbol{\theta}_s$.  For a single snapshot, the posterior is given by Bayes's theorem,

\begin{equation}\label{eqn:SnapshotBayes}
\begin{aligned}
    P_s(\boldsymbol{\theta}_s | \boldsymbol{D}_s) 
          &= \frac{\mathcal{L}_s(\boldsymbol{D}_s | \boldsymbol{\theta}_s)\pi_s(\boldsymbol{\theta}_s)}{\mathcal{Z}_s(\boldsymbol{D}_s)} ,
\end{aligned}
\end{equation}

\noindent where $\boldsymbol{D}_s$ denotes the data available on snapshot $s$, $\mathcal{L}_s$ is the likelihood, $\pi_s$ is the prior distribution, and $\mathcal{Z}$ is the Bayesian evidence.

In our snapshot modeling analyses we make use of three different classes of interferometric data products: visibility amplitudes $|V|$, log closure amplitudes $\ln A$, and closure phases $\psi$.  Each analysis uses only a single amplitude data product (either visibility amplitudes or log closure amplitudes) along with the closure phases.  For analyses that use visibility amplitudes and closure phases, the likelihood is given by

\begin{equation}
\mathcal{L}_s = \mathcal{L}_{|V|,s} \mathcal{L}_{\psi, s} , \label{eqn:ShapshotLikelihoodVisamp}
\end{equation}

\noindent while for those that use log closure amplitudes, we instead have

\begin{equation}
\mathcal{L}_s = \mathcal{L}_{A, s} \mathcal{L}_{\psi, s} . \label{eqn:ShapshotLikelihoodClosures}
\end{equation}

\noindent Here $\mathcal{L}_{|V|,s}$, $\mathcal{L}_{\psi, s}$, and $\mathcal{L}_{A, s}$ are components of the likelihood on snapshot $s$ associated with the visibility amplitudes, closure phases, and log closure amplitudes, respectively.  We assume Gaussian likelihood functions for the amplitude data components and a von Mises likelihood function for the closure phases; the detailed expressions for each likelihood function are provided in \autoref{app:SnapshotLikelihoods}.

\subsection{Averaging the snapshot results} \label{sec:snapshot:averaging}

\begin{figure}[t]
    \centering
    \includegraphics[width=\linewidth]{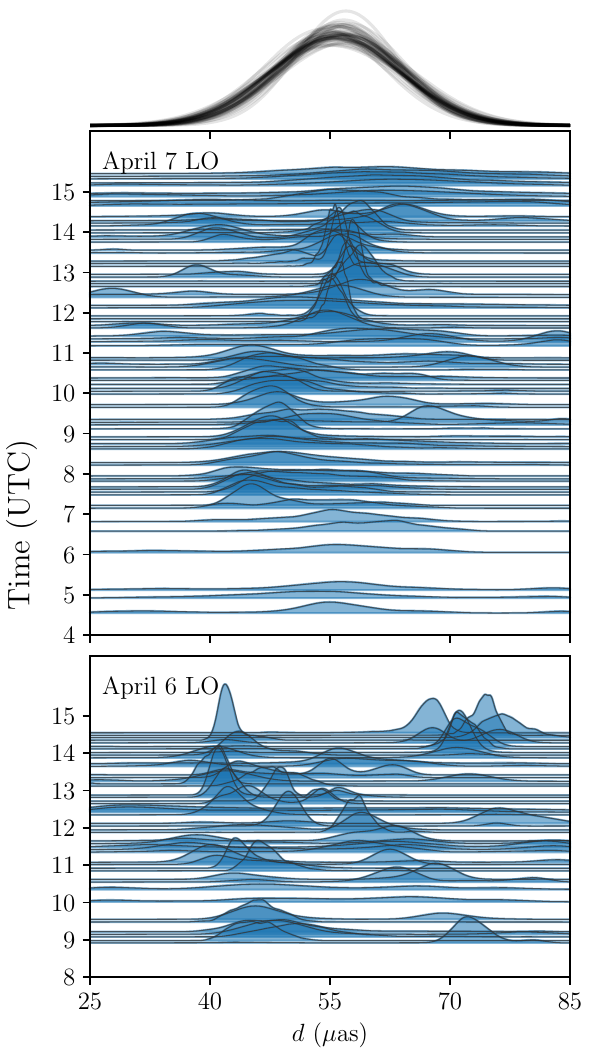}
    \caption{Example snapshot modeling results and averaging scheme applied to the \sgra April 6 and 7 low-band HOPS data sets. The blue filled regions show the posterior distribution of the \model diameter parameter for an $m=4$ model fit to each 120\,s snapshot. We find the tightest diameter posteriors from 12.5 to 14 UTC, which corresponds to the best time region from \citetalias{PaperIII}. For visual clarity we only show the distributions for every second snapshot. The black curves at the top show the diameter distribution corresponding to 100 random draws from the hypermodel posterior (\autoref{eq:snapshot:avgposterior}).}
    \label{fig:snapshot:average_model}
\end{figure}

The output of a snapshot fitting analysis is a set of posterior samples for the model parameters from each individual snapshot; \autoref{fig:snapshot:average_model} shows an example set of posterior distributions for the \model diameter parameter on each snapshot in the April 6 and April 7 data sets.  To arrive at a single posterior on these parameters that combines the information from all snapshots across both days, we use a Bayesian hierarchical model similar to the one used in \cite{Baronchelli_2020}. This approach treats the model fit to each snapshot as a realization from some average model or ``hypermodel.'' 

\subsubsection{Averaging procedure}

We denote the parameters of the average model as $\boldsymbol{\bar{\theta}}$ and the distribution of the snapshot model conditioned on the average model by $\pi_s(\boldsymbol{\theta_s} | \boldsymbol{\bar{\theta}})$. Given this conditional probability, the joint snapshot and average parameter posterior is given by

\begin{equation}\label{eq:snapshot:topset}
\begin{aligned}
	P(\boldsymbol{\Theta},\bm{\bar{\theta}}| \bm{\mathcal{D}}) 
	    &= \frac{\pi(\bm{\bar{\theta}})}{\mathcal{Z}(\bm{\mathcal{D})}}\prod_s \mathcal{L}_s(\bm{D}_s|\bm{\theta}_s) \pi_s(\bm{\theta}_s|\bm{\bar{\theta}}),
\end{aligned}
\end{equation}

\noindent where $\pi(\bm{\bar{\theta}})$ is the prior distribution for the hypermodel parameters (the ``hyperprior''), $\boldsymbol{\Theta} = (\boldsymbol{\theta}_1, \boldsymbol{\theta}_2, \dots, \boldsymbol{\theta}_{N_s})$ is the parameter vector across all snapshots, and $\boldsymbol{\mathcal{D}} = (\boldsymbol{D}_1, \boldsymbol{D}_2, \dots, \boldsymbol{D}_{N_s})$ is the data vector across all snapshots. To find the marginal average parameter posterior, we integrate this expression over all the snapshot parameters,

\begin{equation}\label{eq:snapshot:topset_marginal}
\begin{aligned}
	P(\bm{\bar{\theta}}| \bm{\mathcal{D}}) 
	    &= \frac{\pi(\bm{\bar{\theta}})}{\mathcal{Z}(\bm{\mathcal{D}})}\int \left( \prod_s \mathcal{L}_s(\bm{D}_s|\bm{\theta}_s) \pi_s(\bm{\theta}_s|\bm{\bar{\theta}}) \right) \mathrm{d}\boldsymbol{\Theta} .
\end{aligned}
\end{equation}

\noindent In general, this integral is analytically intractable.  However, a bit of manipulation permits us to use the posterior samples from the individual snapshot fits to make headway. Because the snapshots are independent, we can swap the order of the integral and product in \autoref{eq:snapshot:topset_marginal} and use Bayes's theorem to substitute in for the snapshot likelihood (\autoref{eqn:SnapshotBayes}), giving

\begin{equation}\label{eq:snapshot:avgposterior}
    	P(\bm{\bar{\theta}}| \bm{\mathcal{D}}) 
	    = \pi(\bm{\bar{\theta}}) \prod_s \left( \int  P(\bm{\theta}_s|\bm{D}_s)\frac{\pi_s(\bm{\theta}_s|\bm{\bar{\theta}})}{\pi_s(\bm{\theta}_s)} \mathrm{d}\bm{\theta}_s \right) .
\end{equation}

\noindent Note that the evidence term from the prefactor denominator in \autoref{eq:snapshot:topset_marginal} has now been subsumed into the posterior term $P(\bm{\theta}_s|\bm{D}_s)$ inside of the integral. To evaluate \autoref{eq:snapshot:avgposterior}, we make use of the fact that the snapshot posterior samples $\bm{\theta}_s^{(i)}$ permit us to approximate the integral by a sum,

\begin{equation}\label{eq:snapshot:avgposterior_approximate}
    	P(\bm{\bar{\theta}}| \bm{\mathcal{D}}) 
	    \approx \pi_s(\bm{\bar{\theta}})\prod_s \sum_{i} \frac{\pi_s(\bm{\theta}^{(i)}_s|\bm{\bar{\theta}})}{\pi(\bm{\theta}^{(i)}_s)} .
\end{equation}

\noindent We can use this expression to sample from the posterior distribution over just the hypermodel parameters $\bm{\bar{\theta}}$, having fully marginalized over the parameters from each individual snapshot.

We note that the averaging procedure described here is simply a generalization of standard inverse-variance weighting.  If we consider a delta-function hypermodel that contains only a single parameter (i.e., the to-be-determined mean value) for each snapshot model parameter, then in the limit where the individual snapshot posteriors $P(\bm{\theta}_s|\bm{D}_s)$ are Gaussian and the priors on $\bar{\bm{\theta}}$ and $\bm{\theta}_s$ are uninformative, the posterior maximum for $P(\bm{\bar{\theta}}| \bm{\mathcal{D}})$ is equal to the mean of the snapshot posterior means weighted by their inverse posterior variances.  However, because the model we employ (described in the following section) does not conform to the necessary conditions (see, e.g., the non-Gaussian snapshot posteriors shown in \autoref{fig:snapshot:average_model}), we proceed with the more general averaging procedure.

\subsubsection{Hypermodel specification}

We now need to specify the hypermodel $\pi(\bm{\theta}_s| \bm{\bar{\theta}})$ that determines the distribution from which the individual snapshot models are drawn; for simplicity, we choose a hypermodel that is approximately Gaussian.  Let $\bm{\bar{\theta}} = (\bm{\mu}, \bm{\sigma})$, where $\bm{\mu}$ is a vector of the mean parameter values and $\bm{\sigma}$ is a vector containing their standard deviations across scans.  We assign most hypermodel parameters to be distributed according to a truncated normal distribution,
\begin{equation}
    \pi_{s, \text{tN}}(\theta_{s,i} | \mu_i, \sigma_i) =  \mathcal{T}(\theta_{s,i} | \mu_i, \sigma_i;\, a_i, b_i) . \label{eqn:hypermodel:nonangular}
\end{equation}
Here $\mathcal{T}(\theta | \mu, \sigma;\, a, b)$ denotes the density for a truncated normal distribution with mean $\mu$ and standard deviation $\sigma$, and whose lower and upper bounds are given by $a$ and $b$, respectively; we index the separate parameters by $i$. 
This truncation is necessary to ensure that the support of the hypermodel parameters matches that of the individual snapshot model parameters.  However, for angular parameters -- i.e., those with values that are periodic in $[0,2\pi)$ -- we instead use a von Mises distribution,
\begin{equation}
    \pi_{s, \text{vM}}(\theta_{s,j} | \mu_j, \sigma_j) = \frac{1}{2\pi I_0(\sigma_j^{-2})}\exp\left[\frac{\cos(\theta_{s,j} - \mu_j)}{\sigma_j^2}\right] . \label{eqn:hypermodel:angular}
\end{equation}
The subscripted ``tN'' in \autoref{eqn:hypermodel:nonangular} indicates that the corresponding parameters use a truncated normal prior, while the subscripted ``vM'' in \autoref{eqn:hypermodel:angular} similarly indicates that the corresponding parameters use a von Mises prior.
The total hypermodel is then given by
\begin{equation}
    \pi(\bm{\theta}_s | \bm{\mu}, \bm{\sigma}) = \prod_{i} \pi_{s, \text{tN}}(\theta_{s,i} | \mu_i, \sigma_i) \prod_{j}\pi_{s, \text{vM}}(\theta_{s,j} | \mu_j, \sigma_j) ,
\end{equation}

\noindent where the first product runs over nonangular parameters and the second runs over angular parameters.

We set the hyperpriors for $\bm{\mu}$ to be equal to the corresponding snapshot priors, which are specified in \autoref{tab:snapshot:model_priors}. For $\bm{\sigma}$ we instead use a half-normal hyperprior,

\begin{equation}
    \pi(\bm{\sigma}) =  \prod_i \mathcal{T}(\sigma_i| 0, L_i/4;\; 0, L_i),
\end{equation}

\noindent where $L_i = (b_i - a_i)$ is the breadth of support for parameter $\theta_{s,i}$.
\autoref{app:HypermodelMarginals} describes the level of consistency between these selected hyperpriors and the priors for the individual snapshot model parameters.

\begin{deluxetable}{l|cc}
\tablecolumns{3}
\tabletypesize{\normalsize}
\tablewidth{0pt}
\tablecaption{Snapshot modeling \model priors \label{tab:snapshot:model_priors}}
\tablehead{\colhead{Parameter} & \colhead{\rose} & \colhead{\texttt{DPI}}}
\startdata
$F_0$ & $\mathcal{U}(0.8, 1.2)$ & $\delta(1)$ \\
$d$ (\uas) & $\mathcal{U}(25, 85)$ & \ldots  \\
$d^{\prime}$ (\uas) & \ldots & $\mathcal{U}(25, 85)$ \\
$W$ (\uas) & $\mathcal{U}(1, 40)$  & \ldots  \\
$W^{\prime}$ (\uas) & \ldots & $\mathcal{U}(1, 40)$ \\
$|\beta_m|$ & $\mathcal{U}(0, 0.5)$ & $\mathcal{U}(0, 0.5)$ \\
$\arg(\beta_m)$ (deg) & $\mathcal{U}(-180, 180)$ & $\mathcal{U}(-180, 180)$ \\
$f_{\rm Gauss}$ & $\mathcal{U}(0, 1)$ & $\mathcal{U}(0, 1)$ \\
$W_{\rm Gauss}$ (\uas) & $\mathcal{U}(40, 200)$ & $\mathcal{U}(40, 200)$\\
\enddata
\tablecomments{Prior distributions for \rose and \texttt{DPI} snapshot geometric modeling analyses. $\mathcal{U}(a,b)$ denotes a uniform prior on the interval $[a,b]$ and $\delta(a)$ denotes a delta-function (i.e., fixed-value) prior, with the parameter value fixed at $a$. For the definitions of the parameters see \autoref{sec:mring}.}
\end{deluxetable}

\subsection{Software implementations}
We use three different software packages to carry out snapshot geometric modeling on the \sgra data and a fourth software to perform the hypermodel sampling.  In this section, we specify the relevant implementation specifics for these different tools.  Cross-validation tests are detailed in \autoref{app:SnapshotValidation}.

\subsubsection{\rose} \label{sec:ROSE}

Our primary snapshot fitting software is the modeling framework \rose \citep{comrade}, which is written in the dynamic programming language Julia \citep{bezanson2017julia}. \rose does not natively include functionality for constructing a joint probability describing both observations and model. Instead, it interfaces with existing probabilistic programming languages present in Julia. For the analyses presented in this paper, we use the probabilistic programming package Soss\footnote{\url{https://github.com/cscherrer/Soss.jl}} to construct the joint probability.  This interface is specified in the package ComradeSoss.jl\footnote{\url{https://github.com/ptiede/ComradeSoss.jl}}. To sample from the posterior, we use the nested sampling package \texttt{dynesty}, which also produces estimates of the Bayesian evidence \citep{Speagle_2020}.

Given a model specification, \rose can fit a variety of interferometric data products, including visibility amplitudes, closure phases, and log closure amplitudes. Unless otherwise specified, for the snapshot modeling analyses performed in this paper, we use \rose to fit to visibility amplitudes and closure phases to the mG-ring model; the snapshot likelihood is thus given by \autoref{eqn:ShapshotLikelihoodVisamp}.  Prior to fitting, all time stamps that contain fewer than four baselines are flagged.

When fitting to visibility amplitudes, we include the station gain amplitudes as model parameters alongside the geometric parameters that describe the \model model. For the gain amplitudes we use a lognormal prior with a log-mean of zero (i.e., corresponding to unit gain amplitude) and a log standard deviation of 0.1 on all stations except for LMT, for which we use a log standard deviation of 0.2 to accommodate its larger variations \citepalias{M87PaperIII,PaperII}.

\subsubsection{\ehtim}

We also utilize the geometric model-fitting tools developed within the \ehtim Python library \citep{Chael_2016,Chael_2018}. This library enables visibility-domain fitting to arbitrary combinations of simple analytic models, including the \model model, and it can do so using a variety of interferometric data products, including visibility amplitudes, closure phases, and log closure amplitudes.  \ehtim is also able to interface with a variety of external packages to perform parameter optimization or posterior exploration.

For the snapshot modeling analyses performed in this paper, we match the operation of \ehtim with that of \rose.  Unless otherwise specified, we use \ehtim to fit to visibility amplitudes and closure phases, so that the snapshot likelihood is given by \autoref{eqn:ShapshotLikelihoodVisamp}, and we use \texttt{dynesty} \citep{Speagle_2020} for posterior exploration and evidence estimation. We also specify the same priors for the station gain parameters as used in the \rose fits.

Given that both \ehtim and \rose use identical model specifications, priors, and samplers, we expect all results produced by these softwares to be identical up to sampling precision.  We thus use only \rose fits for all \sgra snapshot geometric analyses in this paper. 

\subsubsection{\texttt{DPI}} \label{sec:DPI}

The third software we use for snapshot geometric modeling is the Python code Deep Probabilistic Imaging/Inference \citep[\texttt{DPI}/\texttt{$\alpha$-DPI}][]{sun2021deep, sun2021alpha}.  
\texttt{DPI} approximates the posterior over all model parameters by fitting a normalizing flow neural network \citep{rezende2015variational} to the data using a R\'enyi $\alpha$-divergence variational inference technique \citep{li2016r}.
\texttt{DPI} is an optimization-based posterior estimation framework, and it uses the auto-differentiation package PyTorch \citep{paszke2017automatic} to optimize the neural network weights. The posterior estimation accuracy is further improved post-optimization through importance reweighting of the samples generated by the normalizing flow neural network.

\texttt{DPI} supports fitting to multiple data products, including visibility amplitudes and closure quantities, but it does not currently support the inclusion of station gain amplitudes as model parameters.  We thus use \texttt{DPI} to fit to closure phases and log closure amplitudes; the snapshot likelihood is given by \autoref{eqn:ShapshotLikelihoodClosures}. Prior to fitting, all time stamps that are unable to form at least one closure phase and at least one closure amplitude are flagged.

\texttt{DPI} differs from both \rose and \ehtim in that it defines geometric models in the image domain rather than in the visibility domain, and it uses a nonuniform fast Fourier transform (NFFT) to compute the necessary data products.  For the analyses carried out in this paper, we discretize the model as an image containing $32 \times 32$ pixels spanning a 160\,\uas field of view.

Because the pixel size is finite, \texttt{DPI} cannot support a model containing infinitesimally thin rings such as that in \autoref{eqn:RingImage}; furthermore, convolutions in the image domain are computationally expensive.  The \texttt{DPI} fits in this paper thus employ a modified version of the \model model specification,

\begin{equation} \label{eq:dpi-mGring}
\begin{split}
    I_{\text{ring}}^{\prime}(r, \phi) &= F_{\text{ring}} \frac{\exp\left[ \frac{4\ln{2}}{{W^{\prime}}^2} \left( r-\frac{d^{\prime}}{2} \right)^2 \right] \sum_{k=-m}^{m} \beta_k e^{ik\phi}}{2 \pi \int_0^{\infty} r \exp\left[ \frac{4\ln{2}}{{W^{\prime}}^2} \left( r-\frac{d^{\prime}}{2} \right)^2 \right] \,dr}, \\
    I^{\prime}(r, \phi) &= I_{\text{ring}}^{\prime}(r, \phi) + I_{\text{Gauss}}(r, \phi) , \\
    F_0 &= F_{\text{ring}} + F_{\text{Gauss}} = 1 ,
\end{split}
\end{equation}

\noindent where we note that $d^{\prime}$ and $W^{\prime}$ are conceptually distinct from $d$ and $W$.  The quantities $d$ and $W$ in the \model model from \autoref{sec:mring} determine the diameter of the infinitesimally thin ring and the FWHM of its convolving kernel, respectively.  In contrast, the quantities $d^{\prime}$ and $W^{\prime}$ determine the intensity peak and FWHM, respectively, of a radial Gaussian function.  These two specifications converge only in the limit of large $d$ and small $W$. In addition, the total flux of the \texttt{DPI} model implementation is fixed to be 1\,Jy because \texttt{DPI} fits only to closure quantities and closure amplitudes are not sensitive to the absolute flux scale.

\subsubsection{Sampling the hypermodel posterior}

To sample from the hypermodel posterior $P(\bm{\bar{\theta}}| \bm{\mathcal{D}})$ (\autoref{eq:snapshot:avgposterior_approximate}), we use the adaptive Metropolis sampler from \citet{ram_sampler} via its implementation in the Julia package RobustAdaptiveMetropolisSampler.jl\footnote{\url{https://github.com/anthofflab/RobustAdaptiveMetropolisSampler.jl}}. The sampler is initialized by first running an adaptive genetic algorithm from the Julia package BlackBoxOptim.jl\footnote{\url{https://github.com/robertfeldt/BlackBoxOptim.jl}}, which provides a starting point near the maximum posterior density. We run the sampler for a minimum of 2 million Markov Chain Monte Carlo (MCMC) steps or until we have effective sample sizes of 500 for all parameters.

\subsection{Model selection}\label{sec:snapshot:model_selection}

\begin{figure}[t]
  \begin{center}
    \includegraphics[width=\linewidth]{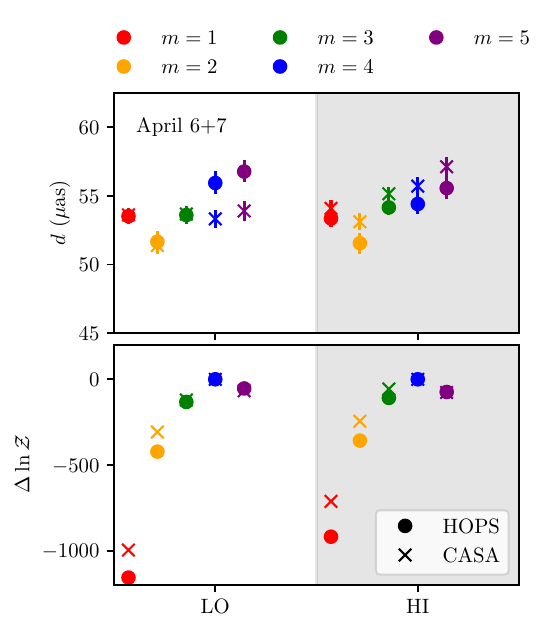}
  \end{center}
\caption{
Relative Bayesian evidence and best-fit \model model diameter vs. $m$-order, from the snapshot+averaging geometric modeling from the \rose pipeline applied to the \sgra data.  Because the absolute values of $\ln\mathcal{Z}$ can be substantially different for each data set (i.e., each combination of observing day, frequency band, and calibration pipeline), and because only the relative values carry information about model specification preferences, we reference all $\ln\mathcal{Z}$ values to the maximum value attained at any $m$ for that data set.}
\label{fig:mring_models}
\end{figure}

\begin{figure*}[t]
    \centering
    \includegraphics[width=\columnwidth]{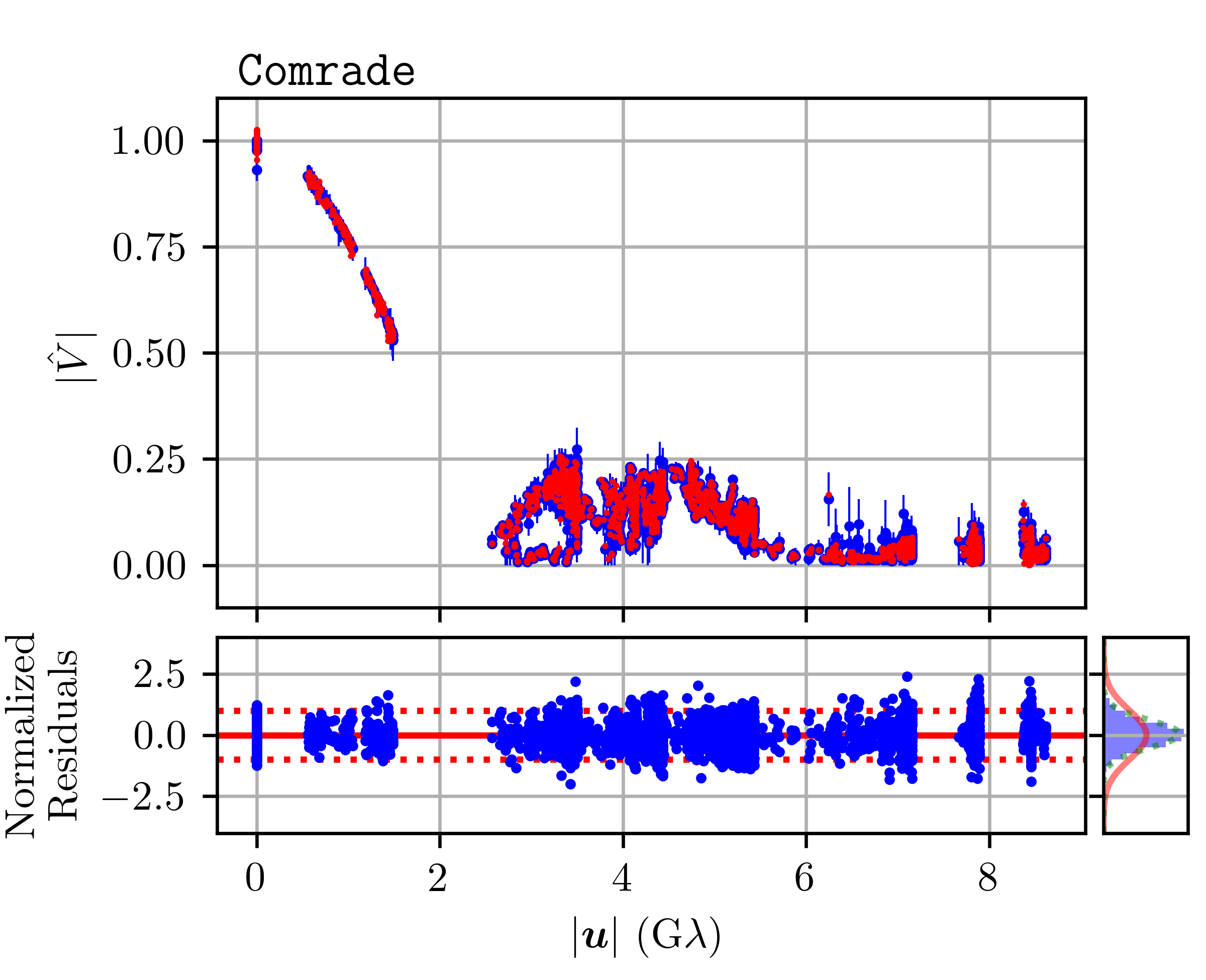}
    \includegraphics[width=\columnwidth]{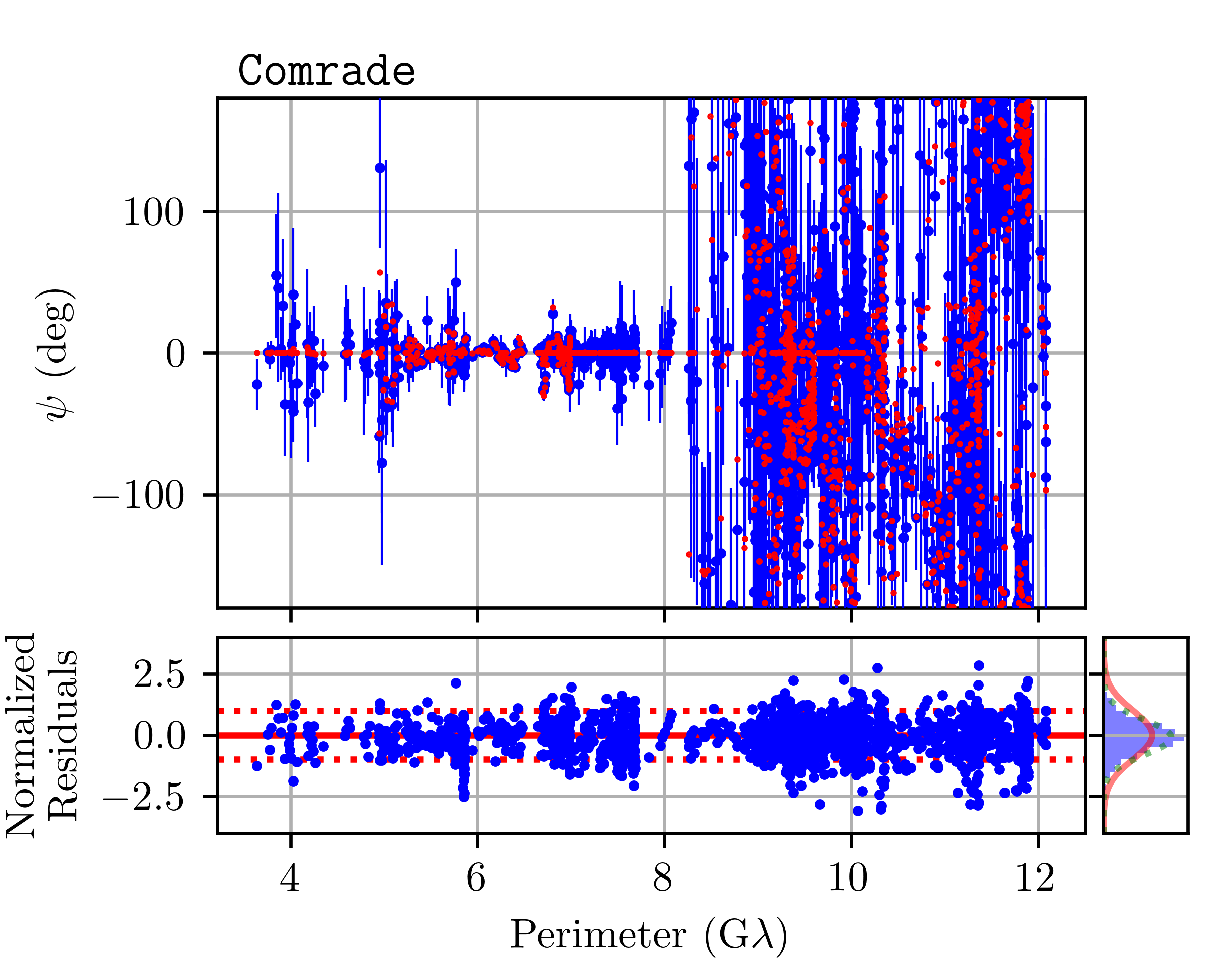} \\
    \includegraphics[width=\columnwidth]{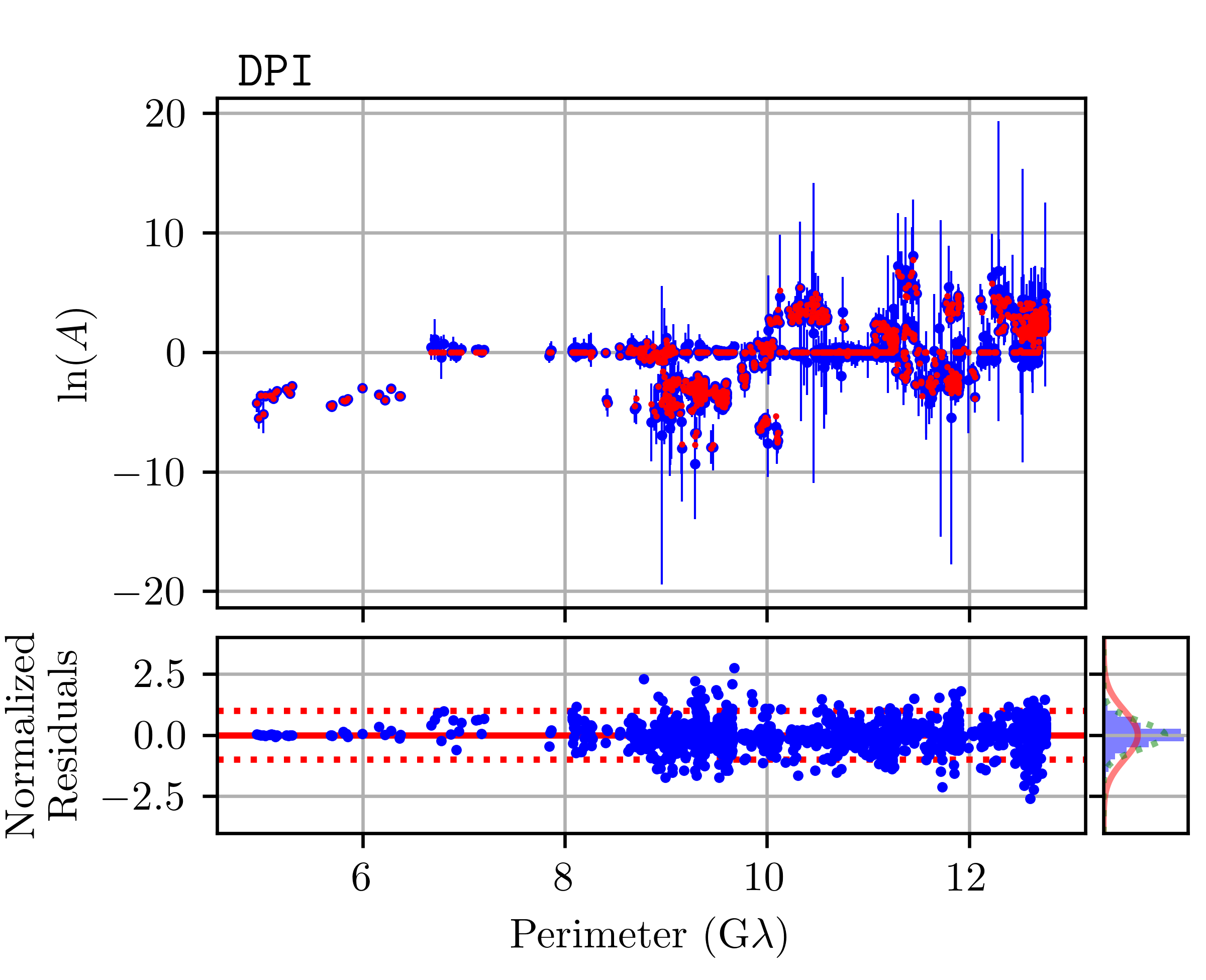}
    \includegraphics[width=\columnwidth]{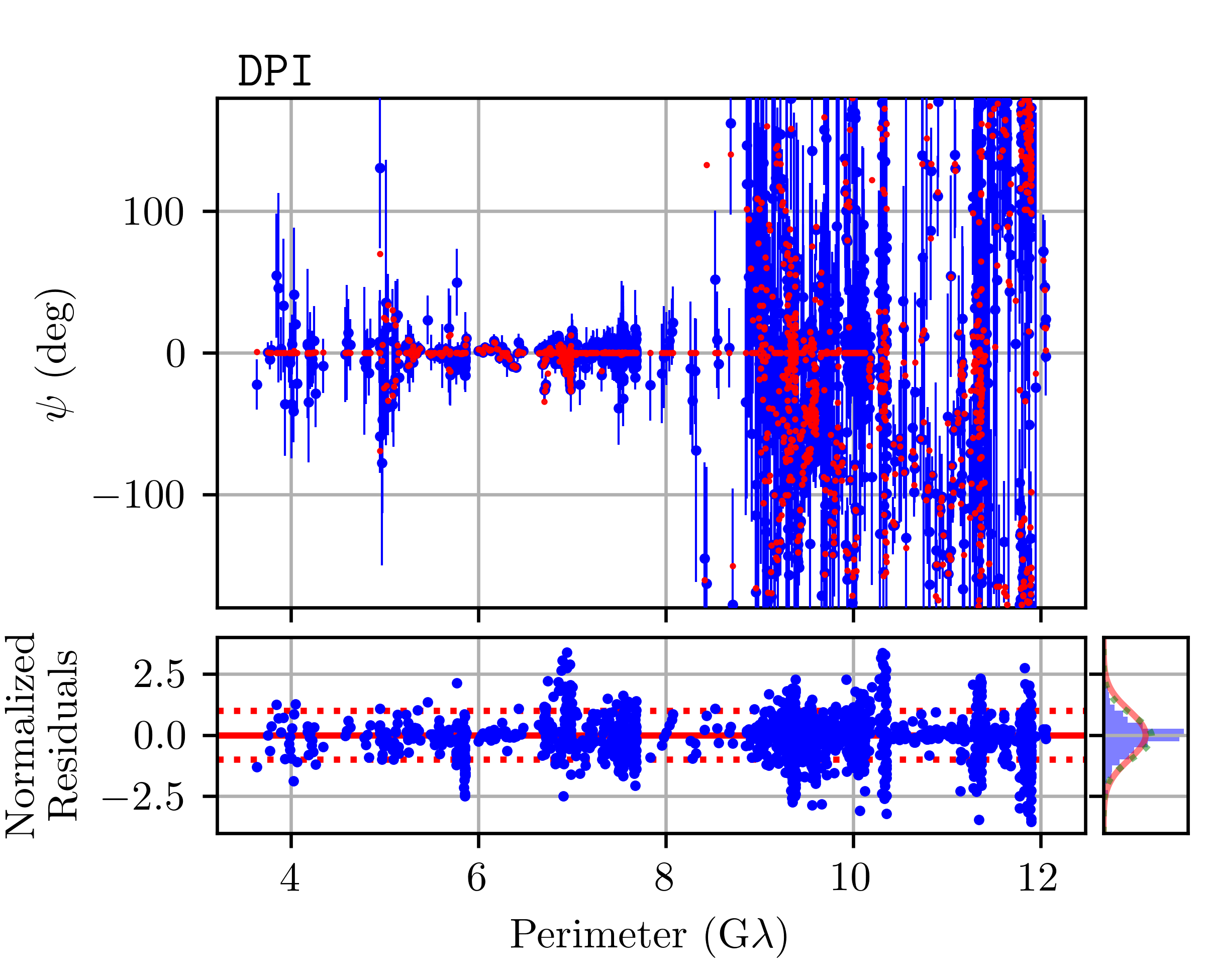}
    \caption{Representative examples of snapshot modeling results for both the \comrade (top row) and \texttt{DPI} (bottom row) pipelines.  The top row of panels shows results from fitting an $m=4$ \model with \comrade to the HOPS low-band \sgra data on both April 6 and 7, while the bottom row of panels shows results from fitting an $m=2$ \model with \texttt{DPI} to the same dataset.  Each panel is arranged analogously to the individual panels of \autoref{fig:imaging_residuals}, though the plotted data products are different.  For the \comrade results, the visibility amplitudes and closure phases are used during fitting, so these are the data products shown in the top left and top right panels, respectively.  For the \texttt{DPI} results, the log closure amplitudes and closure phases are used during fitting, so these are the data products shown in the bottom left and bottom right panels, respectively.  All closure quantities are plotted as a function of the perimeter of the relevant polygon (i.e., triangles for closure phases, quadrangles for log closure amplitudes); the evident zero-valued closure phases and log closure amplitudes primarily correspond to so-called ``trivial'' polygons, i.e., those with near-zero area \citepalias{M87PaperIII,PaperII}.  We note that the data-flagging procedures differ slightly between the two pipelines (see Sections \ref{sec:ROSE} and \ref{sec:DPI}), resulting in small differences in the fitted data sets.}
    \label{fig:snapshot_residuals}
\end{figure*}

The \model model described in \autoref{sec:mring} is not actually a single model but rather a class of models, delineated by the order $m$ (\autoref{eqn:RingImage}). To determine the $m$-order that is preferred by the \sgra data, we carry out a series of snapshot \model fits to the \sgra using different values of $m$ and compare Bayesian evidence estimates.  Given a set of log-evidences $\ln \mathcal{Z}_s$ computed for every snapshot $s$ in a single observation, the total evidence for the entire observation is simply given by their sum,
\begin{equation}
    \ln \mathcal{Z} = \sum_s \ln \mathcal{Z}_s . \label{eqn:EvidenceCombination}
\end{equation}

The \rose snapshot fitting analyses directly estimate the Bayesian evidence on every snapshot, and so the total evidence across an entire observation can be computed directly using \autoref{eqn:EvidenceCombination}.  The results of a \rose $m$-order survey covering $m=\{1,2,3,4,5\}$ are shown in \autoref{fig:mring_models}.  We find that the $m=4$ order is preferred in both bands and across both calibration pipelines.  We thus use the $m=4$ \model as our fiducial model for all \rose \sgra analyses in this paper.

Unlike \rose, \texttt{DPI} does not directly estimate the Bayesian evidence during each fit.  Instead, we use the evidence lower bound (ELBO) to determine the $m$-order preference. The ELBO is a combination of the true evidence modified by a relative entropy term that encodes the performance of the variational approximation,
\begin{equation}
\label{eq:elbo}
\text{ELBO}(m) = \log p(\bm{D}_s \| m ) - D_{KL}[q(\bm{\theta}_s)\|p(\bm{\theta}_s|\bm{D}_s, m)],
\end{equation}
where $D_{\rm{KL}}[A\|B]$ is the Kullback-Leibler divergence of $A$ from $B$, and $q(\bm{\theta}_s)$ is the optimized \texttt{DPI} normalizing flow distribution.
The relative entropy term is zero when the \texttt{DPI} distribution $q(\bm{\theta}_s)$ and the true posterior $p(\bm{\theta}_s|\bm{D}_s)$ are identical, so the ELBO provides a rough estimate of the log-evidence. The results of a \texttt{DPI} $m$-order survey covering $m=\{1,2,3,4\}$ indicate that either $m=1$ or $m=2$ is preferred, depending on the day and band.  We choose to err on the side of increased model flexibility and use the $m=2$ \model as our fiducial model for all \texttt{DPI} \sgra analyses in this paper.

\autoref{fig:snapshot_residuals} shows representative \model fits to the \sgra HOPS low-band data for both the \comrade and \texttt{DPI} pipelines.  In all cases, we find that the normalized residuals are distributed around a value of zero with a subunity variance, and there is no evidence of systematic structure.  The $\chi^2$ statistics for each of these fits are provided in \autoref{app:Chisq}.

%% file: fulltrack_geometric_modeling.tex
\section{Full-track geometric modeling}\label{sec:FullTrackGeometricModeling}

The snapshot modeling analysis presented in the previous section addresses the variability of the \sgra data by explicitly permitting the source structure to vary in time.  As described in \autoref{sec:mitigation}, an alternative approach to fitting variable data is to statistically capture the impact of variability, treating it as an additional source of uncertainty modifying data that otherwise describe a static (or average) source structure.  We pursue such an approach here in the form of ``full-track'' geometric modeling, whereby we fit the \model model (see \autoref{sec:mring}) to an entire data set at once and account for the variability by simultaneously fitting a parameterized noise model.  In this section we detail our formalism for fitting the \model geometric model alongside a model that captures the noise budget inflation associated with source variability.

\subsection{Data preparation} \label{sec:fulltrack:dataprep}

The data preparation for the full-track geometric modeling analyses is similar to that used for snapshot geometric modeling analyses (see \autoref{sec:snapshot:dataprep}).  The data are first processed through the pre-imaging pipeline described in \citetalias{PaperIII}, which applies light-curve normalization and performs some a priori gain calibration.  However, unlike in \citetalias{PaperIII} and \autoref{sec:snapshot:dataprep}, we do not modify the data uncertainties at all beyond their thermal noise values; neither a systematic error term nor a refractive scattering term is added to the error budget. Additionally, no ``deblurring'' is applied to the data; instead, the blurring is applied directly to the model as described in the next section.

Following the application of the pre-imaging pipeline, we coherently average the visibilities from each baseline on a per-scan basis.  A scan length ($\sim$10 minutes) is approximately the amount of time over which we expect structural variability to be subdominant to other sources of uncertainty (see \autoref{sec:variability_theory}, in particular \autoref{fig:GRMHD_PSD}).  Furthermore, the station gains are expected to be constant in time across a single scan but not from one scan to the next \citepalias{M87PaperIII,PaperII}, meaning that a scan length is also the longest coherent integration time that the a priori calibration can support.

While the full-track modeling is necessarily focused on reconstructing a time-averaged image structure, the underlying data remain a collection of complex visibilities that sample different instantaneous realizations of the intrinsic \sgra source structure.  As a consequence, the \sgra data exhibit subhour correlations that over a single day are localized in the \uv-plane.  As previously noted in \autoref{sec:mitigation} and detailed in \autoref{app:SingleDayBias}, these unmodeled correlations can result in significant biases in the reconstructed properties of \sgra.  However, by fitting to multiple days of \sgra data, and thus combining multiple samplings of the variable source structure at each location in the \uv-plane, we better match the statistical properties of the data to those assumed by the full-track analysis.  An additional benefit of combining days is that the multiday analyses more clearly emphasize the static signatures of gravitational lensing from the spurious astrophysical variability. For these reasons, all full-track analyses presented in \autoref{sec:results} make use of the combined April 6 and April 7 \sgra data.  For comparison we provide single-day analysis results in \autoref{app:SingleDayFits}.

\subsection{Model specification and implementation}

The goal of the full-track geometric modeling procedure is to determine the posterior distribution for the parameters of the static \model model and parameterized noise model that best describe an entire \sgra data set.  Our specification for the \model model is described in \autoref{sec:mring}, and we retain the same notation and terminology in this section. Additionally, we incorporate the blurring effects of scattering in the same manner described in Section 4 of \citetalias{PaperIII}, through multiplication of the \model visibilities by the Fourier transform of the scattering kernel.  We note, however, that all images shown in the figures in this paper correspond to the underlying (i.e., nonscattered) image.

\subsubsection{Parameterized noise model}

Our parameterized noise model for a complex visibility $V_{i}$ measured on a baseline $\boldsymbol{u}_i$ is given by

\begin{equation}
\sigma_{i}^2 = \sigma_{\text{th},i}^2 + f^2 |V|_{i}^2 + \sigma_{\text{ref}}^2 + \sigma_{\text{var}}^2(|\boldsymbol{u}_i|) . \label{eqn:FullTrackNoise}
\end{equation}

\noindent Here the first term is the thermal noise in the measurement (see \autoref{eqn:VisibilityCorruptions}), the second term is a component that is multiplicative in the visibility amplitude $|V|_{i}$ and which is intended to capture residual (nongain) calibration errors (e.g. residual polarization leakage), the third term is a component that is additive and which is intended to account for refractive scattering noise, and the fourth term is a component that is a function of the baseline length $|\boldsymbol{u}_i|$ and which is intended to capture the effects of source variability.  With the exception of the variability term, \autoref{eqn:FullTrackNoise} is similar to the noise budget used in the snapshot modeling (see \autoref{eqn:SnapshotNoise}); the only difference is that now $f$ and $\sigma_{\text{ref}}$ enter into the model as free parameters.

The variability noise $\sigma_{\text{var}}$ is described in \autoref{sec:Variability} (see \autoref{eq:noise-model}) and consists of a broken power law in $|\boldsymbol{u}|$ specified by four parameters: an overall amplitude $a_4$ specified at a baseline length of 4\,G$\lambda$, a falling long-baseline power-law index $b$, a rising short-baseline power-law index $c$, and a baseline length $u_0$ at which the power-law breaks.  Informative prior bounds for each of these parameters are determined from the model-agnostic variability quantification described in \autoref{sec:pre-modeling}, and these bounds are listed in \autoref{tab:fulltrack:model_priors}.

\begin{deluxetable}{l|c}
\tablecolumns{2}
\tabletypesize{\normalsize}
\tablewidth{0pt}
\tablecaption{Full-track modeling \model priors \label{tab:fulltrack:model_priors}}
\tablehead{\colhead{Parameter} & \colhead{Prior}}
\startdata
$f_{\text{ring}}$ & $\mathcal{U}(0.05, 4.0)$ \\
$d$ (\uas) & $\mathcal{U}(20, 85)$  \\
$W$ (\uas) & $\mathcal{U}(1, 40)$  \\
$|\beta_m|$ & $\mathcal{U}(0.0, 0.5)$ \\
$\arg(\beta_m)$ (deg) & $\mathcal{U}(-180, 180)$ \\
$f_{\rm Gauss}$ & $\mathcal{U}(0.05, 4.0)$ \\
$W_{\rm Gauss}$ (\uas) & $\mathcal{U}(20, 200)$ \\
\midrule
$\sigma_{\text{ref}}$ & $\mathcal{N}_L(\log(0.004), 1)$ \\
$f$ & $\mathcal{N}_L(\log(0.01), 1)$ \\
$a_4$ & $\mathcal{U}_L(10^{-4.39784},10^{-4.27339})$ \\
$b$ & $\mathcal{U}(2.35213,3.37849)$ \\
$c$ & $\mathcal{U}(1.5, 2.5)$ \\
$u_0$ (G$\lambda$) & $\mathcal{U}_L(10^{-1.09771},10^{0.236534})$ \\
\enddata
\tablecomments{Prior distributions for \themis full-track geometric modeling analyses.  The top section lists priors for the \model model parameters, and the bottom section lists priors for the parameterized noise model. $\mathcal{U}(a,b)$ denotes a uniform prior on the interval $[a,b]$, $\mathcal{U}_L(a,b)$ denotes a log-uniform prior on the interval $[a,b]$, and $\mathcal{N}_L(\mu,\sigma^2)$ denotes a lognormal prior with mean $\mu$ and variance $\sigma^2$. Priors for the variability noise parameters $a_4$, $b$, $c$, and $u_0$ are informed by the model-agnostic variability quantification analysis described in \autoref{sec:pre-modeling}.}
\end{deluxetable}

\begin{figure}[t]
    \centering
    \includegraphics[width=\linewidth]{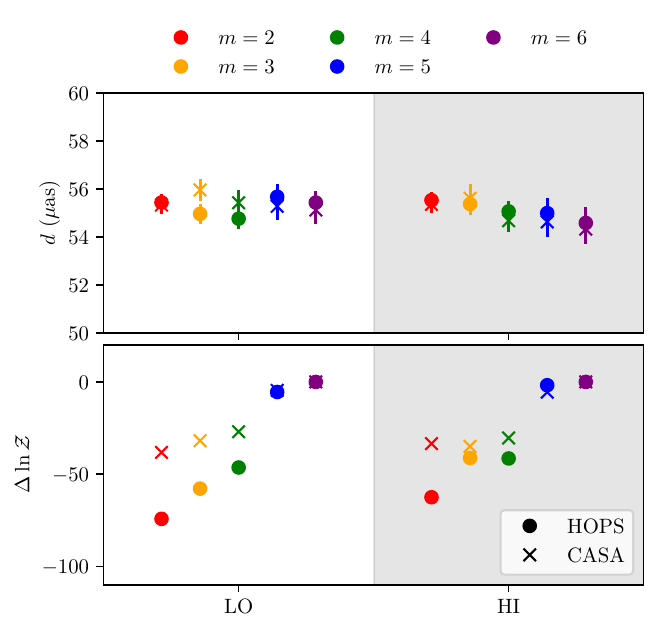}
    \caption{Relative Bayesian evidence and median posterior \model model diameter vs. $m$-order, from the full-track geometric modeling applied to the \sgra data; error bars on the diameter measurements show $1\sigma$ credible intervals.  Because the absolute values of $\ln\mathcal{Z}$ can be substantially different for each data set (i.e., each combination of frequency band and calibration pipeline), and because only the relative values carry information about model specification preferences, we reference all $\ln\mathcal{Z}$ values to the maximum value attained at any $m$ for that data set.}
    \label{fig:fulltrack-morder-survey}
\end{figure}

\subsubsection{\themis implementation}

We have implemented the combined \model plus noise parameterization as a model within the sampling-based parameter estimation framework \themis developed for the EHT \citep{Broderick_2020a,Broderick_2020b}.  Given a model specification and a data set, \themis works within a Bayesian formalism to produce a set of samples from the posterior distribution of the model parameters.  \themis uses a MCMC sampling scheme to explore the posterior space, employing a parallel tempering scheme \citep{Syed_2019} to ensure traversal over the entire prior volume and the Hamiltonian Monte Carlo sampling kernel from the
Stan package \citep{carpenter2017stan} to efficiently sample within each tempering level.  A detailed description of the \themis sampling framework can be found in \cite{TiedeThesis}.

The full-track geometric modeling analyses carried out in this paper fit to complex visibility data.  Given a vector of geometric model parameters $\boldsymbol{p}$ and a vector of noise model parameters $\boldsymbol{n}$, the \themis likelihood function for complex visibilities is Gaussian,
\begin{equation}
\ln \mathcal{L} = - \frac{1}{2} \sum_i \left( \frac{| V_i - \hat{V}(\boldsymbol{u}_i ; \boldsymbol{p}) |^2}{\sigma_i^2(\boldsymbol{n})} + \ln\left[ 2 \pi \sigma_i^2(\boldsymbol{n}) \right] \right) .
\end{equation}
\noindent Here $V_i$  is a measured visibility, $\boldsymbol{u}_i$ is its baseline vector, $\hat{V}$ is the corresponding modeled visibility,
and the sum is taken over all data points $i$.  The noise $\sigma_i(\boldsymbol{n})$ in each visibility is specified as in \autoref{eqn:FullTrackNoise}, with the noise model parameters $\boldsymbol{n} = \{ f, \sigma_{\text{ref}}, a, b, c, u_0 \}$.  \themis internally solves for and marginalizes over the full set of complex gain parameters (i.e., one complex gain per station per time stamp) at every sampling step using a Laplace approximation \citep[see][]{Broderick_2020a}.  It also applies the \citet{Johnson_2018} diffractive scattering kernel directly to the model prior to computing visibilities.  A validation test of the \themis \model plus noise model implementation is described along with other tests in a dedicated paper on the noise modeling approach  \citep{NoiseModeling}.

We assess MCMC convergence through both visual inspection of the traces and a number of quantitative chain statistics, including the integrated autocorrelation time, split-$\hat{R}$, and parameter rank distributions \citep{vehatri2019}.  The number of tempering levels is selected to ensure efficient communication between the highest and lowest levels \citep[per][]{Syed_2019}, which typically requires about 20 levels.  We run the sampler for between $5 \times 10^4$ and $10^5$ steps per tempering level.

To compute the Bayesian evidence, \themis uses thermodynamic integration \citep[e.g.,][]{LartillotTI}, which computes the log-evidence through
\begin{equation}\label{eq:TI}
    \ln \mathcal{Z} = \int_{0}^1{\rm d}\beta \left<\ln \mathcal{L}\right>_{\beta},
\end{equation}
where $\left<\ln \mathcal{L}\right>_{\beta}$ is the expectation of the log-likelihood taken over the distribution 
\begin{equation}
    P^{(\beta)}(\bm{p},\bm{n}) = \left(\frac{P(\bm{p},\bm{n})}{\pi_{\rm ref}(\bm{p},\bm{n})}\right)^\beta \pi_{\rm ref}(\bm{p}, \bm{n}).
\end{equation}
Note that \themis does not take $\pi_{\rm ref}$ to be the prior distribution. Instead, \themis uses a uniform distribution whose support matches the support of the priors given in \autoref{tab:fulltrack:model_priors}. To compute \autoref{eq:TI}, we compute the average log-likelihood for each tempering level and then use trapezoidal integration to numerically compute the integral.

Priors for all \model model and noise model parameters are listed in \autoref{tab:fulltrack:model_priors}.  We impose mean-zero lognormal priors for all station gain amplitudes, with a log standard deviation of 0.01 for all network calibrated stations (ALMA, APEX, JCMT, SMA), 0.2 for the LMT, and 0.1 for the remaining stations (PV, SMT, SPT); these gain priors are motivated by the expected performance of each station after the post-processing described in \autoref{sec:fulltrack:dataprep} \citepalias[see also][]{PaperII}.  All station gain phase priors are uniform on the unit circle.

\subsection{Model selection}

As with the snapshot geometric analyses (see \autoref{sec:snapshot:model_selection}), the \model model used for the full-track analyses is really a class of models that increases in complexity with $m$.  The results of a \themis $m$-order survey covering $m=\{2,3,4,5,6\}$ are shown in \autoref{fig:fulltrack-morder-survey}.  In contrast to the snapshot geometric modeling (see \autoref{sec:snapshot:model_selection}), we find that the full-track analysis is able to support more complex model specifications, exhibiting a strong preference for $m>4$ over $m=4$.
However, we find that increasing the $m$-order does not significantly impact the values of the primary morphological parameters of choice (see the similar diameter measurements and uncertainties for $m \geq 4$ in \autoref{fig:fulltrack-morder-survey}).  Thus, to maintain consistency among model specifications and to facilitate comparison with the snapshot geometric modeling analyses, we proceed with the $m=4$ \model as our fiducial model for all full-track \sgra analyses in this paper.

\begin{figure}[t]
    \centering
    \includegraphics[width=\columnwidth]{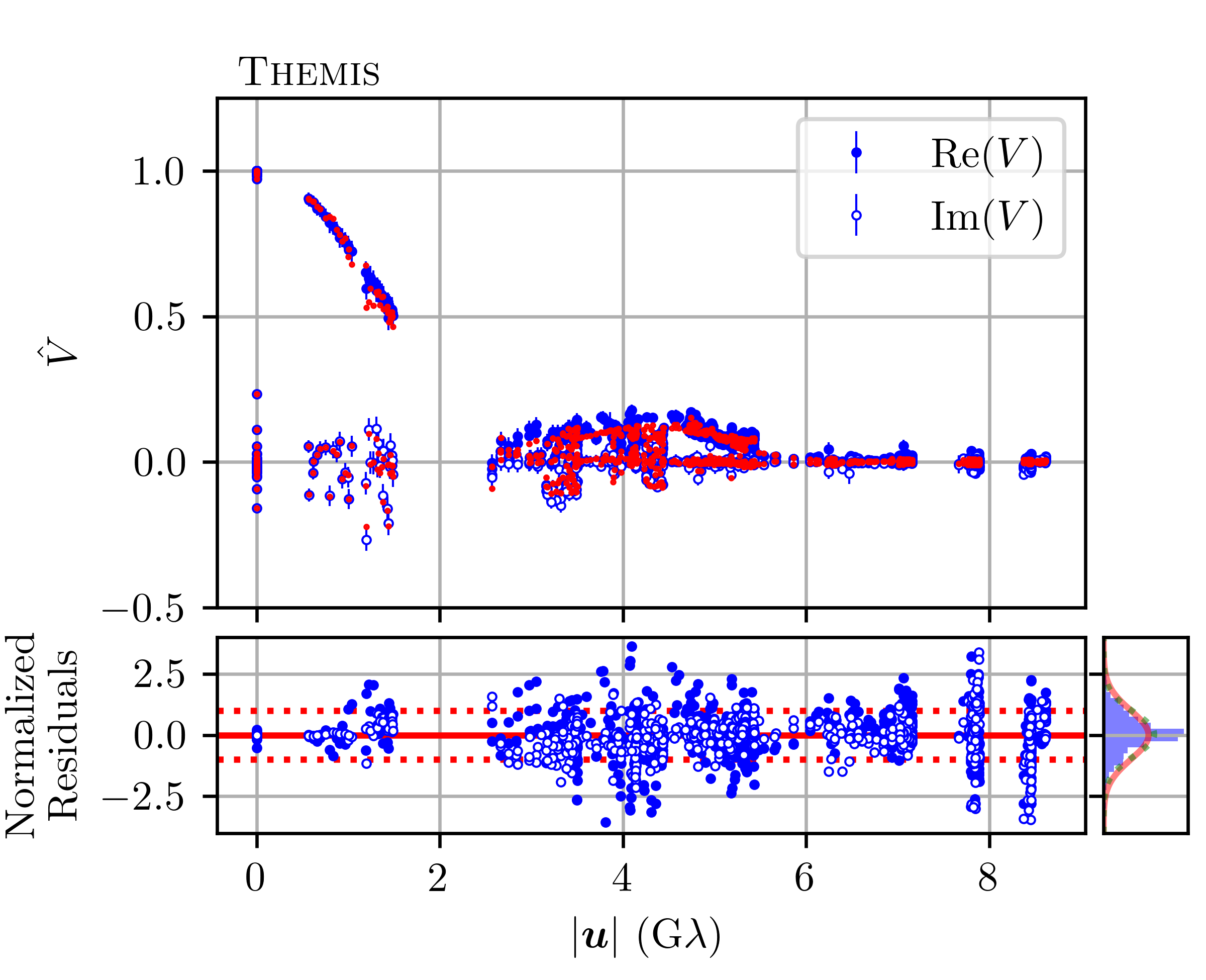}
    \caption{Results of full-track modeling using an $m=4$ \model fit to the \sgra HOPS low-band data on April 6 and 7, arranged analogously to the individual panels of \autoref{fig:imaging_residuals}.  As in \autoref{fig:imaging_residuals}, $\hat{V}$ denotes light-curve-normalized complex visibilities.}
    \label{fig:fulltrack_residuals}
\end{figure}

A representative $m=4$ \model fit to the \sgra HOPS low-band data is shown in \autoref{fig:fulltrack_residuals}.  We find that the normalized residuals are distributed around a value of zero with near-unity variance and that there is no evidence of systematic structure.  The $\chi^2$ statistics for this fit are discussed in \autoref{app:Chisq}.

%% file: results.tex
\section{Results}\label{sec:results}

In this section we aggregate and present the results from the analyses described in Sections \ref{sec:Variability} through \ref{sec:FullTrackGeometricModeling} as applied to the 2017 EHT \sgra data.  

\begin{figure}[t]
  \begin{center}
    \includegraphics[width=\columnwidth]{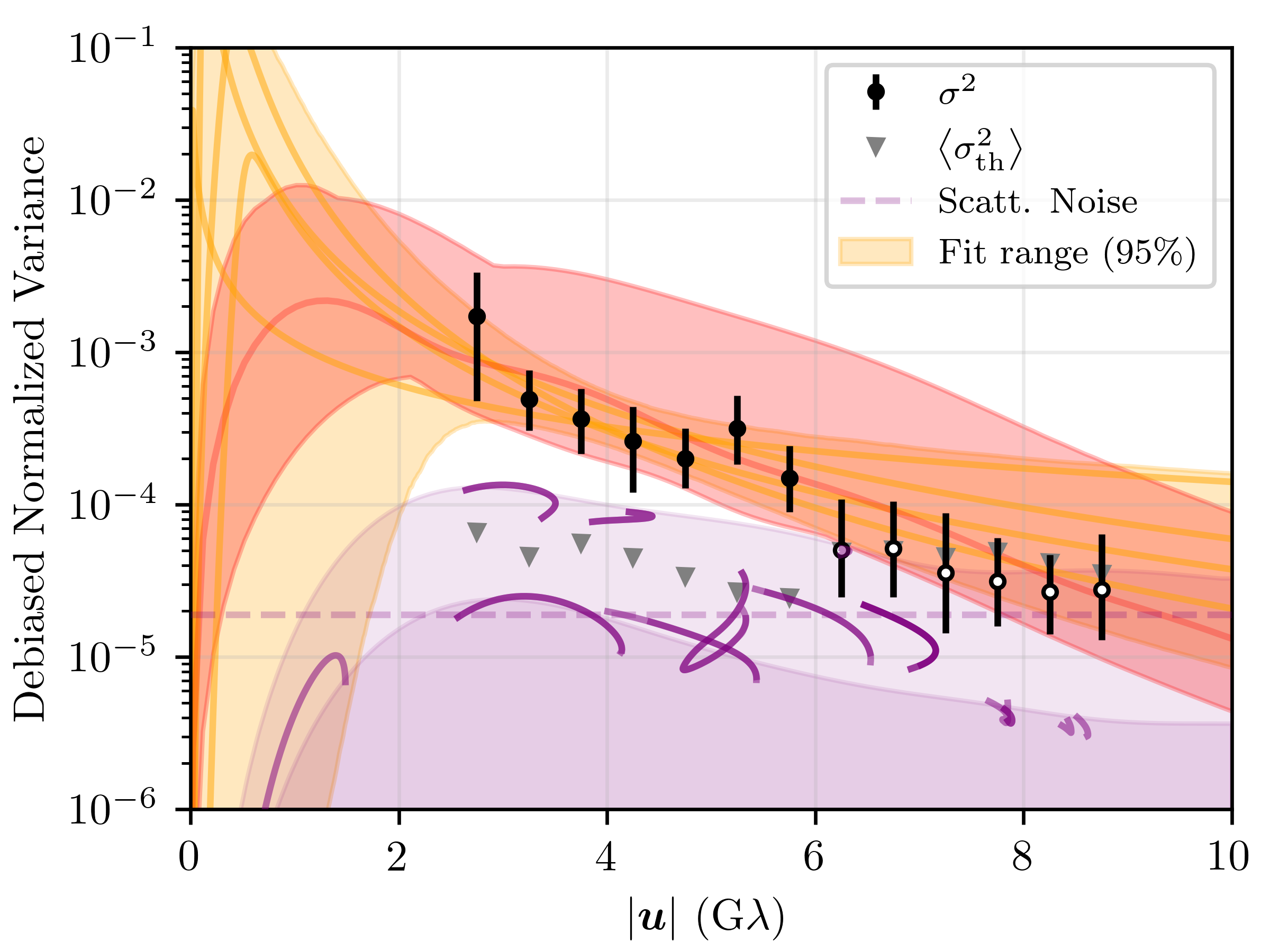}
  \end{center}
\caption{Similar to \autoref{fig:sevv1}, but after the direct visibility estimates have been debiased to account for the short-time temporal correlations as described in \citet{NoiseModeling}.  These estimates are directly comparable to the power spectra implied by GRMHD simulations \citep{Georgiev_2022}. A single example and the range associated with the library presented in \citetalias{PaperV} are shown by the red line and band, respectively.}
\label{fig:sevvcf}
\end{figure}

\begin{figure}
  \begin{center}
    \includegraphics[width=\columnwidth]{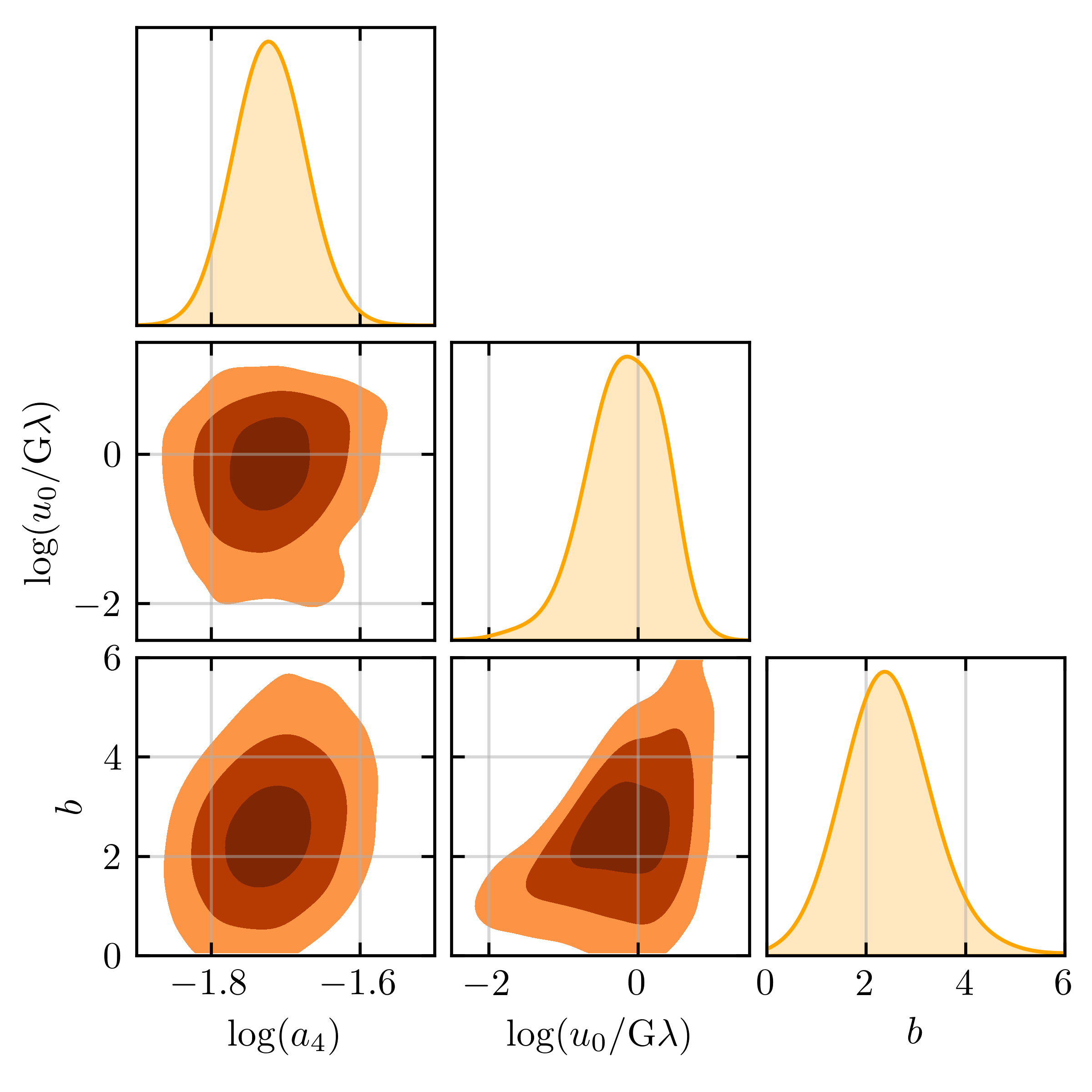}
  \end{center}
\caption{Joint posteriors of the constrained parameters after fitting a broken power law to the debiased model-agnostic normalized variance estimates, specifically the normalization at $|\boldsymbol{u}|=4\,{\rm G}\lambda$, denoted by $a_4$, the break location $u_0$, and the long-baseline power-law index $b$.
Contours enclose 50\%, 90\%, and 99\% probability.}
\label{fig:debiased_bpltri}
\end{figure}

\begin{deluxetable}{lcc}
\tablecolumns{3}
\tabletypesize{\small}
\tablewidth{0pt}
\tablecaption{Debiased variability parameters \label{tab:debiased_noise}}
\tablehead{
\colhead{Baseline length (\Gl)} & \colhead{Normalized variance} & \colhead{Fit}
}
\startdata
\multicolumn{1}{c}{2.75} & $(17.2 \pm 14.3)\times10^{-4}$ & Y \\
\multicolumn{1}{c}{3.25} & $(4.9 \pm 2.3)\times10^{-4}$ & Y \\
\multicolumn{1}{c}{3.75} & $(3.7 \pm 1.8)\times10^{-4}$ & Y \\
\multicolumn{1}{c}{4.25} & $(2.6 \pm 1.6)\times10^{-4}$ & Y \\
\multicolumn{1}{c}{4.75} & $(2.0 \pm 0.9)\times10^{-4}$ & Y \\
\multicolumn{1}{c}{5.25} & $(3.2 \pm 1.7)\times10^{-4}$ & Y \\
\multicolumn{1}{c}{5.75} & $(1.5 \pm 0.8)\times10^{-4}$ & Y \\
\multicolumn{1}{c}{6.25} & $(0.5 \pm 0.4)\times10^{-4}$ & N \\
\multicolumn{1}{c}{6.75} & $(0.5 \pm 0.4)\times10^{-4}$ & N \\
\multicolumn{1}{c}{7.25} & $(0.4 \pm 0.4)\times10^{-4}$ & N \\
\multicolumn{1}{c}{7.75} & $(0.3 \pm 0.2)\times10^{-4}$ & N \\
\multicolumn{1}{c}{8.25} & $(0.3 \pm 0.2)\times10^{-4}$ & N \\
\multicolumn{1}{c}{8.75} & $(0.3 \pm 0.3)\times10^{-4}$ & N \\
\hline\hline
\colhead{Quantity} & \colhead{Symbol} & \colhead{Estimate\tablenotemark{a}}\\
\hline
Excess noise at $|\boldsymbol{u}|=4$\,\Gl  & $a_4$   & $1.90\pm0.2~\%$\\
Long-baseline power-law index & $b$     & $2.4\pm0.8$\\
Break baseline length         & $u_0$   & $<1.3~{\rm G}\lambda$\\
\enddata
\tablecomments{Measurements of intrinsic variability from the EHT \sgra data.  The top portion of the table lists the visibility amplitude variance measurements from \autoref{sec:VariabilityMeasurement} after applying the debiasing procedure described in \cite{NoiseModeling}; these values correspond to the black points plotted in \autoref{fig:sevvcf}.  The bottom portion of the table lists the best-fit broken power-law model parameters for the variability measurements.}
\tablenotetext{a}{Quoted are the median values and $1\sigma$ ranges.  Upper limits are $1\sigma$ limits.}
\end{deluxetable}

\subsection{Structural variability measurements}\label{sec:VariabilityMeasurement}

The model-agnostic variability quantification analysis carried out in \autoref{sec:Variability} demonstrates that the \sgra data exhibit variability -- quantified here in terms of a normalized visibility amplitude variance -- that is significantly in excess of that expected from thermal noise, station gains, and refractive scattering.  As illustrated in \autoref{fig:sevv1}, the measured variability can be broken down into three regions with qualitatively distinct behavior:

\begin{enumerate}
    \item On short baselines with lengths $|\boldsymbol{u}| \lesssim 2.5~{\rm G}\lambda$, corresponding to spatial scales $\gtrsim$100\,\uas, limitations in our calibration and the subsequent choices made in our preprocessing procedure preclude meaningful constraints on the variability.  The light-curve normalization procedure removes all variability on intrasite baselines and suppresses variability on short intersite baselines that are highly correlated with the light curve; the variability of the light curve itself is thoroughly characterized in \cite{Wielgus2022}.  The source size constraint used to perform gain calibration of the LMT-SMT baseline \citepalias[see][]{PaperII,PaperIII} imposes a further, more artificial suppression of the variability on this baseline.  We thus do not obtain any variability measurements for baselines shorter than $2.5~{\rm G}\lambda$.
    \item On intermediate baselines with lengths between $2.5~{\rm G}\lambda \lesssim |\boldsymbol{u}| \lesssim 6~{\rm G}\lambda$, corresponding to spatial scales between $\sim$30 and $\sim$100\,\uas, we measure significant variability that exhibits an approximately power-law decline with increasing baseline length.  The power-law index is between $\sim$2 and 3, and the magnitude of the variability ranges from a peak RMS of $\sim$5\% of the total flux density ($\sim$120~mJy) near $2.5~{\rm G}\lambda$ down to $\sim$1\% of the total flux density ($\sim$25~mJy) near $6~{\rm G}\lambda$.
    \item On long baselines with lengths $|\boldsymbol{u}| \gtrsim 6~{\rm G}\lambda$, corresponding to spatial scales $\lesssim$30\,\uas, the measured variability is comparable in magnitude to that expected from statistical errors and refractive scattering.  These measurements thus do not contain statistically significant detections of structural variability.
\end{enumerate}

\noindent These measurements describe the level of excess variance that the data exhibit about an underlying average source model.  The parameters describing a broken power-law noise model fit to these measurements are thus used as a variability noise budget during image reconstruction \citepalias{PaperIII} and to define priors on the corresponding parameters in the full-track modeling analyses (\autoref{sec:FullTrackGeometricModeling}).

\begin{figure*}[t]
    \centering
    \includegraphics[width=\textwidth]{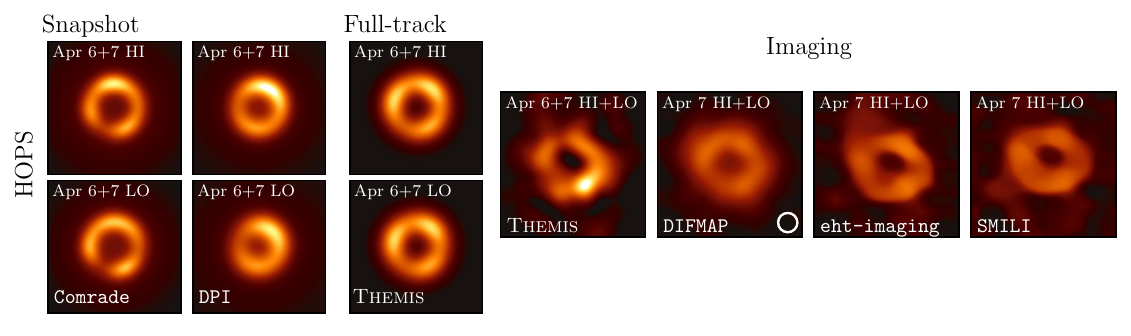}\\
    \includegraphics[width=\textwidth]{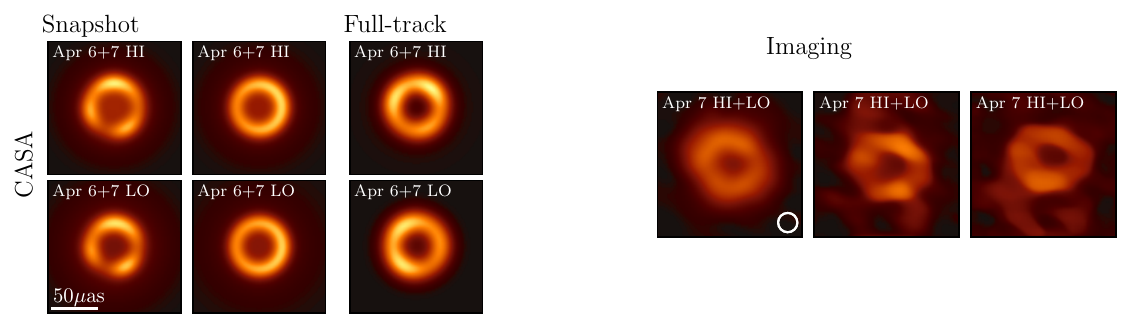}\\
    \caption{Mean images for each geometric modeling and image reconstruction pipelines applied to the \sgra data, showing both low band and high band separately for the geometric modeling and the combined bands for image reconstruction \citepalias[from][]{PaperIII}.  The geometric modeling and \themis imaging pipelines have been applied to the combined April 6 and 7 data, while the \difmap, \ehtim, and \smili imaging pipelines have been applied to the April 7 descattered data; \autoref{fig:morphology_april6} and \autoref{fig:morphology_april7} show single-day results for all pipelines on April 6 and 7, respectively.  The upper group of images have been produced from the HOPS calibration pipeline \sgra data, while the bottom group of images correspond to the CASA calibration pipeline. All of the images share a common brightness color scale; the absolute brightness scale is arbitrary because each image has been normalized to have unit total flux density. 
    }
    \label{fig:morphology}
\end{figure*}

Determining the intrinsic (i.e., infinite-time) source variability from these measurements requires an additional debiasing step to remove the impact of correlations between data points that are closely spaced in time.  The analysis carried out in \autoref{sec:Variability} involves binning the visibility data in the \uv-plane for the purpose of computing variances.  However, many data points within a single bin are from measurements taken close in time, which can introduce correlations that bias the computed variance.  A procedure for removing this bias is detailed in \cite{NoiseModeling}, whereby the factor relating the measured variability to the intrinsic variability at every baseline length is calibrated using synthetic measurements of GRMHD simulations.  In practice, \cite{NoiseModeling} derive the debiasing function using the same 90 GRMHD simulations we use in this paper for $\theta_g$ calibration (see \autoref{sec:Calibration} and \autoref{sec:MoDSyntheticData}), and the resulting debiasing factor is close to unity everywhere except between $\sim$6 and 7.5~${\rm G}\lambda$ (see Fig. 5 in \citealt{NoiseModeling}).  Applying this debiasing function to the variability measurements from the \sgra data yields the results shown in \autoref{fig:sevvcf} and reported in \autoref{tab:debiased_noise}.  

Due to the near-unit debiasing factor, the variability measurements shown in \autoref{fig:sevvcf} are similar to those from \autoref{fig:sevv1}. Quantitative constraints on the parameters of the noise model that are well constrained are presented in \autoref{fig:debiased_bpltri} and \autoref{tab:debiased_noise}. Where strong constraints on the the excess noise exist (i.e., on baseline lengths between $\sim$2 and 6\,G$\lambda$), it continues to be well described by a single power law with index 
$b=2.4^{+0.8+2.1}_{-0.8-1.6}$
and amplitude (evaluated at a baseline length of $4\,{\rm G}\lambda$) of 
$a_4 = 1.9^{+0.2+0.5}_{-0.2-0.3}\%$,
where in each value the $1\sigma$ and $2\sigma$ ranges are indicated.  The measured variability magnitude is between $\sim$2 and 10 times higher than that expected from refractive scattering alone. The lack of an observable break places an upper limit on its location of $u_0<1.3\,{\rm G}\lambda$ at $1\sigma$ and $u_0<3.1\,{\rm G}\lambda$ at $2\sigma$.

The excess variability is broadly consistent with that due to structural fluctuations anticipated by the GRMHD simulations discussed in \citetalias{PaperV} and \citet{Georgiev_2022}.  
The magnitude of the excess variability lies within the range of that predicted by GRMHD simulations for $|\boldsymbol{u}|>2~{\rm G}\lambda$, though it does appear to marginally favor less variable models.
The long-baseline power-law index is consistent with all GRMHD simulations.  A detailed discussion of the implications for GRMHD models is contained within \citetalias{PaperV}.

\movetabledown=60mm
\begin{rotatetable*}
\begin{deluxetable*}{lcccCCCCCcCCCCC}
\tablecolumns{15}
\tabletypesize{\scriptsize}
\tablewidth{0pt}
\tablecaption{Parameters describing \sgra image morphology \label{tab:SgrAMorphology}}
\tablehead{ & & & & \multicolumn{5}{c}{CASA} & & \multicolumn{5}{c}{HOPS} \\ \cline{5-9} \cline{11-15}
\colhead{Analysis class} & \colhead{Software} & \colhead{Day} & \colhead{Band} & \colhead{$\hat{d}$ (\uas)} & \colhead{$W$ (\uas)} & \colhead{$A$} & \colhead{$\eta$ (deg)} & \colhead{$f_c$} & & \colhead{$\hat{d}$ (\uas)} & \colhead{$W$ (\uas)} & \colhead{$A$} & \colhead{$\eta$ (deg)} & \colhead{$f_c$}}
\startdata
\multirow{8}{*}{Imaging} & \difmap + \rex & April 7 & HI+LO & 50.6_{-3.0}^{+1.1} & 31.9_{-3.1}^{+4.2} & 0.09_{-0.04}^{+0.07} & 48.4 \pm 57.1 & 0.43_{-0.10}^{+0.12} & & 50.1_{-2.2}^{+1.2} & 31.8_{-2.7}^{+3.3} & 0.05_{-0.02}^{+0.09} & -22.8 \pm 69.6 & 0.42_{-0.09}^{+0.10} \\
 & \difmap + \vida & April 7 & HI+LO & 51.2_{-2.8}^{+1.2} & 33.4_{-1.8}^{+4.1} & 0.10_{-0.04}^{+0.11} & 49.8 \pm 61.1 & 0.40_{-0.09}^{+0.14} & & 51.3_{-1.4}^{+1.3} & 33.1_{-1.1}^{+2.2} & 0.07_{-0.03}^{+0.18} & -28.1 \pm 57.1 & 0.39_{-0.08}^{+0.15} \\ \cline{2-15}
 & \ehtim + \rex & April 7 & HI+LO & 53.0_{-2.5}^{+1.9} & 25.7_{-3.2}^{+3.6} & 0.13_{-0.06}^{+0.08} & -33.9 \pm 56.3 & 0.33_{-0.18}^{+0.12} & & 53.6_{-1.1}^{+1.8} & 27.2_{-2.6}^{+1.6} & 0.11_{-0.03}^{+0.05} & -81.0 \pm 62.6 & 0.21_{-0.10}^{+0.11} \\
 & \ehtim + \vida & April 7 & HI+LO & 52.7_{-2.1}^{+5.6} & 28.0_{-3.2}^{+16.3} & 0.19_{-0.11}^{+0.26} & -40.5 \pm 65.2 & 0.30_{-0.16}^{+0.22} & & 53.5_{-1.5}^{+3.6} & 27.4_{-2.1}^{+2.5} & 0.11_{-0.05}^{+0.23} & -94.4 \pm 60.9 & 0.19_{-0.05}^{+0.13} \\ \cline{2-15}
 & \smili + \rex & April 7 & HI+LO & 52.0_{-3.0}^{+5.5} & 25.1_{-4.4}^{+1.1} & 0.16_{-0.06}^{+0.04} & 130.9 \pm 70.3 & 0.29_{-0.20}^{+0.09} & & 51.4_{-4.8}^{+4.0} & 26.7_{-3.0}^{+1.2} & 0.12_{-0.03}^{+0.04} & 135.3 \pm 90.0 & 0.23_{-0.13}^{+0.08} \\
 & \smili + \vida & April 7 & HI+LO & 49.3_{-1.7}^{+13.6} & 28.5_{-3.7}^{+34.0} & 0.15_{-0.04}^{+0.21} & 124.4 \pm 77.0 & 0.26_{-0.07}^{+0.48} & & 50.6_{-1.1}^{+6.1} & 27.2_{-1.3}^{+0.8} & 0.07_{-0.01}^{+0.11} & 163.0 \pm 90.4 & 0.19_{-0.05}^{+0.09} \\ \cline{2-15}
 & \themis + \rex & April 6+7 & HI+LO & \ldots & \ldots & \ldots & \ldots & \ldots & & 51.7_{-0.3}^{+0.4} & 23.6_{-0.3}^{+0.3} & 0.07_{-0.01}^{+0.08} & -152.7 \pm 20.9 & 0.02_{-0.01}^{+0.01} \\
 & \themis + \vida & April 6+7 & HI+LO & \ldots & \ldots & \ldots & \ldots & \ldots & & 53.7_{-1.7}^{+0.4} & 24.7_{-0.4}^{+0.4} & 0.08_{-0.01}^{+0.05} & -137.1 \pm 15.7 & 0.08_{-0.01}^{+0.01} \\ \cline{2-15} \hline
\multirow{4}{*}{Snapshot} & \multirow{2}{*}{\rose} & April 6+7 & HI & 53.9_{-0.5}^{+0.6} & 16.4_{-0.2}^{+0.3} & 0.23_{-0.02}^{+0.01} & 3.8 \pm 5.3 & 0.23_{-0.01}^{+0.01} & & 52.3_{-0.6}^{+0.7} & 17.7_{-0.2}^{+0.2} & 0.23_{-0.04}^{+0.03} & -8.1 \pm 6.7 & 0.15_{-0.01}^{+0.01} \\
 &  & April 6+7 & LO & 51.4_{-0.6}^{+0.5} & 17.0_{-0.3}^{+0.2} & 0.22_{-0.03}^{+0.02} & -11.4 \pm 9.0 & 0.20_{-0.01}^{+0.01} & & 53.9_{-0.7}^{+0.8} & 17.7_{-0.2}^{+0.2} & 0.22_{-0.02}^{+0.02} & -0.2 \pm 6.6 & 0.18_{-0.01}^{+0.01} \\ \cline{2-15}
 & \multirow{2}{*}{\texttt{DPI}} & April 6+7 & HI & 50.9_{-1.0}^{+1.0} & 16.1_{-0.7}^{+0.7} & 0.21_{-0.14}^{+0.18} & -75.6 \pm 43.2 & 0.18_{-0.02}^{+0.02} & & 47.7_{-1.1}^{+1.1} & 19.1_{-0.7}^{+0.7} & 0.31_{-0.20}^{+0.13} & -44.9 \pm 11.7 & 0.12_{-0.01}^{+0.02} \\
 &  & April 6+7 & LO & 51.6_{-0.9}^{+0.9} & 16.2_{-0.7}^{+0.7} & 0.14_{-0.10}^{+0.07} & -97.5 \pm 9.5 & 0.14_{-0.01}^{+0.02} & & 45.3_{-1.7}^{+1.6} & 21.6_{-0.9}^{+0.9} & 0.31_{-0.19}^{+0.13} & -73.2 \pm 8.2 & 0.10_{-0.01}^{+0.01} \\ \cline{2-15} \hline
\multirow{2}{*}{Full-track} & \multirow{2}{*}{\themis} & April 6+7 & HI & 52.1_{-0.4}^{+0.4} & 19.7_{-0.4}^{+0.4} & 0.16_{-0.02}^{+0.02} & -4.0 \pm 11.8 & 0.09_{-0.02}^{+0.02} & & 52.0_{-0.5}^{+0.4} & 21.5_{-0.5}^{+0.5} & 0.14_{-0.02}^{+0.02} & 8.9 \pm 8.0 & 0.05_{-0.02}^{+0.02} \\
 &  & April 6+7 & LO & 52.7_{-0.5}^{+0.5} & 20.3_{-0.4}^{+0.4} & 0.15_{-0.02}^{+0.02} & -74.0 \pm 10.5 & 0.16_{-0.03}^{+0.03} & & 51.4_{-0.5}^{+0.4} & 22.5_{-0.5}^{+0.4} & 0.14_{-0.02}^{+0.02} & 1.6 \pm 7.5 & 0.03_{-0.01}^{+0.02} \\ \hline
\enddata
\tablecomments{Median values and 68\% credible intervals for the morphological quantities of interest, measured from the EHT \sgra data. Because medians and quantiles are not well defined for angular variables, for the position angle $\eta$ we instead quote the circular mean and standard deviation.}
\end{deluxetable*}
\end{rotatetable*}

\subsection{Image morphology measurements}\label{sec:Measurements}

Both the snapshot (\autoref{sec:SnapshotGeometricModeling}) and full-track (\autoref{sec:FullTrackGeometricModeling}) geometric modeling analyses produce reconstructions of the \sgra emission structure, and the posterior distributions determined for the parameters describing these geometric model reconstructions provide a quantification of the morphological properties directly from the EHT interferometric data.  Similarly, the IDFE analyses carried out in \autoref{sec:ImageDomain} quantify the morphological properties of the top-set and posterior images reconstructed in \citetalias{PaperIII}.

\autoref{fig:morphology} compares the geometric models and image reconstructions determined for \sgra; for the geometric modeling analyses we show posterior mean images (i.e., the mean of many images sampled from the posterior distribution),
while for the image reconstructions we show averages over the top sets (for \ehtim, \smili, and \difmap) or posterior means (for \themis).  We see that both the snapshot and full-track geometric modeling analyses each recover a grossly similar overall structure across frequency bands and that this structure is also similar between the snapshot and full-track analyses.  The image reconstructions permit much more flexibility in the permitted image structure, and so we see correspondingly more variation both within the imaging methods and between the imaging and geometric modeling.  The primary point of consistency between the ring-like structures recovered from imaging and the rings fit via geometric modeling seems to be their sizes.

\begin{figure*}[t]
    \centering
    \includegraphics[width=\textwidth]{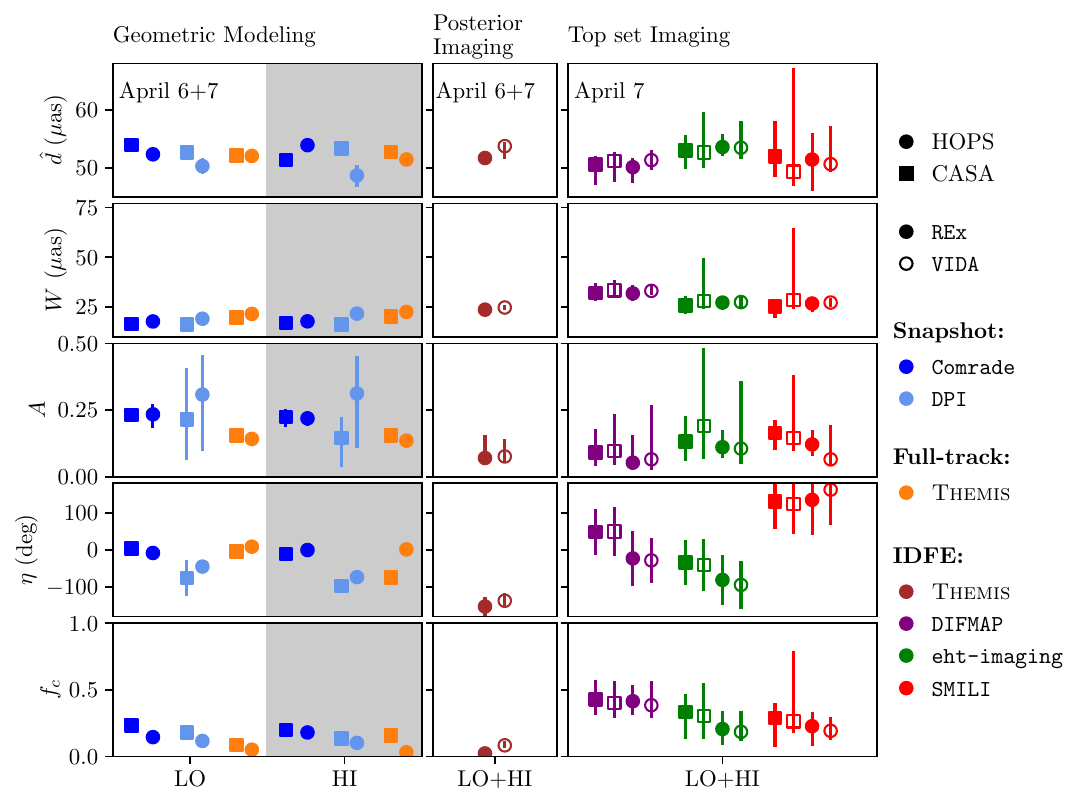}
    \caption{Morphological parameters measured from \sgra data using the geometric modeling (left columns), posterior imaging IDFE (middle column), and top-set imaging IDFE (right columns) analyses; each marker denotes a median value, and the error bars indicate 68\% credible intervals. For geometric and posterior imaging we show the combined April 6 and 7 results, while for top-set imaging we show just descattered April 7. Each row of panels shows the results for a single morphological parameter, with the markers colored according to the method used to make the measurement (per the legend on the right).  Measurements made using \sgra data from both calibration pipelines are shown using circular and square markers for HOPS and CASA, respectively.  IDFE measurements made using \rex and \vida are indicated by filled and open markers, respectively; no metronization cuts have been applied to either top-set or posterior images.  The IDFE analyses have been applied to images reconstructed using both frequency bands simultaneously (LO+HI), while the geometric modeling analyses have been applied to each frequency band separately (LO and HI).}
    \label{fig:quant_struct}
\end{figure*}

We use the geometric modeling and IDFE analyses to quantify a number of morphological parameters of interest, which are shown in \autoref{fig:quant_struct} and listed in \autoref{tab:SgrAMorphology}.

\subsubsection{Ring size} \label{sec:RingSize}

The parameter of most interest for gravitational studies \citepalias[e.g.,][]{M87PaperVI,PaperVI} is the ring size, which we quantify using its diameter.  For the \model modeling results from both snapshot and full-track analyses, we report a debiased diameter $\hat{d}$, given by

\begin{equation}\label{eq:moran_bias}
    \hat{d} = d - \frac{1}{4\ln(2)}\frac{W^2}{d} ,
\end{equation}

\noindent where $d$ and $W$ are the \model ring diameter and thickness, respectively (see \autoref{sec:mring}).  This debiasing corrects for the lowest-order impact of the Gaussian blurring kernel on the radial location of the peak intensity -- which is shifted inward with respect to the radius of the pre-convolved ring \citepalias[see Appendix G of][]{M87PaperIV} -- and thus aids a more direct comparison of the geometric modeling diameter values with those obtained from IDFE.

The diameter measurements from the geometric modeling and IDFE analyses are compared in the top row of \autoref{fig:quant_struct} across frequency bands and calibration pipelines.  We find that the diameter is the most well constrained of the geometric parameters we attempt to quantify, with both a typical measurement uncertainty and a scatter between measurement types that is substantially smaller than the magnitude of the value itself. An average of the geometric modeling results across both frequency bands and calibration pipelines yields a debiased diameter of $51.9$\,\uas, with a corresponding symmetrized uncertainty of $2.0$\,\uas. The quoted error is the  68\% (i.e., approximately $1\sigma$) probability and corresponds to the samples from each measurement weighted equally. The IDFE measurements are broadly consistent with the results from geometric modeling, yielding $\hat{d} = 51.8 \pm 2.6$\,\uas. The corresponding joint constraint from both geometric modeling and IDFE analyses yields $\hat{d} = 51.8 \pm 2.3$\,\uas.

We note that even after debiasing the diameter measurements from different analysis pathways remain interpretationally distinct quantities.  To ensure mutual consistency between different measurement methods, in \autoref{sec:MassMeasurement} we calibrate the diameter measurements to a common physical scale using the GRMHD synthetic data sets generated for this purpose (see \autoref{sec:Calibration}).

\subsubsection{Ring thickness}

The thickness of the ring is of interest for its ability to constrain the location and size of the emitting region near the black hole \citep[e.g.,][]{Lockhart_2021}.  The ring thickness measurements from the geometric modeling and IDFE analyses are compared in the second row of \autoref{fig:quant_struct} across frequency bands and calibration pipelines.  We find that the geometric modeling methods recover similar ring thicknesses, with the snapshot analyses obtaining $\sim$16--22\,\uas and the full-track analyses obtaining $\sim$19--23\,\uas.  The IDFE analyses obtain consistently thicker rings, with $\sim$30\,\uas being a more typical value and a somewhat larger scatter (from $\sim$25 to 35\,\uas) seen both across and within pipelines.  The increased ring thickness recovered from the IDFE analyses likely arises in part from image smoothing introduced by some reconstruction algorithms (e.g., the CLEAN algorithm used by \difmap).

The thickness of the \m87 ring was a parameter that the geometric modeling analyses carried out in \citetalias{M87PaperVI} had difficulty constraining, and it showed substantial variation across days and between modeling methods; only a relatively weak upper limit for the fractional thickness of $W/\hat{d} \lesssim 0.5$ was obtained.  In contrast, we find for \sgra that the thickness parameter is relatively well constrained by geometric modeling approaches. The fractional thickness is $W/\hat{d} = 0.35 \pm 0.05$, where the uncertainties quoted are symmetrized $1\sigma$. IDFE analyses obtain systematically larger fractional thicknesses, finding $W/\hat{d} = 0.53 \pm 0.1$.

Unlike with the diameter measurements, we do not debias the ring thicknesses obtained from different analysis pathways or attempt to calibrate them to a common scale.  The ring thicknesses from geometric modeling and IDFE thus represent two interpretationally distinct quantities, and we do not produce an analysis-agnostic measurement of the ring thickness.

\subsubsection{Position angle and asymmetry}

The magnitude and orientation of any asymmetry in the azimuthal brightness distribution around the ring is of interest because it can be related to the spin and inclination of the black hole \citepalias[e.g.,][]{M87PaperV,PaperV}.  As described in \autoref{sec:mring}, the \model position angle ($\eta$) and degree of azimuthal asymmetry ($A$) are both determined by the coefficient of the $m=1$ mode as specified in \autoref{eqn:PositionAngle} and \autoref{eqn:Asymmetry}, respectively.  These definitions match closely the corresponding IDFE quantities defined in \autoref{subsec:ImageDomain_FEM}.  The asymmetry and position angle measurements we obtain for \sgra are shown in the third and fourth rows of \autoref{fig:quant_struct}, respectively.

Unlike for \m87, where the image morphology exhibits a clearly defined asymmetry axis whose magnitude and orientation can be consistently quantified using either geometric modeling or IDFE analyses \citepalias{M87PaperIV,M87PaperVI}, the image structure for \sgra is less amenable to such a description.  The asymmetry magnitude measurements show a large scatter across methods, spanning $\sim$0.15--0.3 for the geometric modeling methods and $\sim$0.04--0.20 for the IDFE methods.  The IDFE methods recover systematically smaller median levels of asymmetry than the geometric modeling methods, but the uncertainties are large; several of the measurement methods have statistical uncertainties that cover nearly the entire 0--0.5 range of the prior distribution for $A$.

The position angle measurements show similarly little consistency between different analysis methods, spanning essentially the full $(-180^{\circ},180^{\circ})$ range when compared across all data sets and measurement techniques.  The geometric modeling analyses find position angles that are loosely confined to a region between ${\sim}-100^{\circ}$ and $0^{\circ}$ across, but the IDFE analyses show a large ($>$100$^{\circ}$) scatter between methods and a similar magnitude of uncertainty for individual measurements.

\subsubsection{Brightness depression}

The depth of the brightness depression interior to the ring is a key signature of the presence of a black hole.  Additionally, it can be used to constrain the presence of an emitting or reflecting surface, as a potential alternative to a horizon (e.g., \citealt{Broderick_2015}, \citetalias{PaperVI}).  For the \model model, the fractional central flux $f_c$ is given by \autoref{eqn:FractionalCentralFlux}.  For the IDFE analyses, we retain the definitions from \autoref{subsec:ImageDomain_FEM} for $f_c$.

The $f_c$ measurements for \sgra are shown in the bottom row of \autoref{fig:quant_struct}.  We find a large spread in values across analysis methods, ranging from $\sim$0.1 to 0.25 for the geometric modeling analyses and from $\sim$0.0 to 0.5 for the IDFE analyses.  Compared against the constraints from geometric modeling of the \m87 ring structure, which consistently found $f_c \lesssim 0.1$ \citepalias{M87PaperVI}, the results obtained here for \sgra allow for the possibility of substantially more emission interior to the ring.

\subsection{Gravitational radius and mass} \label{sec:MassMeasurement}

The ring size measurements presented in \autoref{sec:RingSize} have been made using a variety of different analysis techniques, with different inherent assumptions and biases.  To bring these otherwise disparate measurement techniques to a common scale, we follow a strategy similar to that developed in \m87 in \citetalias{M87PaperVI} and calibrate the diameter measurements using simulations from the \citetalias{PaperV} GRMHD library.  As described in \autoref{sec:Calibration}, our calibration suite consists of synthetic data sets constructed from 90 GRMHD simulations spanning a range of accretion flow and black hole parameters, and for which an absolute reference size scale (i.e., the angular size of the gravitational radius $\theta_g$) is known.  We apply the same data processing and ring diameter measurement strategies as used for the \sgra data to each of these synthetic data sets, and we use the resulting distribution of diameter measurements to derive the value and uncertainty in the scaling factor ($\alpha$) between $\theta_g$ and $d$ for every method (\autoref{eqn:Alpha}).  We note that a conceptually similar calibration is carried out in the companion \citetalias{PaperVI}, in which the GRMHD assumption is relaxed and a more diverse set of spacetimes and accretion flow models is used to calibrate ring size measurements.

\subsubsection{Calibrated scaling factors}

When applied to the calibration suite datasets, each diameter measurement technique produces a discrete distribution of $\alpha$ scaling factors.  We use a kernel density estimator (KDE) from the \texttt{scikit-learn} package \citep{scikit-learn} to produce a nonparametric estimate of the continuous distribution corresponding to these discrete samples, and we use this KDE to construct the $\theta_g$ distribution for \sgra described below (\autoref{sec:MoDResults}).

\autoref{tab:alphacal} lists the derived $\alpha$ value and its uncertainty -- as computed from the KDE distribution -- for each of the ring size measurement methods used in this paper.  The uncertainty in $\alpha$ contains two main components: a statistical uncertainty $\sigma_{\alpha}^{\text{(stat)}}$ associated with the fidelity of the ring measurement from each data set, and a theoretical uncertainty $\sigma_{\alpha}^{\text{(theory)}}$ associated with the intrinsic scatter in $\alpha$ as measured across different GRMHD calibration data sets.  The total uncertainty $\sigma_{\alpha}^{\text{(tot)}}$ is a combination of both the statistical and theoretical uncertainties, meaning that in practice we do not have access to $\sigma_{\alpha}^{\text{(theory)}}$ in isolation, and thus we do not report it in \autoref{tab:alphacal}.  Nevertheless, the relative values of $\sigma_{\alpha}^{\text{(tot)}}$ and $\sigma_{\alpha}^{\text{(stat)}}$ indicate that the theoretical uncertainty is typically the dominant component.

The calibrated $\alpha$ values span a range of $\sim$10--12, depending on the specific measurement technique.  Statistical uncertainties in the $\alpha$ values determined using the geometric modeling techniques are a few percent, while for the IDFE techniques the statistical uncertainty is typically larger and reaches as high as $\sim$20\% in the worst cases.  Folding in both the statistical and the theoretical components, the total uncertainties are more comparable across methods, though they still exhibit a large range spanning $\sim$15--35\%.  Overall, the calibrated $\alpha$ values show similar magnitudes to what \citetalias{M87PaperVI} derived from fits to \m87, but the calibration uncertainty in the case of \sgra is substantially larger.  This increased uncertainty reflects the increased flexibility that has been built into the ring size measurement techniques to capture structural variability in the source, as well as the increased morphological diversity of the GRMHD calibration suite that is necessary to accommodate the a priori unknown inclination of \sgra.

\begin{deluxetable*}{cccCCC}
\tablecolumns{5}
\tabletypesize{\normalsize}
\tablewidth{0pt}
\tablecaption{$\alpha$ calibration parameters \label{tab:alphacal}}
\tablehead{\colhead{Analysis Class} & \colhead{Software} & \colhead{Day} & \colhead{$\alpha$} & \colhead{$\sigma^{\rm (stat)}_\alpha$} & \colhead{$\sigma^{\rm (tot)}_\alpha$}}
\startdata
\multirow{2}{*}{Snapshot} & 
\multirow{1}{*}{\rose}
&April 6+7 & 12.0 & (+0.2, -0.2) &(+1.6, -1.4)\\
&\multirow{1}{*}{\dpi}
&April 6+7 & 11.0 & (+0.8, -0.8) &(+2.2, -4.3)\\
\hline
\multirow{1}{*}{Full-track} & 
\multirow{1}{*}{\themis}
&April 6+7 & 11.7 & (+0.1, -0.1) &(+1.3, -1.3)\\
\hline
\multirow{8}{*}{Imaging} & 
\multirow{1}{*}{\difmap + \rex}
&April 7 & 10.5 & (+0.9, -1.4) &(+2.0, -2.3)\\
& \multirow{1}{*}{\difmap + \vida}
&April 7 & 10.6 & (+1.0, -1.3) &(+1.7, -3.1)\\
\cline{2-6}
& \multirow{1}{*}{\ehtim + \rex}
&April 7 & 11.0 & (+1.4, -1.3) &(+2.1, -2.5)\\
& \multirow{1}{*}{\ehtim + \vida}
&April 7 & 11.0 & (+1.2, -1.3) &(+1.7, -3.2)\\
\cline{2-6}
& \multirow{1}{*}{\smili + \rex}
&April 7 & 10.3 & (+2.4, -2.1) &(+2.8, -4.4)\\
& \multirow{1}{*}{\smili + \vida}
&April 7 & 10.4 & (+1.4, -1.4) &(+1.8, -3.7)\\
\cline{2-6}
& \multirow{1}{*}{\themis + \rex}
&April 6+7 & 10.3 & (+0.5, -0.4) &(+1.5, -2.7)\\
& \multirow{1}{*}{\themis + \vida}
&April 6+7 & 10.6 & (+0.4, -0.4) &(+1.2, -3.9)\\
\cline{2-6}
\enddata
\tablecomments{Median values and 68\% credible intervals for the calibrated $\alpha$ values, averaged over frequency bands and calibration pipelines.}
\end{deluxetable*}

\begin{figure}[t]
    \centering
    \includegraphics[width=\linewidth]{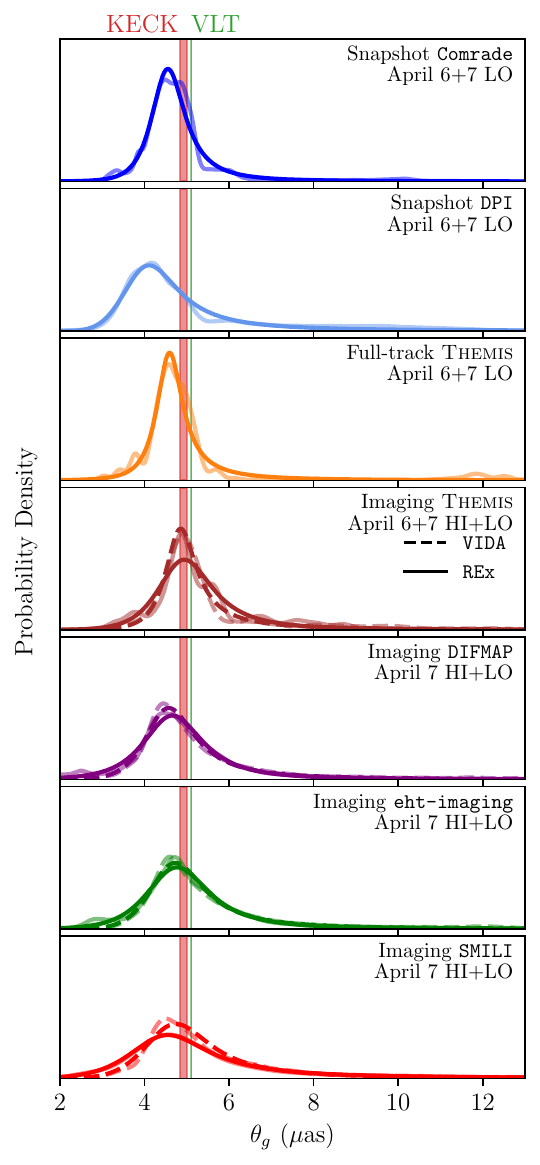}
    \caption{\sgra $\theta_g$ distributions for all of the analyses in this paper. The solid color curves show the fitted generalized lambda distribution. The transparent curves show the kernel density estimate for the distribution. The plotted results have been obtained using the HOPS calibration pipeline. For reference, we overlay the $\theta_g$ constraints measured from stellar orbits with the Keck \citep{Do_2019} and Very Large Telescope \citep[VLT;][]{Gravity_2019,Gravity_2020} facilities.}
    \label{fig:mod_cal}
\end{figure}

\subsubsection{\texorpdfstring{\sgra}{Sgr A*} angular gravitational radius} \label{sec:MoDResults}

We apply the calibrated $\alpha$ values to the measured \sgra ring diameters for each measurement technique, which produces a distribution of $\theta_g$ values that captures the uncertainties in both the ring measurements and in the GRMHD calibration. The resulting $\theta_g$ distribution exhibits sampling noise from KDE of the $\alpha$ distribution (due to the finite size of the calibration suite), and it can exhibit secondary low-probability modes at large values of $\theta_g$ (see \autoref{app:Validation}). To provide a smooth unimodal estimate of the $\theta_g$ distribution, we fit a generalized lambda distribution (G$\lambda$D) to the KDE distribution for each measurement technique. The G$\lambda$D is a unimodal distribution representing a diverse family of probability density functions.  We use the G$\lambda$D parameterization from \cite{freimer1988study}, and we use the GLDEX package in R \citep{su2007numerical, su2007fitting} to carry out the fitting.

The resulting $\theta_g$ distributions are shown in \autoref{fig:mod_cal} and listed in \autoref{tab:thetag}.  An average across all methods and datasets yields a joint constraint of $\theta_g = $ \MoDMeasurement, where the uncertainties are quoted at the 68\% (i.e., $1\sigma$) probability level and the systematic uncertainty is taken to be the standard deviation across all measurement methods.  This value is consistent with the considerably more precise constraints obtained from measurements of stellar orbits \citep{Do_2019,Gravity_2019,Gravity_2020}, and the gravitational implications of this consistency are explored in \citetalias{PaperVI}.

\begin{deluxetable*}{cccCCcCC}
\tablecolumns{8}
\tabletypesize{\normalsize}
\tablewidth{0pt}
\tablecaption{$\theta_g$ values and uncertainties across analyses \label{tab:thetag}}
\tablehead{ &  &  & \multicolumn{2}{c}{HOPS} & &\multicolumn{2}{c}{CASA}\\
            \cline{4-5}\cline{7-8}
          \colhead{Analysis} & \colhead{Software} & \colhead{Day} & \colhead{LO} & \colhead{HI} & & \colhead{LO} & \colhead{HI}}
\startdata
\multirow{2}{*}{Snapshot} & 
\multirow{1}{*}{\rose}
&April 6+7 &$ 4.6_{-0.5}^{+0.7} $ &$ 4.5_{-0.5}^{+0.7} $ &  &$ 4.4_{-0.4}^{+0.7} $ &$ 4.6_{-0.5}^{+0.7} $\\\cline{2-8}
 & \multirow{1}{*}{\dpi}
&April 6+7 &$ 4.5_{-0.8}^{+1.7} $& \ldots  &  &$ 4.9_{-0.8}^{+1.8} $& \ldots \\\hline
\multirow{1}{*}{Full-track} & 
\multirow{1}{*}{\themis}
&April 6+7 &$ 4.7_{-0.4}^{+0.8} $ &$ 4.7_{-0.4}^{+0.8} $ &  &$ 4.7_{-0.4}^{+0.8} $ &$ 4.7_{-0.4}^{+0.8} $\\\hline
\multirow{8}{*}{Imaging} & 
\multirow{1}{*}{\difmap + \rex}
&April 7& \multicolumn{2}{c}{$4.8_{-0.9}^{+1.2}$} &  & \multicolumn{2}{c}{$4.8_{-0.9}^{+1.2}$} \\&\multirow{1}{*}{\difmap + \vida}
&April 7& \multicolumn{2}{c}{$4.8_{-0.7}^{+1.4}$} &  & \multicolumn{2}{c}{$4.8_{-0.8}^{+1.5}$} \\\cline{2-8}
&\multirow{1}{*}{\ehtim + \rex}
&April 7& \multicolumn{2}{c}{$4.9_{-0.8}^{+1.2}$} &  & \multicolumn{2}{c}{$4.9_{-0.9}^{+1.3}$} \\&\multirow{1}{*}{\ehtim + \vida}
&April 7& \multicolumn{2}{c}{$5.0_{-0.8}^{+1.6}$} &  & \multicolumn{2}{c}{$5.0_{-0.9}^{+1.7}$} \\\cline{2-8}
&\multirow{1}{*}{\smili + \rex}
&April 7& \multicolumn{2}{c}{$5.0_{-1.1}^{+2.8}$} &  & \multicolumn{2}{c}{$5.2_{-1.2}^{+2.8}$} \\&\multirow{1}{*}{\smili + \vida}
&April 7& \multicolumn{2}{c}{$5.1_{-0.9}^{+1.9}$} &  & \multicolumn{2}{c}{$5.2_{-1.3}^{+2.6}$} \\\cline{2-8}
&\multirow{1}{*}{\themis + \rex}
&April 6+7& \multicolumn{2}{c}{$5.1_{-0.7}^{+1.1}$} &  & \multicolumn{2}{c}{\ldots}\\&\multirow{1}{*}{\themis + \vida}
&April 6+7& \multicolumn{2}{c}{$5.0_{-0.5}^{+1.3}$} &  & \multicolumn{2}{c}{\ldots}\\
\enddata
\tablecomments{The angular gravitational radius, $\theta_g$, measured across analysis schemes and data sets; all values are quoted as medians and 68\% credible intervals in units of \uas. Entries that straddle both low-band and high-band columns have been computed from fits to band-combined data sets.}
\end{deluxetable*}

\subsubsection{\texorpdfstring{\sgra}{Sgr A*} mass}

The angular size of the gravitational radius is proportional to the ratio of the mass and distance to \sgra, per \autoref{eqn:ThetaG}.  Our constraints on $\theta_g$ can thus be mapped directly to constraints on the black hole mass $M$ by incorporating an independent distance measurement to \sgra.  \cite{Reid_2019} report a distance of $D = 8.15 \pm 0.15$ (at $1\sigma$ probability) to the Galactic center, measured using trigonometric VLBI parallaxes of a large number ($\sim$200) of masers.  Using this distance measurement along with our measurement of $\theta_g$ from \autoref{sec:MoDResults} yields a constraint on the mass of \sgra of $M = $ \MassMeasurement, where we again quote $1\sigma$ uncertainties and the systematic component is taken to be the weighted standard deviation across methods.  This measurement is once again consistent with the more precise constraints obtained from stellar orbits \citep{Do_2019,Gravity_2019,Gravity_2020}, and the uncertainty in the mass remains dominated by our uncertainty in $\theta_g$.

%% file: summary.tex
\section{Summary and conclusions}\label{sec:Summary}

In this paper we quantify the temporal variability and morphological properties of the horizon-scale emission from \sgra, using data taken by the EHT in 2017 April.  Our primary morphological quantity of interest is the diameter of the observed ring of emission, which we quantify using multiple independent analysis pathways.  We then use the ring diameter to place constraints on the angular size of the gravitational radius ($\theta_g$) and on the mass ($M$) of \sgra.  The analyses presented here have been carried out using data taken on April 6 and April 7, across two frequency bands and two data calibration pipelines.

Motivated by theoretical expectations that the dynamical timescales in \sgra should be much shorter than the duration of EHT observing tracks, we employ a new method developed in \citet{NoiseModeling}
for quantifying the time variability observed in the visibilities in a manner that is agnostic to the specifics of the average underlying source structure.
We find that the visibility amplitudes
exhibit a light-curve-normalized variance that is in excess of that expected from thermal noise, station gains, or refractive scattering effects, and we attribute this excess variance to intrinsic structural changes in the source.  The detected variability is most statistically significant on baselines with lengths between 2.5 and 6\,G$\lambda$, where it exhibits an approximately power-law decline with increasing baseline length, with a power-law index of $\sim$2--3.  The magnitude of this variability on baselines near 3\,G$\lambda$ in length exceeds 0.1\,Jy, which is roughly equal to the value of the correlated flux density on these same baselines.

Through an exploration of potential simple geometric source structures, we demonstrate that the EHT \sgra data statistically prefer ring-like morphologies over other morphologies with comparable complexity.
We develop and deploy two new methods for fitting the time-variable \sgra data using static geometric ring models with azimuthally modulated brightness structures.  In the first method, called ``snapshot'' geometric modeling, we first fit the models to short segments of data over which the source variability is subdominant to other sources of uncertainty.  The fits from individual segments are then combined via a hierarchical model averaging scheme to provide parameter constraints across the entire observation.  In the second method, called ``full-track'' geometric modeling, we fit a static geometric model to the entire data set alongside parameters that describe the statistical fluctuations that time variability induces in the data.  Our parameterization for the variability ``noise'' is motivated by the work of \citet{Georgiev_2022} and takes the form of a broken power law in baseline length that contributes to the data uncertainties.

We compare the results from snapshot and full-track geometric modeling, both with one another and with the results of IDFE from the images reconstructed in \citetalias{PaperIII}, to constrain the horizon-scale morphology of \sgra.  The ring diameter is well constrained and stable across both frequency bands and calibration pipelines, with geometric modeling and IDFE techniques jointly determining a value of \DiameterMeasurement (68\% credible intervals).  We find that the magnitude and orientation of the ring asymmetry, as well as the depth of its central brightness depression, are poorly constrained and have values that can depend sensitively on the measurement method employed.  The thickness of the ring is well measured by individual analysis methods but takes on a value that depends on the specifics of each method; geometric modeling methods find a FWHM ring thickness of $35 \pm 5\%$ of the ring diameter, while image-domain feature extraction methods find a FWHM of $53 \pm 10\%$.

Using a suite of synthetic data sets generated from the \citetalias{PaperV} GRMHD simulation library, we calibrate the diameter measurements from both geometric modeling and image-domain feature extraction methods to a common physical scale.  The resulting constraint on the angular size of the \sgra gravitational radius, combined across all methods and data sets, is $\theta_g = $ \MoDMeasurement.  This large uncertainty arises from both the model flexibility necessary to capture structural variability in the source and the broad morphological diversity of the GRMHD calibration suite that reflects the a priori unknown inclination of \sgra.  Combining our $\theta_g$ constraint with an independent distance measurement from \cite{Reid_2019}, we determine the mass of \sgra to be $M = $ \MassMeasurement.  Though the uncertainties are large compared to those derived using other techniques (e.g., stellar orbit modeling), our measurement represents the first time that the mass of \sgra has been constrained by observations of light bending near the horizon.

%% file: EHT-paper4-ackn.tex
We thank an anonymous referee for insightful and constructive comments that helped improve the quality of this paper.

The Event Horizon Telescope Collaboration thanks the following
organizations and programs: the Academia Sinica; the Academy
of Finland (projects 274477, 284495, 312496, 315721); the Agencia Nacional de Investigaci\'{o}n 
y Desarrollo (ANID), Chile via NCN$19\_058$ (TITANs) and Fondecyt 1221421, the Alexander
von Humboldt Stiftung; an Alfred P. Sloan Research Fellowship;
Allegro, the European ALMA Regional Centre node in the Netherlands, the NL astronomy
research network NOVA and the astronomy institutes of the University of Amsterdam, Leiden University and Radboud University;
the ALMA North America Development Fund; the Black Hole Initiative, which is funded by grants from the John Templeton Foundation and the Gordon 
and Betty Moore Foundation (although the opinions expressed in this work are those of the author(s) 
and do not necessarily reflect the views of these Foundations);
Chandra DD7-18089X and TM6-17006X; the China Scholarship
Council; China Postdoctoral Science Foundation fellowship (2020M671266); Consejo Nacional de Ciencia y Tecnolog\'{\i}a (CONACYT,
Mexico, projects  U0004-246083, U0004-259839, F0003-272050, M0037-279006, F0003-281692,
104497, 275201, 263356);
the Consejer\'{i}a de Econom\'{i}a, Conocimiento, 
Empresas y Universidad 
of the Junta de Andaluc\'{i}a (grant P18-FR-1769), the Consejo Superior de Investigaciones 
Cient\'{i}ficas (grant 2019AEP112);
the Delaney Family via the Delaney Family John A.
Wheeler Chair at Perimeter Institute; Direcci\'{o}n General
de Asuntos del Personal Acad\'{e}mico-Universidad
Nacional Aut\'{o}noma de M\'{e}xico (DGAPA-UNAM,
projects IN112417 and IN112820); 
the Dutch Organization for Scientific Research (NWO) VICI award
(grant 639.043.513) and grant OCENW.KLEIN.113; the Dutch National Supercomputers, Cartesius and Snellius  
(NWO Grant 2021.013); 
the EACOA Fellowship awarded by the East Asia Core
Observatories Association, which consists of the Academia Sinica Institute of Astronomy and
Astrophysics, the National Astronomical Observatory of Japan, Center for Astronomical Mega-Science,
Chinese Academy of Sciences, and the Korea Astronomy and Space Science Institute; 
the European Research Council (ERC) Synergy
Grant ``BlackHoleCam: Imaging the Event Horizon
of Black Holes" (grant 610058); 
the European Union Horizon 2020
research and innovation programme under grant agreements
RadioNet (No 730562) and 
M2FINDERS (No 101018682);
the Generalitat
Valenciana postdoctoral grant APOSTD/2018/177 and
GenT Program (project CIDEGENT/2018/021); MICINN Research Project PID2019-108995GB-C22;
the European Research Council for advanced grant `JETSET: Launching, propagation and 
emission of relativistic jets from binary mergers and across mass scales' (Grant No. 884631); 
the Institute for Advanced Study; the Istituto Nazionale di Fisica
Nucleare (INFN) sezione di Napoli, iniziative specifiche
TEONGRAV; 
the International Max Planck Research
School for Astronomy and Astrophysics at the
Universities of Bonn and Cologne; 
DFG research grant ``Jet physics on horizon scales and beyond'' (Grant No. FR 4069/2-1);
Joint Princeton/Flatiron and Joint Columbia/Flatiron Postdoctoral Fellowships, 
research at the Flatiron Institute is supported by the Simons Foundation; 
the Japan Ministry of Education, Culture, Sports, Science and Technology (MEXT; grant JPMXP1020200109); the Japanese Government (Monbukagakusho:
MEXT) Scholarship; 
the Japan Society for the Promotion of Science (JSPS) Grant-in-Aid for JSPS
Research Fellowship (JP17J08829); the Joint Institute for Computational Fundamental Science, Japan; the Key Research
Program of Frontier Sciences, Chinese Academy of
Sciences (CAS, grants QYZDJ-SSW-SLH057, QYZDJSSW-SYS008, ZDBS-LY-SLH011); 
the Leverhulme Trust Early Career Research
Fellowship; the Max-Planck-Gesellschaft (MPG);
the Max Planck Partner Group of the MPG and the
CAS; the MEXT/JSPS KAKENHI (grants 18KK0090, JP21H01137,
JP18H03721, JP18K13594, 18K03709, JP19K14761, 18H01245, 25120007); the Malaysian Fundamental Research Grant Scheme (FRGS) FRGS/1/2019/STG02/UM/02/6; the MIT International Science
and Technology Initiatives (MISTI) Funds; 
the Ministry of Science and Technology (MOST) of Taiwan (103-2119-M-001-010-MY2, 105-2112-M-001-025-MY3, 105-2119-M-001-042, 106-2112-M-001-011, 106-2119-M-001-013, 106-2119-M-001-027, 106-2923-M-001-005, 107-2119-M-001-017, 107-2119-M-001-020, 107-2119-M-001-041, 107-2119-M-110-005, 107-2923-M-001-009, 108-2112-M-001-048, 108-2112-M-001-051, 108-2923-M-001-002, 109-2112-M-001-025, 109-2124-M-001-005, 109-2923-M-001-001, 110-2112-M-003-007-MY2, 110-2112-M-001-033, 110-2124-M-001-007, and 110-2923-M-001-001);
the Ministry of Education (MoE) of Taiwan Yushan Young Scholar Program;
the Physics Division, National Center for Theoretical Sciences of Taiwan;
the National Aeronautics and
Space Administration (NASA, Fermi Guest Investigator
grant 80NSSC20K1567, NASA Astrophysics Theory Program grant 80NSSC20K0527, NASA NuSTAR award 
80NSSC20K0645); 
NASA Hubble Fellowship 
grant HST-HF2-51431.001-A awarded 
by the Space Telescope Science Institute, which is operated by the Association of Universities for 
Research in Astronomy, Inc., for NASA, under contract NAS5-26555; 
the National Institute of Natural Sciences (NINS) of Japan; the National
Key Research and Development Program of China
(grant 2016YFA0400704, 2017YFA0402703, 2016YFA0400702); the National
Science Foundation (NSF, grants AST-0096454,
AST-0352953, AST-0521233, AST-0705062, AST-0905844, AST-0922984, AST-1126433, AST-1140030,
DGE-1144085, AST-1207704, AST-1207730, AST-1207752, MRI-1228509, OPP-1248097, AST-1310896, AST-1440254, 
AST-1555365, AST-1614868, AST-1615796, AST-1715061, AST-1716327,  AST-1716536, OISE-1743747, AST-1816420, AST-1935980, AST-2034306); 
NSF Astronomy and Astrophysics Postdoctoral Fellowship (AST-1903847); 
the Natural Science Foundation of China (grants 11650110427, 10625314, 11721303, 11725312, 11873028, 11933007, 11991052, 11991053, 12192220, 12192223); 
the Natural Sciences and Engineering Research Council of
Canada (NSERC, including a Discovery Grant and
the NSERC Alexander Graham Bell Canada Graduate
Scholarships-Doctoral Program); the National Youth
Thousand Talents Program of China; the National Research
Foundation of Korea (the Global PhD Fellowship
Grant: grants NRF-2015H1A2A1033752, the Korea Research Fellowship Program:
NRF-2015H1D3A1066561, Brain Pool Program: 2019H1D3A1A01102564, 
Basic Research Support Grant 2019R1F1A1059721, 2021R1A6A3A01086420, 2022R1C1C1005255); 
Netherlands Research School for Astronomy (NOVA) Virtual Institute of Accretion (VIA) postdoctoral fellowships; 
Onsala Space Observatory (OSO) national infrastructure, for the provisioning
of its facilities/observational support (OSO receives
funding through the Swedish Research Council under
grant 2017-00648);  the Perimeter Institute for Theoretical
Physics (research at Perimeter Institute is supported
by the Government of Canada through the Department
of Innovation, Science and Economic Development
and by the Province of Ontario through the
Ministry of Research, Innovation and Science); the Spanish Ministerio de Ciencia e Innovaci\'{o}n (grants PGC2018-098915-B-C21, AYA2016-80889-P,
PID2019-108995GB-C21, PID2020-117404GB-C21); 
the University of Pretoria for financial aid in the provision of the new 
Cluster Server nodes and SuperMicro (USA) for a SEEDING GRANT approved towards these 
nodes in 2020;
the Shanghai Pilot Program for Basic Research, Chinese Academy of Science, 
Shanghai Branch (JCYJ-SHFY-2021-013);
the State Agency for Research of the Spanish MCIU through
the ``Center of Excellence Severo Ochoa'' award for
the Instituto de Astrof\'{i}sica de Andaluc\'{i}a (SEV-2017-
0709); the Spinoza Prize SPI 78-409; the South African Research Chairs Initiative, through the 
South African Radio Astronomy Observatory (SARAO, grant ID 77948),  which is a facility of the National 
Research Foundation (NRF), an agency of the Department of Science and Innovation (DSI) of South Africa; 
the Toray Science Foundation; Swedish Research Council (VR); 
the US Department
of Energy (USDOE) through the Los Alamos National
Laboratory (operated by Triad National Security,
LLC, for the National Nuclear Security Administration
of the USDOE (Contract 89233218CNA000001); and the YCAA Prize Postdoctoral Fellowship.

We thank
the staff at the participating observatories, correlation
centers, and institutions for their enthusiastic support.
This paper makes use of the following ALMA data:
ADS/JAO.ALMA\#2016.1.01154.V. ALMA is a partnership
of the European Southern Observatory (ESO;
Europe, representing its member states), NSF, and
National Institutes of Natural Sciences of Japan, together
with National Research Council (Canada), Ministry
of Science and Technology (MOST; Taiwan),
Academia Sinica Institute of Astronomy and Astrophysics
(ASIAA; Taiwan), and Korea Astronomy and
Space Science Institute (KASI; Republic of Korea), in
cooperation with the Republic of Chile. The Joint
ALMA Observatory is operated by ESO, Associated
Universities, Inc. (AUI)/NRAO, and the National Astronomical
Observatory of Japan (NAOJ). The NRAO
is a facility of the NSF operated under cooperative agreement
by AUI.
This research used resources of the Oak Ridge Leadership Computing Facility at the Oak Ridge National
Laboratory, which is supported by the Office of Science of the U.S. Department of Energy under Contract
No. DE-AC05-00OR22725. We also thank the Center for Computational Astrophysics, National Astronomical Observatory of Japan.
The computing cluster of Shanghai VLBI correlator supported by the Special Fund 
for Astronomy from the Ministry of Finance in China is acknowledged.

APEX is a collaboration between the
Max-Planck-Institut f{\"u}r Radioastronomie (Germany),
ESO, and the Onsala Space Observatory (Sweden). The
SMA is a joint project between the SAO and ASIAA
and is funded by the Smithsonian Institution and the
Academia Sinica. The JCMT is operated by the East
Asian Observatory on behalf of the NAOJ, ASIAA, and
KASI, as well as the Ministry of Finance of China, Chinese
Academy of Sciences, and the National Key Research and Development
Program (No. 2017YFA0402700) of China
and Natural Science Foundation of China grant 11873028.
Additional funding support for the JCMT is provided by the Science
and Technologies Facility Council (UK) and participating
universities in the UK and Canada. 
The LMT is a project operated by the Instituto Nacional
de Astr\'{o}fisica, \'{O}ptica, y Electr\'{o}nica (Mexico) and the
University of Massachusetts at Amherst (USA). The
IRAM 30-m telescope on Pico Veleta, Spain is operated
by IRAM and supported by CNRS (Centre National de
la Recherche Scientifique, France), MPG (Max-Planck-Gesellschaft, Germany) 
and IGN (Instituto Geogr\'{a}fico
Nacional, Spain). The SMT is operated by the Arizona
Radio Observatory, a part of the Steward Observatory
of the University of Arizona, with financial support of
operations from the State of Arizona and financial support
for instrumentation development from the NSF.
Support for SPT participation in the EHT is provided by the National Science Foundation through award OPP-1852617 
to the University of Chicago. Partial support is also 
provided by the Kavli Institute of Cosmological Physics at the University of Chicago. The SPT hydrogen maser was 
provided on loan from the GLT, courtesy of ASIAA.

This work used the
Extreme Science and Engineering Discovery Environment
(XSEDE), supported by NSF grant ACI-1548562,
and CyVerse, supported by NSF grants DBI-0735191,
DBI-1265383, and DBI-1743442. XSEDE Stampede2 resource
at TACC was allocated through TG-AST170024
and TG-AST080026N. XSEDE JetStream resource at
PTI and TACC was allocated through AST170028.
This research is part of the Frontera computing project at the Texas Advanced 
Computing Center through the Frontera Large-Scale Community Partnerships allocation
AST20023. Frontera is made possible by National Science Foundation award OAC-1818253.
This research was carried out using resources provided by the Open Science Grid, 
which is supported by the National Science Foundation and the U.S. Department of 
Energy Office of Science. 
Additional work used ABACUS2.0, which is part of the eScience center at Southern Denmark University. 
Simulations were also performed on the SuperMUC cluster at the LRZ in Garching, 
on the LOEWE cluster in CSC in Frankfurt, on the HazelHen cluster at the HLRS in Stuttgart, 
and on the Pi2.0 and Siyuan Mark-I at Shanghai Jiao Tong University.
The computer resources of the Finnish IT Center for Science (CSC) and the Finnish Computing 
Competence Infrastructure (FCCI) project are acknowledged. This
research was enabled in part by support provided
by Compute Ontario (http://computeontario.ca), Calcul
Quebec (http://www.calculquebec.ca) and Compute
Canada (http://www.computecanada.ca).

The EHTC has
received generous donations of FPGA chips from Xilinx
Inc., under the Xilinx University Program. The EHTC
has benefited from technology shared under open-source
license by the Collaboration for Astronomy Signal Processing
and Electronics Research (CASPER). The EHT
project is grateful to T4Science and Microsemi for their
assistance with Hydrogen Masers. This research has
made use of NASA's Astrophysics Data System. We
gratefully acknowledge the support provided by the extended
staff of the ALMA, both from the inception of
the ALMA Phasing Project through the observational
campaigns of 2017 and 2018. We would like to thank
A. Deller and W. Brisken for EHT-specific support with
the use of DiFX. We thank Martin Shepherd for the addition of extra features in the Difmap software 
that were used for the CLEAN imaging results presented in this paper.
We acknowledge the significance that
Maunakea, where the SMA and JCMT EHT stations
are located, has for the indigenous Hawaiian people.


%% file: appendix.tex
\begin{appendix}

\section{Comparison of \texorpdfstring{\sgra}{Sgr A*} variability in 2013 and 2017}
\label{app:premodeling_2013}
\begin{figure}
  \begin{center}
    \includegraphics[width=\columnwidth]{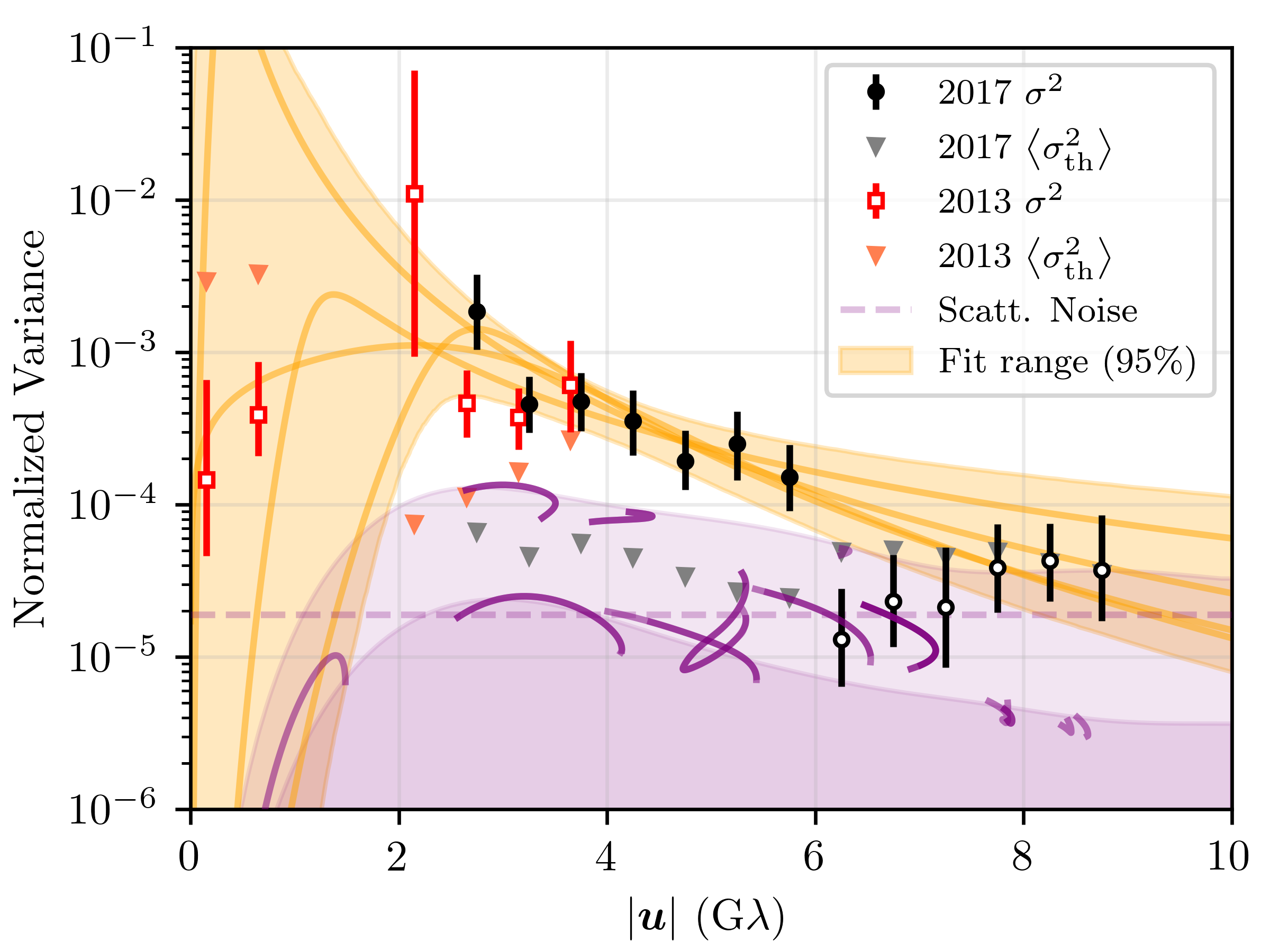}
  \end{center}
\caption{Estimate of the variance of the visibility amplitudes from the 2013 1.3~mm VLBI observations reported in \citet{SgrA2013}, indicated by the red open circles.  The associated statistical errors are shown by the red triangles.  For comparison, all of the elements of \autoref{fig:sevv1} are reproduced as shown there, including the 2017 EHT variance estimates (filled and open black circles).}
\label{fig:premod-2013}
\end{figure}

Additional prior epochs of millimeter-VLBI observations of \sgra provide a means to explore a wider range of baseline lengths and assess the consistency of the variability across many years.  \citet{SgrA2013} reported 1.3\,mm VLBI observations of \sgra from 2013 March 21, 22, 23, 26, and 27 with an array that included JCMT, SMA, SMT, and the Combined Array for Research in Millimeter-wave Astronomy (CARMA, which has since been decommissioned).  By virtue of the small number of participating stations, these observations were much more limited in their \uv-coverage than the 2017 EHT observations.  Nevertheless, the 2013 observations provide a second multiday data set for which normalized variance estimates may be produced and compared to those reported in this paper.

\citet{SgrA2013} reported only the visibility amplitudes, which we average in time on a per-scan basis.  This averaging presumes that there is no intrinsic phase evolution in the visibilities over the approximately 10-minute scan lengths; this approximation is well justified for the source sizes inferred from the 2017 EHT observations.  The visibility data are normalized by the intrasite SMA-JCMT, SMA-SMA, and CARMA-CARMA baselines.  Note that this normalization eliminates the need to perform a phase calibration like that applied to JCMT and LMT for the 2017 EHT data.

\autoref{fig:premod-2013} shows the reconstructed normalized variances from the 2013 observations in comparison to those from the 2017 EHT observations.  
Baselines between 2.5 and 4\,\Gl provide variability estimates that are broadly consistent between the two sets of observations.  This agreement suggests that the degree of structural variability exhibited by \sgra during the 2017 EHT campaign is not anomalous.

Similarly, because the station gains were well characterized for all stations during the 2013 observations, it is not necessary to make assumptions about the source size on short baselines.  Therefore, the 2013 observations provide estimates of the normalized variance on baselines shorter than 2\,\Gl. The large statistical errors of these measurements preclude strong constraints on the variability below 1\,\Gl, but there are nevertheless hints of a turnover in the variability power between 1 and 2\,\Gl.

\section{Origin and mitigation of biases in single-day analyses} \label{app:SingleDayBias}

\begin{figure*}
\begin{center}
\includegraphics[width=\textwidth]{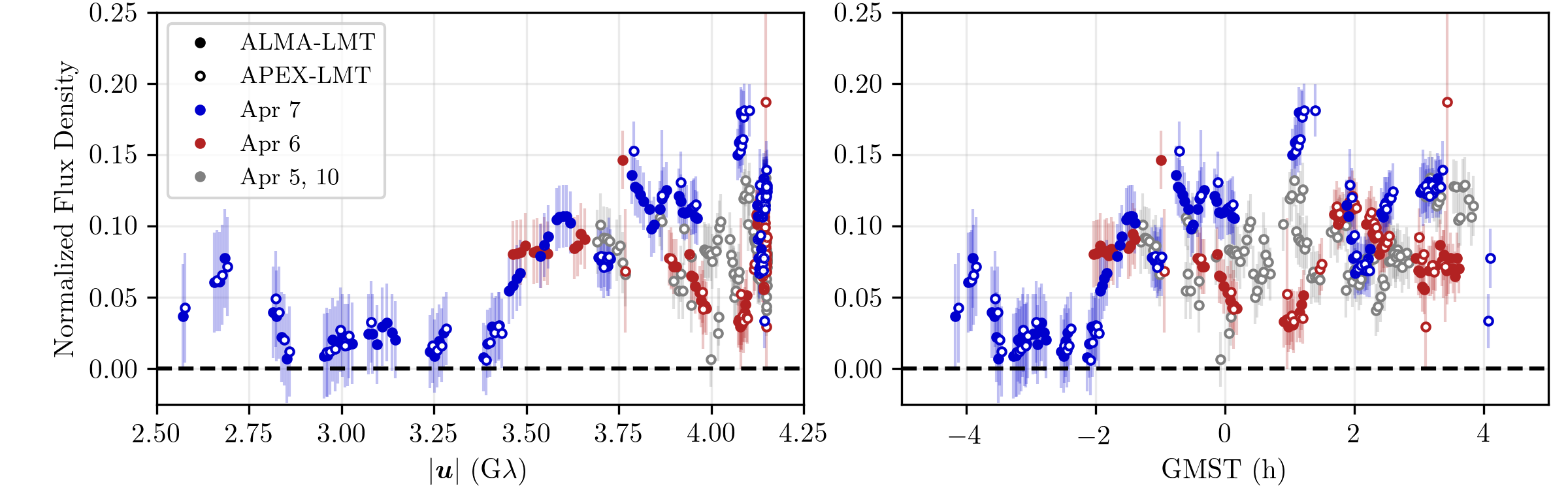}
\end{center}
\caption{Visibility amplitudes from the HOPS low-band \sgra data set, averaged coherently over 120\,s segments, on Apr 6 (red), April 7 (blue), and April 5 and 10 (gray) on the Chile-LMT baselines as functions of baseline length (left) and observing time (right).  Error bars indicate the error implied by the mean noise model and are intended to account for fluctuations due to variability in addition to statistical and known systematic error components.}
\label{fig:amp_vut_sgra}
\end{figure*}

\begin{figure*}
\begin{center}
\includegraphics[width=\textwidth]{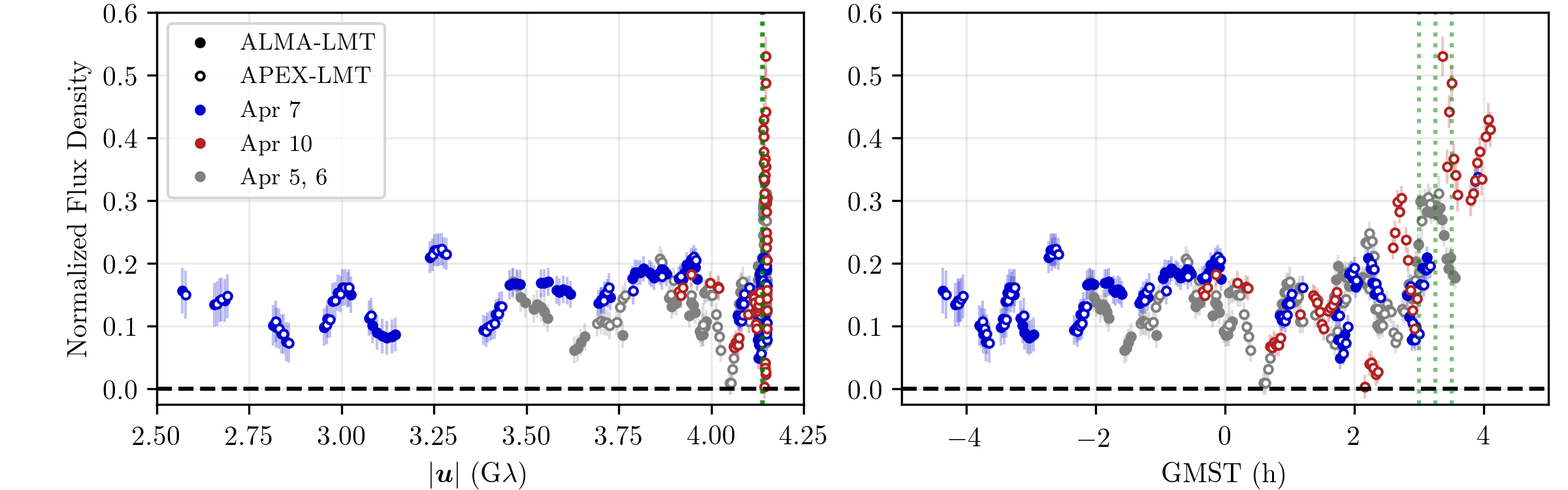}
\end{center}
\caption{Visibility amplitudes from the HOPS low-band GRMHD validation synthetic data set (data set 092 in \autoref{tab:GRMHDSynthData}), averaged coherently over 120\,s segments, on April 10 (red), April 7 (blue), and April 5 and 10 (gray) on the Chile-LMT baselines as functions of baseline length (left) and observing time (right).  Error bars indicate the error implied by the mean noise model and are intended to account for fluctuations due to variability in addition to statistical and known systematic error components.  Vertical dotted green lines indicate the positions at which frames are shown in \autoref{fig:test002_frames}.}
\label{fig:amp_vut_grmhd}
\end{figure*}

\begin{figure*}
\begin{center}
\includegraphics[width=\textwidth]{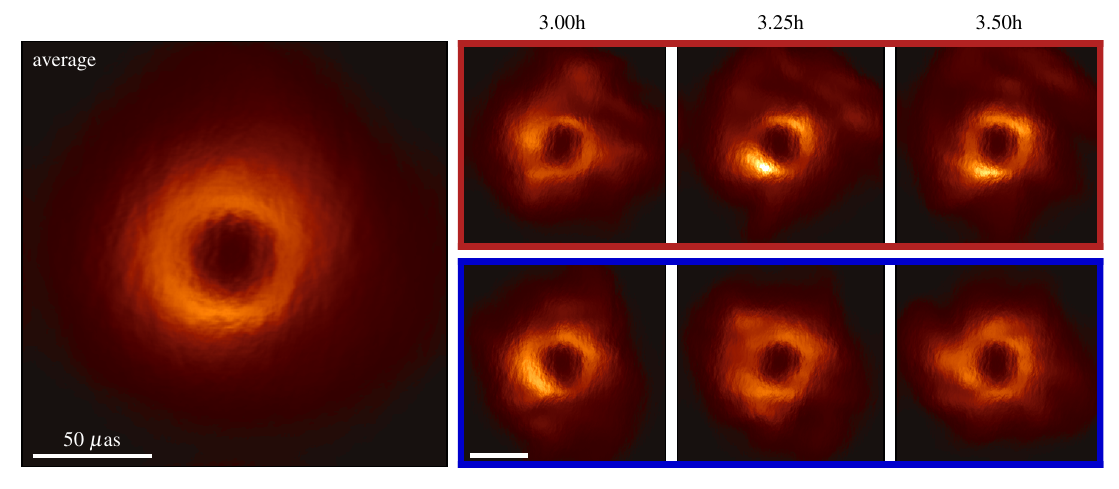}
\end{center}
\caption{Average (left) and snapshot (right) images from the GRMHD simulation used to generate the validation synthetic data set shown in \autoref{fig:amp_vut_grmhd}.  Three frames from the portion of the simulation used to generate the April 10 (top, outlined in red) and April 7 (bottom, outlined in blue) synthetic data sets are shown at the GMST times specified (corresponding to the times indicated by the vertical dotted green lines in \autoref{fig:amp_vut_grmhd}).  All of the images share a common brightness color scale; the absolute brightness scale is arbitrary because each image has been normalized to have unit total flux density, and a modest amount of saturation has been permitted in the brightest regions to enhance the visibility of low-brightness features.}
\label{fig:test002_frames}
\end{figure*}

As discussed in \autoref{sec:Variability}, there is compelling evidence that \sgra exhibits structural variability on timescales ranging from minutes to a full observation night.  The degree of variability is estimated in \autoref{sec:pre-modeling} and \autoref{sec:VariabilityMeasurement}, and found to be the dominant contribution to the difference between the observed visibilities and those associated with a mean image for baselines with lengths between $\sim$3 and 6\,\Gl.  Underlying the noise modeling mitigation method (see \autoref{sec:mitigation}) is the assumption that the added ``variability noise'' modifies the data in a stochastic manner, i.e., coherent deviations do not persist throughout large patches of the \uv-plane.  However, Earth-rotation aperture synthesis naturally results in correlated variability between visibilities that are nearby in the \uv-plane, because visibilities that have small \uv separations tend to also have small temporal separations.  Furthermore, the sparsity of the EHT array prevents most locations in the \uv-plane from being sampled more than once in a single observation (though see \autoref{sec:CrossFollow} for several exceptions), meaning that multiple observing days must be combined to access more than a single instantiation of the source variability.

The impact of structural variability can be seen most prominently on the Chile-LMT baselines, which exhibit coherent deviations on $\sim$1\,hr timescales that are evident in the visibility amplitudes presented in \autoref{fig:amp_vut_sgra}.  These deviations are most pronounced near 1\,h GMST ($\sim$12 UTC), where they are evident on both April 6 and April 7.  The residual gain uncertainties of $\sim$10\%-20\% \citepalias{PaperII} are insufficient to explain the dramatic drop near 4\,\Gl around 1\,h GMST on April 6.  At the same GMST on April 7, the visibility amplitudes fluctuate upward by a similar amount. On the remaining days the amplitudes at this GMST lie between the April 6 and April 7 values, indicating that the variations are associated with a process that is uncorrelated on interday timescales.

Similar coherent deviations are also observed in the synthetic data sets produced from GRMHD simulations for the purposes of calibrating ring size measurements to a common physical scale (see \autoref{sec:Calibration} and \autoref{sec:MoDSyntheticData}). \autoref{fig:amp_vut_grmhd} shows a similar set of visibility amplitudes on the Chile-LMT baselines for one of these synthetic data sets (data set 092 in \autoref{tab:GRMHDSynthData}).  While the date and time of the spurious feature differ from those seen in the real \sgra data -- arising in the simulated data set near 3\,h GMST ($\sim$14 UTC) -- a dramatic deviation is present and persists for $\sim$0.5\,hr.  In both the \sgra data and GRMHD simulations, these observed deviations are ${\gtrsim}3\sigma$ outliers (per the variance expected from the variability quantification scheme detailed in \autoref{sec:pre-modeling}), they are nearly exclusively confined to the Chile-LMT baseline, and they are rare.  However, due to their coherent nature -- i.e., many data points on a given day are similarly displaced -- these fluctuations violate the assumption of statistical independence made by the noise modeling mitigation scheme on a single day.

The origin of the visibility amplitude deviations for the GRMHD simulations can be identified with coherent variable structures moving about the ring.  As shown in \autoref{fig:test002_frames}, instantaneous images from the GRMHD simulations can differ qualitatively from the average image, with the former sometimes dominated by small, bright patches of emission.  When aligned in the NW-SE direction and separated by the $\sim$50\,\uas ring diameter, these bright emission regions significantly impact the visibilities on the $|\boldsymbol{u}| \approx 4$\,\Gl Chile-LMT baselines.  Days that do not exhibit large variations correspond to periods less impacted by such patchy emission structures.

\begin{figure*}
\begin{center}
\includegraphics[width=\textwidth]{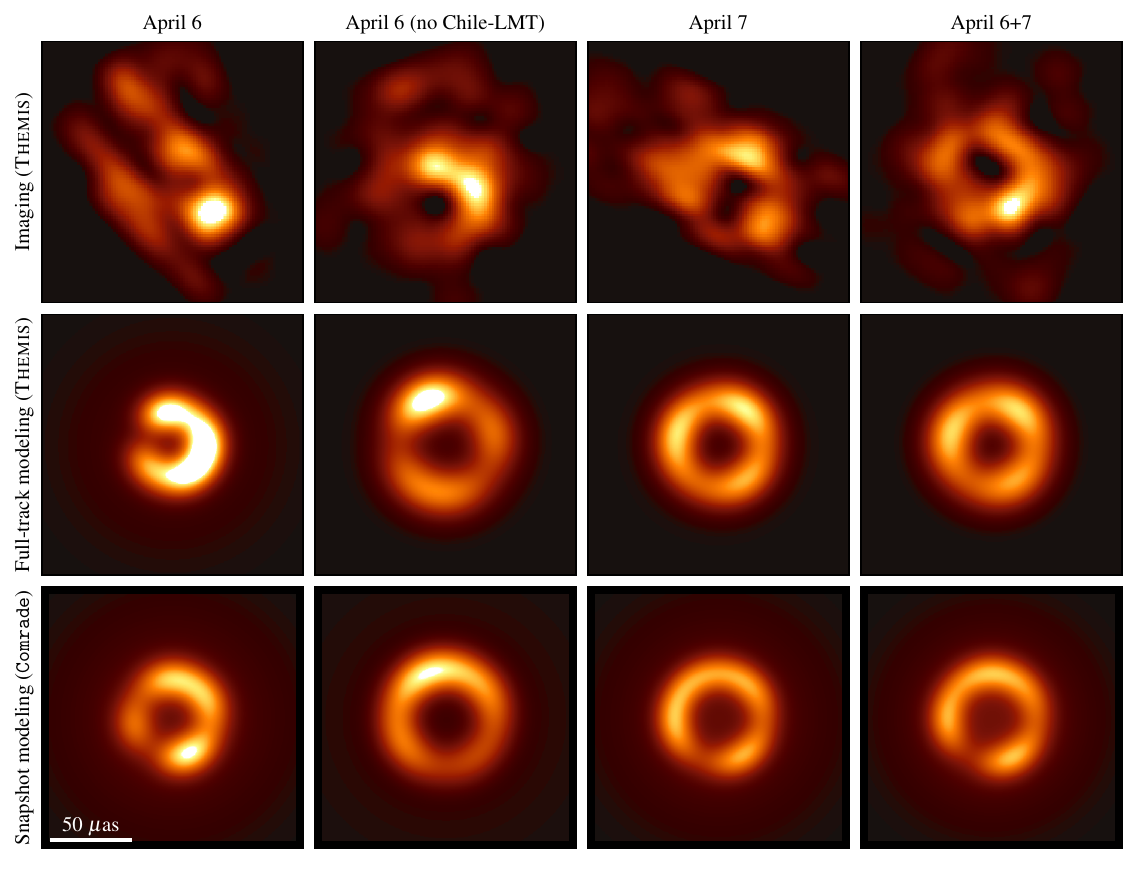}
\end{center}
\caption{Comparison of the results from \themis imaging (top row), full-track geometric modeling (middle row), and snapshot modeling (bottom row) across different combinations of April 6 and 7 data sets.  The first column shows results from fitting to the April 6 data, the second column shows results from fitting to the April 6 data after flagging baselines between Chile and LMT, the third column shows results from fitting to the April 7 data, and the fourth column shows results from fitting to the combined April 6 and 7 data.  In all panels, we show an image corresponding to the posterior mean; for the \themis imaging results, each sample image has been shifted during averaging so as to maximize the normalized cross-correlation computed with respect to a reference sample.  The full-track and snapshot modeling results show fits to the HOPS pipeline low-band data, while the imaging results show fits to the HOPS pipeline combined low- and high-band data.  All of the images share a common brightness color scale; the absolute brightness scale is arbitrary because each image has been normalized to have unit total flux density, and a modest amount of saturation has been permitted in the brightest regions to enhance the visibility of low-brightness features.}
\label{fig:multiday_comparison}
\end{figure*}

The coherent deviations seen on the Chile-LMT baselines can manifest as pathological behaviors in the single-day \sgra analyses carried out in this paper and in \citetalias{PaperIII}, particularly when applied to the sparser April 6 data set (see left column of \autoref{fig:multiday_comparison}).  The specific manner in which the visibility fluctuations impact the reconstructed source structure depends on the details of the analysis scheme and the freedom each underlying image model has to accommodate a subset of visibilities that are discrepant with the time-averaged structure.

Images reconstructed from the April 6 \sgra data exhibit parallel NE-SW streaks, typical of baseline artifacts associated with miscalibration of a single baseline, and a natural consequence of visibility amplitudes on the Chile-LMT baseline that are discrepant with the time-averaged image.  The top left panel of \autoref{fig:multiday_comparison} shows such artifacts in a \themis image reconstruction; similar image artifacts are observed in the RML (\ehtim and \smili) and, to a lesser extent, the CLEAN (\difmap) reconstructions presented in \citetalias{PaperIII}.

By virtue of their specification, the \model source models are not capable of introducing streak-like features into the image structure.  When applied to \sgra, full-track \model fits to the April 6 data instead exhibit smaller ring sizes than those applied to the April 7 data (see the first and third panels in the second row of \autoref{fig:multiday_comparison}), with the April 6 fits preferring rings with a $\sim$40\,\uas diameter and the April 7 fits preferring rings with a $\sim$55\,\uas diameter.  This discrepancy in ring size may be associated with a shift in the location of the visibility minimum from $\sim$3\,\Gl to $\sim$4\,\Gl between April 7 and April 6, respectively.  However, we note that the behavior of the full-track \model model when applied to the April 6 data is a sensitive function of the $m$-order; changing the $m$-order can cause the model to prefer a $\sim$55\,\uas diameter.  Snapshot \model fits exhibit qualitatively similar behavior to the full-track fits, as shown in the third row of \autoref{fig:multiday_comparison}.  We again find that the April 6 data prefer a somewhat smaller ring diameter than the April 7 fits, though for snapshot fits the posterior distributions for the diameter parameter are consistent between the 2 days.

Despite their disparate forms, the various artifacts observed in April 6 reconstructions are effectively ameliorated by flagging the Chile-LMT baselines prior to carrying out the analyses (see the second column in \autoref{fig:multiday_comparison}), providing further evidence that the origins of the artifacts are confined to (or at least dominated by) the Chile-LMT baselines.  However, though flagging of these baselines is successful in preventing the specific analysis pathologies discussed above, such flagging is not otherwise motivated; there is no evidence for atypical data calibration issues on these baselines, and thus no reason to believe that the observed variability excess is anything other than intrinsic to the source.  Rather than flagging data, we proceed instead with the noise modeling scheme described in \autoref{sec:mitigation}, which is itself intended to mitigate the effects of intrinsic variability on the source reconstructions.

As described above, the apparent failure of the noise modeling method to produce consistent results on some individual single-day analyses can be attributed to the coherent (rather than stochastic) nature of the variability sampled at any single location in the \uv-plane, which violates the key assumption underlying the noise modeling approach that each data point represent an independent sample of the source variability.  This assumption would be more faithfully adhered to by a data set containing a larger number of independent variability realizations; within the context of the EHT \sgra observations, additional variability realizations are most naturally acquired by combining data sets across observing days.  Multiday analyses are also more consistent in spirit with the model-agnostic variability estimates in \autoref{sec:pre-modeling} and \autoref{sec:VariabilityMeasurement}.  Combining the independent realizations of the structural variability from multiple days improves the estimate of the mean visibilities, as evident in \autoref{fig:amp_vut_sgra}, in which the multiday mean of the visibility amplitudes is both smoother than and intermediate between those of April 6 and 7 individually.

The fourth column in \autoref{fig:multiday_comparison} shows reconstructions made using the combined April 6 and 7 data sets.  The improved behavior of the visibility means is reflected in better consistency across methods and a reduction in image artifacts.

\section{Single-day fits} \label{app:SingleDayFits}

\begin{figure*}[t]
    \centering
    \includegraphics[width=\textwidth]{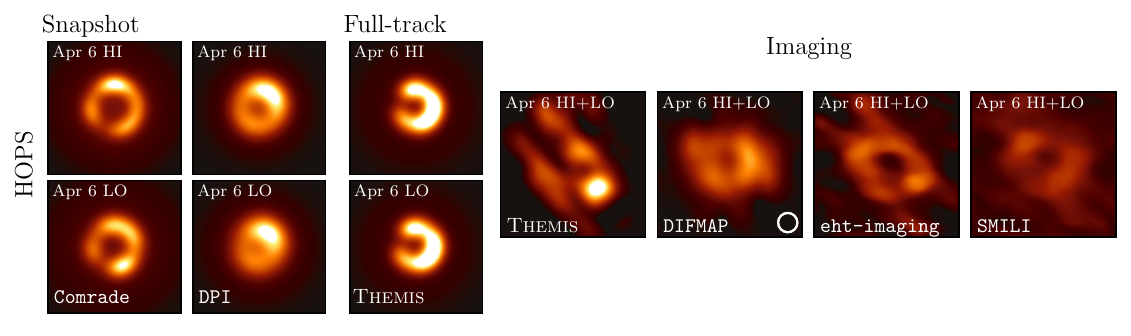}\\
    \includegraphics[width=\textwidth]{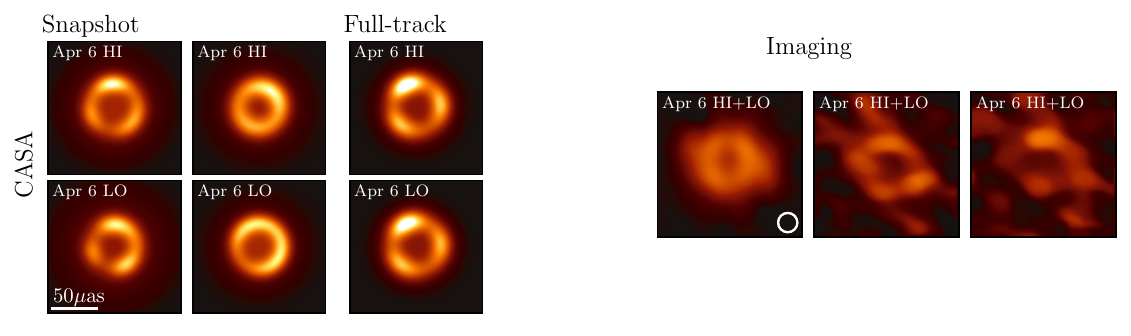}\\
    \caption{Same as \autoref{fig:morphology}, but showing results obtained from using only the April 6 data.}
    \label{fig:morphology_april6}
\end{figure*}

\begin{figure*}[t]
    \centering
    \includegraphics[width=\textwidth]{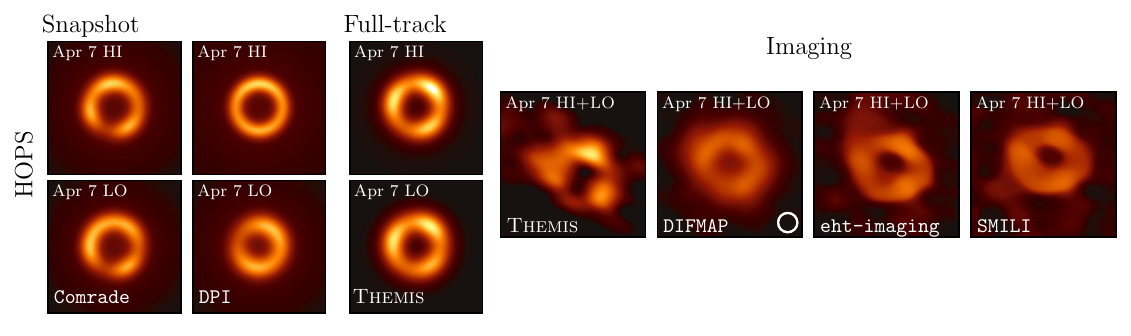}\\
    \includegraphics[width=\textwidth]{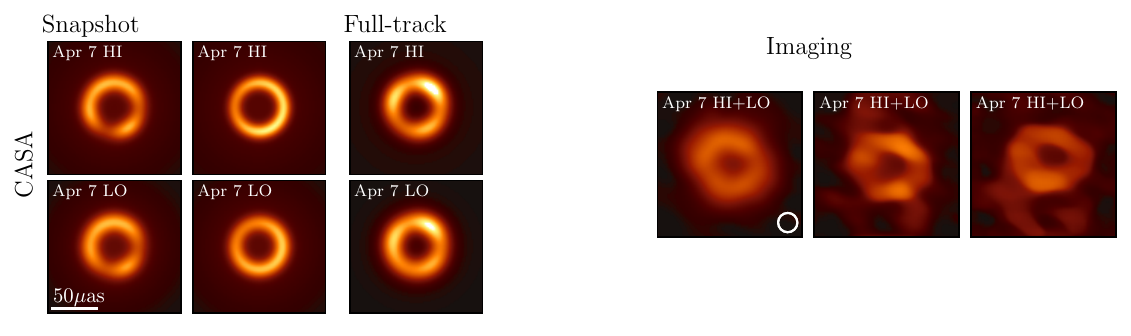}\\
    \caption{Same as \autoref{fig:morphology}, but showing results obtained from using only the April 7 data.}
    \label{fig:morphology_april7}
\end{figure*}

\begin{figure*}[t]
    \centering
    \includegraphics[width=\textwidth]{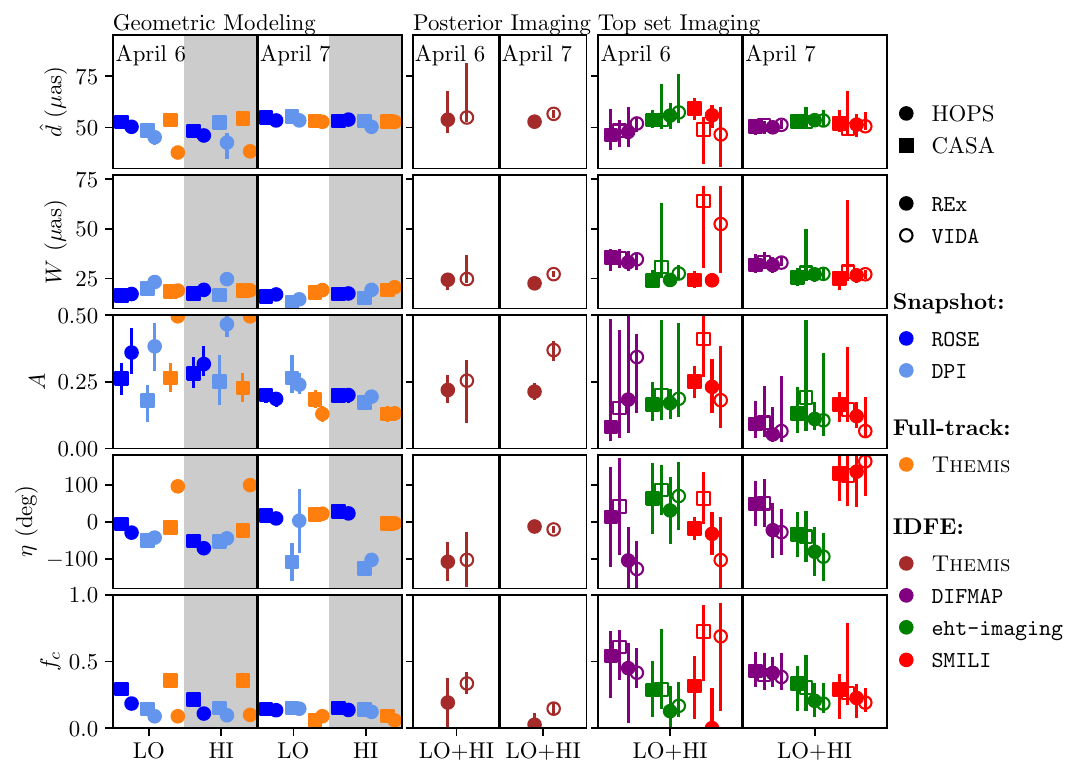}
    \caption{Similar to \autoref{fig:quant_struct}, but showing results obtained from fitting to the April 6 and 7 data sets separately.  Each column from \autoref{fig:quant_struct} has now been split into a pair of columns showing April 6 results on the left and April 7 results on the right. }
    \label{fig:quant_struct_singledays}
\end{figure*}

In this section we present the results from analyses carried out on the April 6 and April 7 data sets individually.  \autoref{fig:morphology_april6} and \autoref{fig:morphology_april7} show single-day images from each of the analysis pipelines (analogous to those shown in \autoref{fig:morphology}), and \autoref{fig:quant_struct_singledays} shows the corresponding measurements of morphological properties (analogous to those shown in \autoref{fig:quant_struct}).

\section{GRMHD synthetic data set generation} \label{sec:MoDSyntheticData}

To calibrate measurements of the angular gravitational radius $\theta_g$ (see \autoref{sec:Calibration}), we rely on the library of GRMHD simulations and associated GRRT synthetic movies produced and described in \citetalias{PaperV}.
In this appendix we provide an overview of our model selection and data generation procedures, which are conceptually similar to the calibration analysis carried out in \citetalias{M87PaperVI}.

\begin{figure*}[t]
  \centering
  \includegraphics[width=1.0\textwidth]{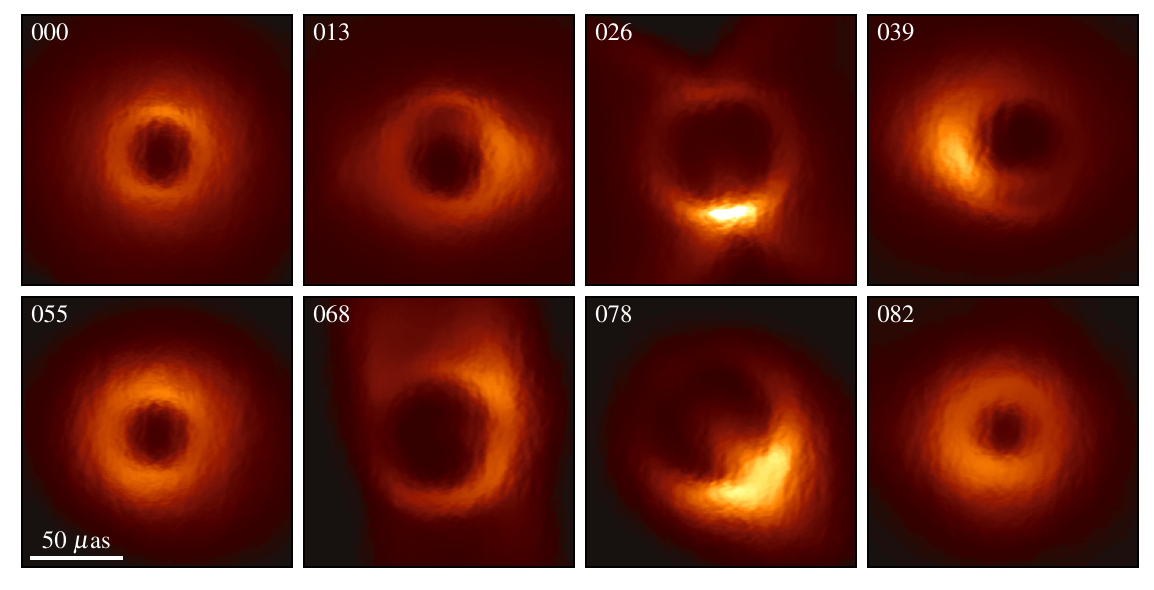} \caption{Example average images from some of the GRMHD movies selected for synthetic data generation; each movie has been light-curve-normalized prior to averaging.
  The data set indices are labeled in the upper left corner of each panel, and the corresponding GRMHD parameters are listed in \autoref{tab:GRMHDSynthData}.  All of the images share a common brightness color scale; the absolute brightness scale is arbitrary because each image has been normalized to have unit total flux density, and a modest amount of saturation has been permitted in the brightest regions to enhance the visibility of low-brightness features.  We note that these average images tend to be have much smoother structure than the individual frames of the movies that were averaged to produce them (see, e.g., \autoref{fig:test002_frames} for several example frames from one movie); the synthetic data sets themselves are produced from the movies and not from the average images.}
  \label{fig:GRMHD_example}
\end{figure*}

We select 90 simulations from the GRMHD library to be used for $\theta_g$ calibration and another 10 simulations to be used to validate this calibration.  The 90 calibration data sets uniformly grid a range of GRMHD parameters: every combination of the two MAD and SANE accretion states, five black hole spin values of $[-0.94, -0.5, 0, 0.5, 0.94]$, three inclinations of $[10^{\circ}, 50^{\circ}, 90^{\circ}]$, and three $R_{\text{high}}$ values of $[10, 40, 160]$ are represented.\footnote{$R_{\text{high}}$ is a parameter that sets the ratio of ion to electron temperatures in the simulated images; see \citetalias{M87PaperV}, \citetalias{PaperV}, and \cite{Moscibrodzka_2016} for details.  For the GRMHD calibration suite, we have not included models with $R_{\text{high}}=1$ because these models tend to produce images with substantial extended emission that is large compared to the black hole shadow size; we note that the $R_{\text{high}}=1$ models are also rejected by the model selection constraints applied in \citetalias{PaperV}.}  The 10 validation data sets are split evenly between MAD and SANE, but the black hole spins are randomly selected from $[-0.94, -0.5, 0, 0.5, 0.94]$, the inclinations are randomly selected from $[30^{\circ}, 70^{\circ}, 110^{\circ}, 130^{\circ}, 150^{\circ}]$, and the $R_{\text{high}}$ values are randomly selected from $[10, 40, 160]$.  The resulting model images contain a large variety of possible image morphologies, along with self-consistent dynamics as governed by the equations of GRMHD and GRRT.  The GRMHD parameters corresponding to each selected calibration and validation model are listed in \autoref{tab:GRMHDSynthData}, and some example average images are shown in \autoref{fig:GRMHD_example}.

\clearpage
\startlongtable
\begin{deluxetable*}{lcCccCc}
\tablecolumns{7}
\tabletypesize{\normalsize}
\tablewidth{0pt}
\tablecaption{GRMHD synthetic data set parameters \label{tab:GRMHDSynthData}}
\tablehead{\colhead{Data set index} & \colhead{Accretion state} & \colhead{Spin} & \colhead{Inclination} & \colhead{$R_{\text{high}}$} & \colhead{Position angle} & \colhead{Input $\theta_g$}}
\startdata
000 & MAD  & -0.94                         & 10  & 10  & 155   & 4.477  \\
001 & MAD  & -0.94                         & 10  & 40  & -114  & 4.140  \\
002 & MAD  & -0.94                         & 10  & 160 & 24    & 5.317  \\
003 & MAD  & -0.94                         & 50  & 10  & 168   & 5.493  \\
004 & MAD  & -0.94                         & 50  & 40  & 90    & 5.493  \\
005 & MAD  & -0.94                         & 50  & 160 & 89    & 6.423  \\
006 & MAD  & -0.94                         & 90  & 10  & -177  & 3.843  \\
007 & MAD  & -0.94                         & 90  & 40  & -73   & 5.503  \\
008 & MAD  & -0.94                         & 90  & 160 & 112   & 6.152  \\
009 & MAD  & -0.5\hphantom{0}              & 10  & 10  & 168   & 5.704  \\
010 & MAD  & -0.5\hphantom{0}              & 10  & 40  & 51    & 5.684  \\
011 & MAD  & -0.5\hphantom{0}              & 10  & 160 & -12   & 4.502  \\
012 & MAD  & -0.5\hphantom{0}              & 50  & 10  & -22   & 5.724  \\
013 & MAD  & -0.5\hphantom{0}              & 50  & 40  & 178   & 5.563  \\
014 & MAD  & -0.5\hphantom{0}              & 50  & 160 & 105   & 4.039  \\
015 & MAD  & -0.5\hphantom{0}              & 90  & 10  & -171  & 5.935  \\
016 & MAD  & -0.5\hphantom{0}              & 90  & 40  & 146   & 3.596  \\
017 & MAD  & -0.5\hphantom{0}              & 90  & 160 & -3    & 5.110  \\
018 & MAD  & \hphantom{-}0.0\hphantom{0}   & 10  & 10  & 35    & 3.677  \\
019 & MAD  & \hphantom{-}0.0\hphantom{0}   & 10  & 40  & 143   & 5.719  \\
020 & MAD  & \hphantom{-}0.0\hphantom{0}   & 10  & 160 & 115   & 5.030  \\
021 & MAD  & \hphantom{-}0.0\hphantom{0}   & 50  & 10  & 112   & 3.813  \\
022 & MAD  & \hphantom{-}0.0\hphantom{0}   & 50  & 40  & -101  & 4.301  \\
023 & MAD  & \hphantom{-}0.0\hphantom{0}   & 50  & 160 & -12   & 4.909  \\
024 & MAD  & \hphantom{-}0.0\hphantom{0}   & 90  & 10  & 76    & 4.059  \\
025 & MAD  & \hphantom{-}0.0\hphantom{0}   & 90  & 40  & 11    & 4.029  \\
026 & MAD  & \hphantom{-}0.0\hphantom{0}   & 90  & 160 & 97    & 6.323  \\
027 & MAD  & \hphantom{-}0.5\hphantom{0}   & 10  & 10  & 40    & 3.974  \\
028 & MAD  & \hphantom{-}0.5\hphantom{0}   & 10  & 40  & -4    & 4.658  \\
029 & MAD  & \hphantom{-}0.5\hphantom{0}   & 10  & 160 & 126   & 6.272  \\
030 & MAD  & \hphantom{-}0.5\hphantom{0}   & 50  & 10  & -42   & 4.472  \\
031 & MAD  & \hphantom{-}0.5\hphantom{0}   & 50  & 40  & 25    & 4.089  \\
032 & MAD  & \hphantom{-}0.5\hphantom{0}   & 50  & 160 & -135  & 5.598  \\
033 & MAD  & \hphantom{-}0.5\hphantom{0}   & 90  & 10  & 108   & 5.251  \\
034 & MAD  & \hphantom{-}0.5\hphantom{0}   & 90  & 40  & 171   & 5.432  \\
035 & MAD  & \hphantom{-}0.5\hphantom{0}   & 90  & 160 & 140   & 5.015  \\
036 & MAD  & \hphantom{-}0.94              & 10  & 10  & -54   & 5.679  \\
037 & MAD  & \hphantom{-}0.94              & 10  & 40  & 1     & 4.200  \\
038 & MAD  & \hphantom{-}0.94              & 10  & 160 & -92   & 5.915  \\
039 & MAD  & \hphantom{-}0.94              & 50  & 10  & -2    & 6.282  \\
040 & MAD  & \hphantom{-}0.94              & 50  & 40  & 161   & 5.131  \\
041 & MAD  & \hphantom{-}0.94              & 50  & 160 & -89   & 5.699  \\
042 & MAD  & \hphantom{-}0.94              & 90  & 10  & -48   & 5.025  \\
043 & MAD  & \hphantom{-}0.94              & 90  & 40  & -99   & 4.587  \\
044 & MAD  & \hphantom{-}0.94              & 90  & 160 & 54    & 4.467  \\
045 & SANE & -0.94                         & 10  & 10  & 97    & 5.880  \\
046 & SANE & -0.94                         & 10  & 40  & 127   & 6.388  \\
047 & SANE & -0.94                         & 10  & 160 & -142  & 6.267  \\
048 & SANE & -0.94                         & 50  & 10  & -59   & 6.016  \\
049 & SANE & -0.94                         & 50  & 40  & 144   & 3.652  \\
050 & SANE & -0.94                         & 50  & 160 & -110  & 4.411  \\
051 & SANE & -0.94                         & 90  & 10  & 46    & 3.783  \\
052 & SANE & -0.94                         & 90  & 40  & -129  & 5.075  \\
053 & SANE & -0.94                         & 90  & 160 & 68    & 6.106  \\
054 & SANE & -0.5\hphantom{0}              & 10  & 10  & 53    & 5.276  \\
055 & SANE & -0.5\hphantom{0}              & 10  & 40  & 76    & 4.281  \\
056 & SANE & -0.5\hphantom{0}              & 10  & 160 & 144   & 4.854  \\
057 & SANE & -0.5\hphantom{0}              & 50  & 10  & -173  & 6.418  \\
058 & SANE & -0.5\hphantom{0}              & 50  & 40  & 55    & 5.070  \\
059 & SANE & -0.5\hphantom{0}              & 50  & 160 & 66    & 5.000  \\
060 & SANE & -0.5\hphantom{0}              & 90  & 10  & 154   & 5.080  \\
061 & SANE & -0.5\hphantom{0}              & 90  & 40  & -154  & 5.236  \\
062 & SANE & -0.5\hphantom{0}              & 90  & 160 & 42    & 6.363  \\
063 & SANE & \hphantom{-}0.0\hphantom{0}   & 10  & 10  & -31   & 4.321  \\
064 & SANE & \hphantom{-}0.0\hphantom{0}   & 10  & 40  & -145  & 4.990  \\
065 & SANE & \hphantom{-}0.0\hphantom{0}   & 10  & 160 & -13   & 3.612  \\
066 & SANE & \hphantom{-}0.0\hphantom{0}   & 50  & 10  & 70    & 5.684  \\
067 & SANE & \hphantom{-}0.0\hphantom{0}   & 50  & 40  & 83    & 4.773  \\
068 & SANE & \hphantom{-}0.0\hphantom{0}   & 50  & 160 & -175  & 6.262  \\
069 & SANE & \hphantom{-}0.0\hphantom{0}   & 90  & 10  & 104   & 4.019  \\
070 & SANE & \hphantom{-}0.0\hphantom{0}   & 90  & 40  & -68   & 5.367  \\
071 & SANE & \hphantom{-}0.0\hphantom{0}   & 90  & 160 & -49   & 3.994  \\
072 & SANE & \hphantom{-}0.5\hphantom{0}   & 10  & 10  & -116  & 6.142  \\
073 & SANE & \hphantom{-}0.5\hphantom{0}   & 10  & 40  & -76   & 5.287  \\
074 & SANE & \hphantom{-}0.5\hphantom{0}   & 10  & 160 & -17   & 4.763  \\
075 & SANE & \hphantom{-}0.5\hphantom{0}   & 50  & 10  & 138   & 5.613  \\
076 & SANE & \hphantom{-}0.5\hphantom{0}   & 50  & 40  & -80   & 3.717  \\
077 & SANE & \hphantom{-}0.5\hphantom{0}   & 50  & 160 & -109  & 6.333  \\
078 & SANE & \hphantom{-}0.5\hphantom{0}   & 90  & 10  & 128   & 6.403  \\
079 & SANE & \hphantom{-}0.5\hphantom{0}   & 90  & 40  & -162  & 5.271  \\
080 & SANE & \hphantom{-}0.5\hphantom{0}   & 90  & 160 & -7    & 3.586  \\
081 & SANE & \hphantom{-}0.94              & 10  & 10  & -46   & 4.768  \\
082 & SANE & \hphantom{-}0.94              & 10  & 40  & 38    & 5.548  \\
083 & SANE & \hphantom{-}0.94              & 10  & 160 & 122   & 5.875  \\
084 & SANE & \hphantom{-}0.94              & 50  & 10  & -36   & 5.920  \\
085 & SANE & \hphantom{-}0.94              & 50  & 40  & 142   & 4.311  \\
086 & SANE & \hphantom{-}0.94              & 50  & 160 & 177   & 6.096  \\
087 & SANE & \hphantom{-}0.94              & 90  & 10  & 84    & 4.592  \\
088 & SANE & \hphantom{-}0.94              & 90  & 40  & 138   & 6.142  \\
089 & SANE & \hphantom{-}0.94              & 90  & 160 & 165   & 3.521  \\
\midrule
090 & MAD  & \hphantom{-}0.0\hphantom{0}   & 150 & 160 & 19    & 5.694  \\
091 & MAD  & \hphantom{-}0.5\hphantom{0}   & 70  & 160 & -32   & 5.271  \\
092 & MAD  & -0.5\hphantom{0}              & 30  & 160 & 46    & 4.497  \\
093 & MAD  & \hphantom{-}0.94              & 30  & 10  & 86    & 4.069  \\
094 & MAD  & \hphantom{-}0.5\hphantom{0}   & 150 & 160 & 11    & 5.437  \\
095 & SANE & -0.94                         & 70  & 10  & -51   & 4.778  \\
096 & SANE & \hphantom{-}0.5\hphantom{0}   & 110 & 40  & 176   & 5.835  \\
097 & SANE & -0.94                         & 130 & 160 & -118  & 3.984  \\
098 & SANE & \hphantom{-}0.0\hphantom{0}   & 150 & 40  & 70    & 5.065  \\
099 & SANE & \hphantom{-}0.5\hphantom{0}   & 110 & 10  & -71   & 5.020  \\
\enddata
\tablecomments{Simulation parameters for each of the GRMHD-based synthetic data sets used for $\theta_g$ calibration (top; indices 000--089) and validation (bottom; indices 090-099).  The sign of the spin follows the convention of \citetalias{M87PaperV}, where negative values indicate that the angular momentum of the accretion flow is antialigned with that of the black hole.  The inclination angle is given in degrees, with $90^{\circ}$ indicating an edge-on system and $0^{\circ}$ indicating a system whose spin vector is pointed toward us. The position angle is given in degrees east of north, and refers to the orientation of the black hole spin vector.  For each simulation, the input value of $\theta_g$ is given in \uas.}
\end{deluxetable*}

After selecting the GRMHD models and prior to generating synthetic data, we first modify their orientations and angular sizes from their default simulation values.  We rotate each simulated movie by a position angle that is a uniformly chosen integer in the range $[-180^{\circ},180^{\circ}]$.  Each of the simulations from \citetalias{PaperV} was produced assuming a mass of $M = 4.14 \times 10^6$\,M$_{\odot}$ and a distance of 8.127\,kpc \citep{Do_2019,Gravity_2019}, corresponding to an angular gravitational radius of $\theta_g = 5.03$\,\uas.  To avoid biasing our calibration in favor of any one value of $\theta_g$, we modify the overall spatial scale of each simulated movie by a random factor that is uniformly drawn from the range $[0.7,1.3]$.  The input position angles and gravitational radii for each movie are listed in \autoref{tab:GRMHDSynthData}.

Once the GRMHD models are selected and their movies rotated and scaled, we generate synthetic data sets in the same manner as for the synthetic data sets used in \citetalias{PaperIII}.  We use the \ehtim software to first apply artificial scattering to the source structure per the scattering model from \cite{Johnson_2018} and then sample the Fourier transform of each movie at a cadence and at $(u,v)$ locations identical to those of the EHT observations of \sgra.  The resulting visibilities are then corrupted with thermal noise and station-based gain and leakage effects at a level that is consistent with the \sgra data \citepalias{PaperII}.  Eight synthetic data sets are generated for each GRMHD model, corresponding to the $(u,v)$-coverage on four observing nights -- 2017 April 5, 6, 7, and 10 -- and two frequency bands (see \autoref{sec:Observations}).

\section{Representative \texorpdfstring{$\chi^2$}{} values for each analysis method} \label{app:Chisq}

In this section we provide some example representative $\chi^2$ values and associated quantities for each of the analysis methods used in this paper.  Specifically, we report values corresponding to the example fits shown in \autoref{fig:imaging_residuals}, \autoref{fig:snapshot_residuals}, and \autoref{fig:fulltrack_residuals}.

For any fitted data quantity $q$ with modeled counterpart $\hat{q}$ and associated measurement uncertainty $\sigma$, we determine the $\chi^2$ as
\begin{equation}
\chi^2 = \sum_i^{N_{\text{data}}} \left( \frac{q_i - \hat{q}_i}{\sigma_i} \right)^2 ,
\end{equation}
\noindent where the sum is understood to be taken over all $N_{\text{data}}$ fitted data points.  We also define a reduced-$\chi^2$ value,
\begin{equation}
\chi_{\text{red}}^2 = \frac{\chi^2}{N_{\text{dof}}} ,
\end{equation}
\noindent where $N_{\text{dof}}$ is the number of degrees of freedom remaining in the data after accounting for the free parameters in the model.  However, we note that despite their familiarity, the interpretation of either of these $\chi^2$ statistics is complicated by several aspects of the analyses presented in this paper.

The first complication is that the number of degrees of freedom is generically unknown, rendering $\chi_{\text{red}}^2$ difficult to define in practice.  The use of informative priors and the presence of correlations among model parameters mean that $N_{\text{dof}}$ cannot be determined as simply the difference between $N_{\text{data}}$ and the number of free parameters in the model.  For example, in the RML imaging methods (\ehtim and \smili), the number of effective image parameters is implicitly reduced -- relative to the number that would be assumed by simply counting the total number of image pixels -- by a large factor by the imposition of regularization terms (e.g., smoothness, sparseness) in the objective function \citepalias{PaperIII}.  Additionally, for all methods that fit to complex visibilities or visibility amplitudes, station gains are reconstructed as part of the fitting process; for those methods that simultaneously reconstruct high- and low-band data, the gains for the two bands are necessarily strongly correlated.  Furthermore, strong priors are imposed by network calibration \citepalias{PaperII,M87PaperIII}, further reducing the number of effective model parameters and growing the effective number of degrees of freedom.

The second complication is that, with the exception of the snapshot modeling presented in \autoref{sec:SnapshotGeometricModeling} (\comrade and \dpi), all models make use of an added uncertainty budget to account for source variability.  For the \themis imaging and full-track modeling analyses, the parameters describing this excess variability noise are simultaneously fit alongside those describing the image structure.  In both cases, the impact is to drive the $\chi_{\text{red}}^2$ value toward unity, rendering the resulting $\chi_{\text{red}}^2$ value not particularly meaningful as a metric of fit quality.

Nevertheless, in \autoref{tab:chisq} we present the $\chi^2$ values and relevant properties of the data sets and models used for representative examples from each analysis pathway.  In lieu of a well-defined number of degrees of freedom, we consider two limits.  An optimistic estimate is given by the procedure adopted in \citetalias{PaperIII}, in which $N_{\rm dof}\approx N_{\rm data}$, appropriate when the total number of effective model parameters is much less than the number of data points.  In this limit, $\chi_{\text{red}}^2$ ranges from 0.35 to 0.79 for the fits listed in \autoref{tab:chisq}.  A more pessimistic accounting is given by $N_{\rm dof}\approx N_{\rm data}-N_{\rm gains}-N_{\rm params}$.  This quantity is negative for some analyses, a consequence of the strong correlations that limit the effective number of model parameters in practice (e.g., for RML imaging and for certain individual snapshot models).  Among those analysis methods that do not exhibit this pathology in the $N_{\rm dof}$, the pessimistic $\chi_{\text{red}}^2$ estimates range from 0.9 to 1.4.

\begin{deluxetable*}{lccccccc}
\tabletypesize{\normalsize}
\tablewidth{0pt}
\tablecaption{Representative $\chi^2$ values and ancillary model information \label{tab:chisq}}
\tablehead{\colhead{Method} & 
\colhead{Day} &
\colhead{Band} &
\colhead{$\chi^2$} & 
\colhead{$N_{\text{data}}$} & 
\colhead{$N_{\text{gains}}$\tablenotemark{\rm\dag}} & 
\colhead{$N_{\text{params}}$\tablenotemark{\rm\dag}}}
\startdata
\ehtim imaging              &  Apr 7   & HI+LO & 4174.5 & 12082 &   6358 &  6400 \\
\smili imaging              &  Apr 7   & HI+LO & 2192.9 &  6806 &   3562 &  5625 \\
\difmap imaging             &  Apr 7   & HI+LO & 4194.5 & 12082 &   6358 &  2262 \\
\themis imaging             &  Apr 6+7 & HI+LO & 1444.5 &  2990 &   1323 &    58 \\
\themis full-track modeling &  Apr 6+7 & LO    & 1237.0 &  1562 &    666 &    19 \\
\comrade snapshot modeling  &  Apr 6+7 & LO    & 1987.2 &  4977 &   1345 &  3328 \\
\dpi snapshot modeling      &  Apr 6+7 & LO    & 1752.7 &  3176 & \ldots &  1664 \\
\enddata
\tablecomments{The $\chi^2$ values, number of data points fit ($N_{\text{data}}$), number of station gain parameters ($N_{\text{gains}}$), and number of parameters describing the image structure ($N_{\text{params}}$) for representative example fits from each of the analysis methods used in this paper.  For the \ehtim and \smili analyses, $N_{\text{params}}$ is equal to the number of pixels in the image \citepalias{PaperIII}; for the \difmap analyses, $N_{\text{params}}$ is determined by the number of CLEAN components, with each component contributing three parameters.  For the \themis imaging and full-track modeling analyses, $N_{\text{params}}$ is equal to the number of model parameters used to specify the image structure (see \citetalias{PaperIII} and \autoref{sec:FullTrackGeometricModeling}, respectively).  For the \comrade and \dpi snapshot modeling analyses, each of the $N_s$ snapshots contributes 13 and 8 parameters, respectively, to the total $N_{\text{params}}$ count (see \autoref{sec:SnapshotGeometricModeling}).  Note that because the \dpi analysis fits only to closure quantities, it does not contain any station gain parameters.}
\tablenotetext{\dag}{Note that the listed numbers of gains and model parameters are solely those necessary to forward-compute the model data values, i.e., these values characterize a property of the model specification.  Importantly, 
due to model nonlinearity and strong correlations between parameters,
these numbers are not generally suitable for determining $N_{\text{dof}}$ (see \autoref{app:Chisq}).}
\end{deluxetable*}

\section{Snapshot modeling likelihood functions} \label{app:SnapshotLikelihoods}

In this appendix we provide specific expressions for the likelihood functions used during the snapshot geometric modeling analyses described in \autoref{sec:SnapshotGeometricModeling}.  We assume the high signal-to-noise ratio limit for all data products, which is not strictly satisfied for the relatively short integration times (120\,s) employed in the snapshot modeling analyses, but which has the benefit of reducing all likelihood functions to Gaussians.

\subsection{Visibility amplitude likelihood}\label{sec:snapshot:likelihood:visamp}

For each snapshot and baseline, the visibility amplitudes are distributed according to a Rice distribution, which in the high signal-to-noise ratio limit becomes Gaussian \citep[e.g.,][]{Wardle_1974, Broderick_2020a}.  We can thus write the visibility amplitude likelihood function as
\begin{equation}\label{eq:ssmod:va}
    \mathcal{L}_{|V|, sb} = \frac{1}{\sqrt{2\pi\sigma_{sb}^2}}
        \exp\left[-\frac{(|\hat{V}_{sb}| - |g_i||g_j||\hat{\mathcal{V}}_{sb}|)^2}{2\sigma_{sb}^2}\right] ,
\end{equation}
\noindent where $b$ is a baseline index that runs over all station pairs $\{i,j\}$ in snapshot $s$.  Here $|\hat{\mathcal{V}}_{sb}|$ is the model visibility amplitude and $|g_i|$ and $|g_j|$ are the individual station gain amplitudes (see \autoref{eqn:VisibilityCorruptions}). We use this Gaussian approximation to the Rice distribution for all of the snapshot geometric modeling analyses. For a snapshot $s$, the joint visibility amplitude likelihood across all baselines is then given by
\begin{equation}
    \mathcal{L}_{|V|,s} = \prod_{b}\mathcal{L}_{|V|, sb} ,
\end{equation}
\noindent where the product is taken over all baselines $b$.

\subsection{Closure phase likelihood}\label{sec:snapshot:likelihood:cphase}

The visibility phases in EHT data sets are heavily corrupted by atmospheric fluctuations \citepalias{M87PaperII,M87PaperIII}, so all of our snapshot geometric modeling analyses work instead with closure phases $\psi$ (see \autoref{eqn:ClosurePhase}).  In the high signal-to-noise ratio limit, the variance in the closure phase on the triangle containing stations $i$, $j$, and $k$ is given by
\begin{equation}\label{eq:cphaseerr}
    \sigma^2_{\psi, ijk} =
    \sigma^2_{\ln |V|, ij} +
    \sigma^2_{\ln |V|, jk} + 
    \sigma^2_{\ln |V|, ki} ,
\end{equation}
\noindent where
\begin{equation}
    \sigma_{\ln |V|,ij} = \frac{\sigma_{ij}}{|V_{ij}|},
\end{equation}
\noindent is the uncertainty in the log visibility amplitude in the same limit and $\sigma_{ij}$ is the uncertainty in $V_{ij}$. Hereafter we replace the triangle indices $ijk$ with a single multi-index $t$ for clarity. We approximate the closure phase likelihood for a single triangle $t$ and snapshot $s$ by a von Mises distribution,
\begin{equation}
    \mathcal{L}_{\psi, st} = \frac{1}{2\pi I_0(\sigma_{\psi, t}^{-2})}\exp\left[\frac{\cos(\psi_{t} - \hat{\psi}_{t})}{\sigma^2_{\psi,t}}\right] ,
\end{equation}
\noindent where $\hat{\psi}$ denotes a measured closure phase and $I_0(x)$ is a modified Bessel function of the first kind of order 0.  In the high signal-to-noise ratio limit, the von Mises becomes a Gaussian distribution with mean $\hat{\psi}_{t}$ and standard deviation $\sigma_{\psi, t}$. Note that up to a normalization $\ln \mathcal{L}_{\psi}$ is equal to the closure phase $\chi^2$ defined in \citetalias{PaperIII}.

Because a full set of closure phases is highly redundant \citep[i.e., the set is not linearly independent; see][]{Blackburn_2020}, we use \ehtim \citep{Chael_2016,Chael_2018} to construct a minimal nonredundant set within every snapshot.  The minimal set of closure phases is constructed by selecting all triangles that contain the most sensitive station in the array (typically ALMA), which ensures that the resulting closure phases are minimally covariant.  We ignore the remaining covariances and approximate the joint closure phase likelihood for a snapshot $s$ as
\begin{equation}
    \mathcal{L}_{\psi, s} = \prod_{t} \mathcal{L}_{\psi, st},
\end{equation}
\noindent where the product is taken over all triangles $t$ in the minimal set.

\subsection{Log closure amplitude likelihood}\label{sec:snapshot:likelihood:logcamp}

Some of our snapshot modeling fits use log closure amplitudes in place of visibility amplitudes.  In the high signal-to-noise ratio limit, the variance in the log closure amplitude on the quadrangle containing stations $i$, $j$, $k$, and $\ell$ is given by
\begin{equation}\label{eq:logclosureerr}
    \sigma^2_{\ln A, ijk\ell} =
    \sigma^2_{\ln |V|, ij} +
    \sigma^2_{\ln |V|, k\ell} + 
    \sigma^2_{\ln |V|, ik} +
    \sigma^2_{\ln |V|, j\ell} .
\end{equation}
\noindent Hereafter we replace the quadrangle indices $ijk\ell$ with a single multi-index $q$ for clarity.
For a single log closure amplitude on quadrangle $q$, we thus have the Gaussian likelihood
\begin{equation}
    \mathcal{L}_{\ln A, sq} =\frac{1}{\sqrt{2\pi\sigma_{\ln A, sq}^2}}
    \exp\left[-\frac{(\ln A_{sq} - \ln\hat{A}_{sq})^2}{2\sigma_{\ln A,sq}^2}\right],
\end{equation}
\noindent in the same limit.  We use \ehtim to construct a minimal nonredundant set of log closure amplitudes within every snapshot, and for the joint likelihood over all of the linearly independent quadrangles within that snapshot we again ignore covariance and treat the data products as being statistically independent.  The joint likelihood is then given by
\begin{equation}
    \mathcal{L}_{A, s} = \prod_{q} \mathcal{L}_{A, sq},
\end{equation}
\noindent where the product is taken over all quadrangles $q$ in the minimal set.

\section{Snapshot modeling prior distribution consistency} \label{app:HypermodelMarginals}

In specifying the hypermodel for combining geometric modeling results from individual snapshots (see \autoref{sec:snapshot:averaging}), we have selected a set of snapshot priors $\pi(\bm{\theta}_s)$, an average model prior $\pi(\bar{\bm{\theta}})$, and a hypermodel $\pi(\bm{\theta}_s | \bar{\bm{\theta}})$.  A self-consistent set of hypermodel and snapshot priors should satisfy the relation
\begin{equation}
    \pi(\bm{\theta}_s) = \int \pi(\bm{\theta}_s | \bm{\mu}, \bm{\sigma}) \pi(\bm{\mu})\pi(\bm{\sigma}) {\rm d}\bm{\mu}{\rm d}\bm{\sigma}. \label{eqn:PriorConsistency}
\end{equation}
However, because we have carried out the hierarchical modeling from \autoref{sec:SnapshotGeometricModeling} in two stages -- i.e., first fitting the \model model to individual snapshots, then combining the results from all snapshots -- the priors on the model parameters during each stage are selected for local convenience, and the relation in \autoref{eqn:PriorConsistency} does not strictly hold.

\begin{figure}[t]
    \centering
    \includegraphics[width=\columnwidth]{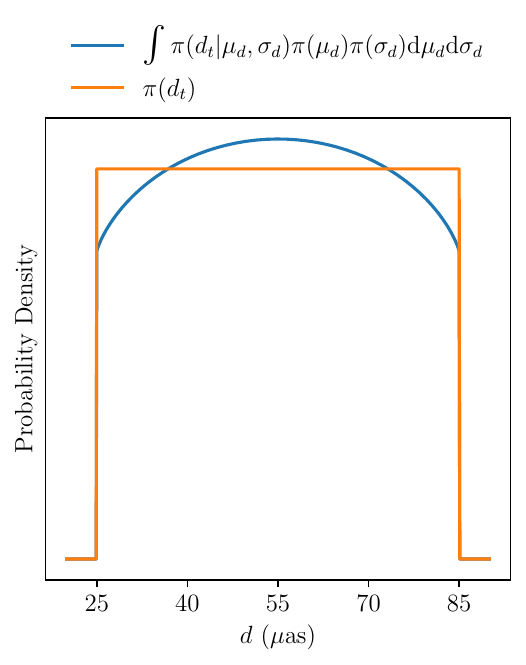}
    \caption{Illustration of the centralizing bias induced by specifying both the snapshot priors and hypermodel priors separately, using the diameter parameter as an example. The orange curve shows the diameter prior specified during \model model fitting of an individual snapshot, and the blue curve shows the effective prior on this parameter after snapshots are combined via the procedure specified in \autoref{sec:snapshot:averaging}.}
    \label{fig:heirarch_marginal}
\end{figure}

As a result, the final posteriors for each parameter behave as though the individual snapshot fits were carried out using effective priors that differ from the priors we actually imposed (see \autoref{tab:snapshot:model_priors}).  \autoref{fig:heirarch_marginal} illustrates how the original priors differ from the effective priors induced by the snapshot-combining procedure.  The net result is a modest ($\lesssim$20\% maximum probability density deviation) bias for parameter values to fall toward the center of the original (flat) prior range and away from the edges of that range.  Because the widths of the individual snapshot posteriors are typically much narrower than the full prior range (see, e.g., \autoref{fig:snapshot:average_model}), we expect the impact of this centralizing bias on our reported values to be negligible.

\section{Snapshot geometric modeling validation tests}\label{app:SnapshotValidation}

\begin{figure*}[t]
  \begin{center}
    \includegraphics[width=\columnwidth]{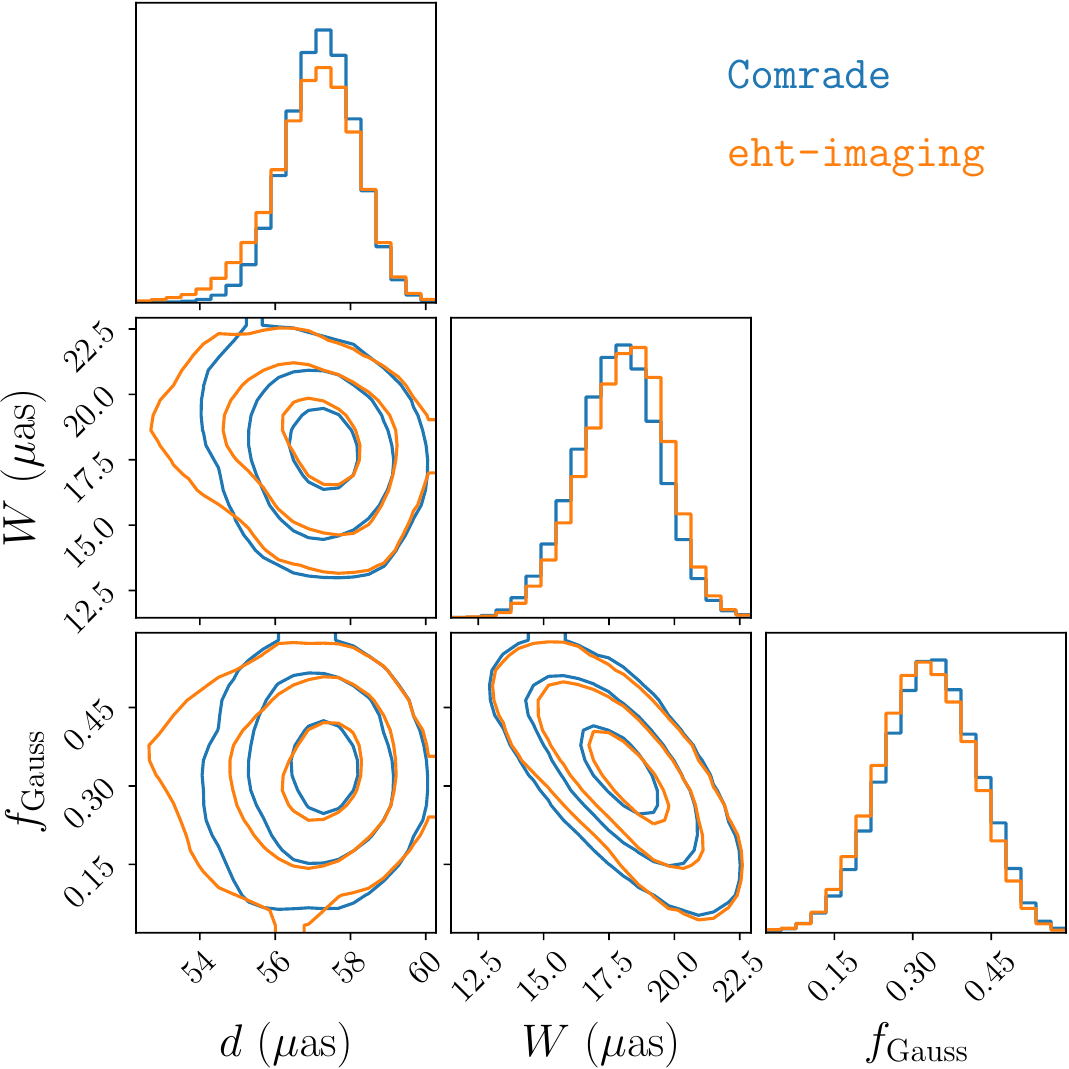}
    \includegraphics[width=\columnwidth]{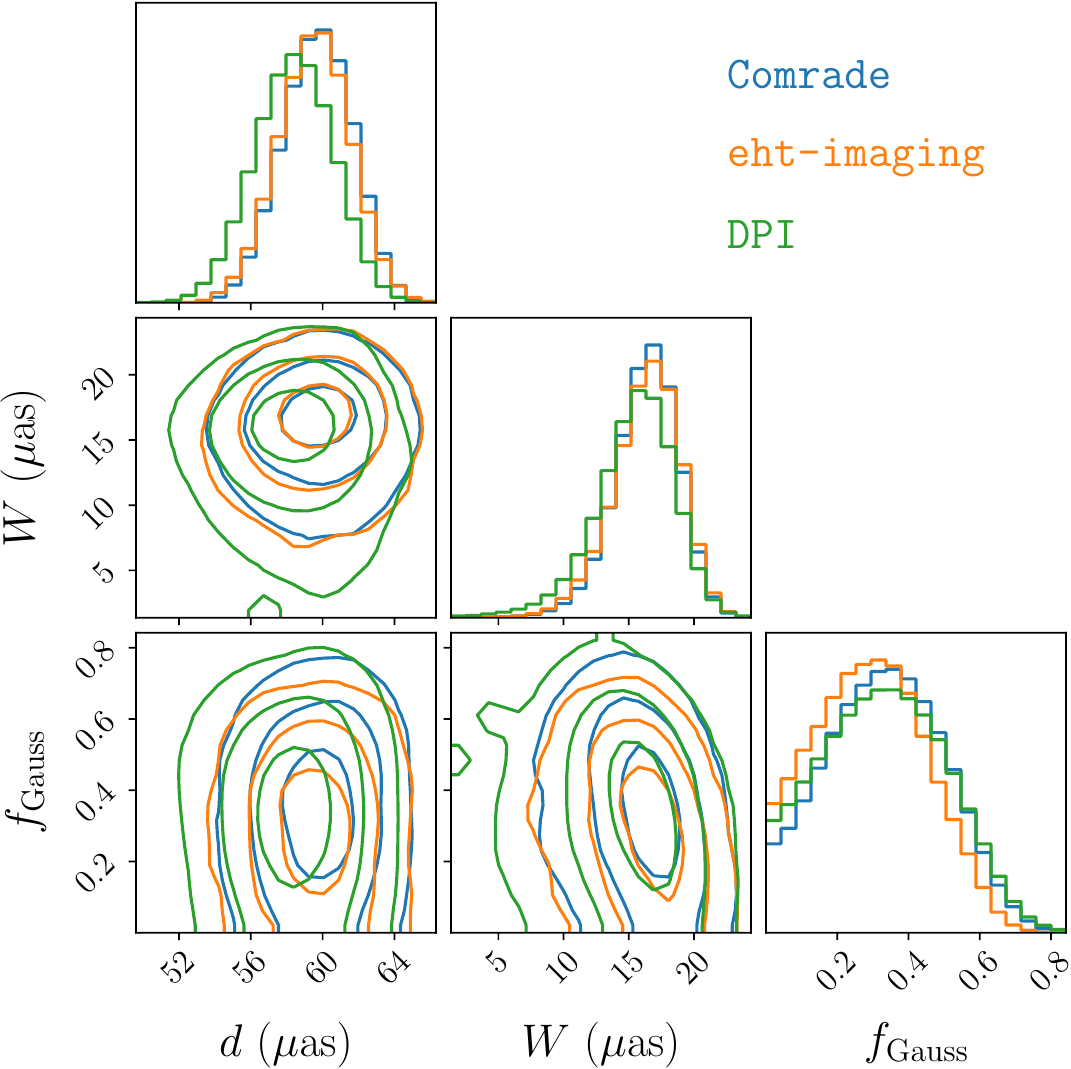}
  \end{center}
\caption{Comparison of the 2D joint posterior distributions obtained from fitting an \model model to a 120\,s snapshot starting at 12.65\,h UT in the \sgra April 7 HOPS low-band data set. The left plot compares the results from \rose (blue) and \ehtim (orange), fitting to visibility amplitudes (including gains) and closure phases with an $m=3$ \model.  The right triangle plot shows the results from \rose (blue), \ehtim (orange), and \dpi (green), fitting to closure amplitudes and closure phases for an $m=2$ \model. In both cases we only show the results for the diameter, width, and fractional Gaussian component flux parameters. Since \dpi fits the diameter of the blurred m-ring $d'$ (\autoref{eq:dpi-mGring}), the \dpi diameter was debiased so that it corresponds to the infinitesimally thin m-ring diameter that is fit by \ehtim and \rose \citepalias[see also][]{M87PaperVI}. The contours show 1$\sigma$, 2$\sigma$, and 3$\sigma$ levels of the posterior distributions.}
\label{fig:mring_triangle}
\end{figure*}

We cross-validate the results of each snapshot modeling software using a single snapshot of the low-band \sgra data from April 7 processed with the HOPS pipeline. The specific snapshot selected begins at 12.65\,h UTC, where the $(u,v)$-coverage of the observation is maximized \citep{Farah_2022}.  Given the differences in model specification and fitted data products between \rose/\ehtim and DPI, we run two separate cross-validation tests:
\begin{enumerate}
    \item The first test compares the results between \rose and \ehtim when fitting an $m=3$ \model model to visibility amplitudes (including gain amplitudes as model parameters) and closure phases. We do not include \dpi in this comparison because it cannot currently fit for station gain parameters.
    \item The second test compares the results between \rose, \dpi, and \ehtim when fitting an $m=2$ \model model to log closure amplitudes and closure phases.
\end{enumerate}

\noindent \autoref{fig:mring_triangle} shows the diameter, thickness, and fractional central flux posteriors obtained from performing the tests described above. The posteriors show generally good agreement across codes.

\section{Analysis specifics and validation of calibration strategy} \label{app:Validation}

Our calibration strategy for determining the scaling factor $\alpha$ that relates measured ring diameters to intrinsic angular gravitational radii $\theta_g$ is described in \autoref{sec:Calibration}.  In this appendix, we describe summarize the elements of this strategy that are specific to the different analysis pathways described in Sections \ref{sec:ImageDomain}, \ref{sec:SnapshotGeometricModeling}, and \ref{sec:FullTrackGeometricModeling}.  We also validate the calibration procedure by applying the calibrated $\alpha$ values to ring diameter measurements from 10 synthetic GRMHD data sets.  For these data sets we know the underlying ground-truth $\theta_g$ values, and so we can use them to verify whether our measurement and calibration strategy is working as intended.

\subsection{IDFE specifics}

To perform an IDFE-based $\theta_g$ calibration, top-set and posterior images are produced for each of the 90 synthetic GRMHD-based data sets described in \autoref{sec:Calibration} (see also \autoref{sec:MoDSyntheticData}).  Each of these images is run through both \rex and \vida in the same manner as described in \autoref{sec:ImageDomain} for the \sgra data.

\subsection{Snapshot geometric modeling specifics}

We carry out snapshot geometric modeling of the GRMHD calibration and validation data sets in the same manner as described in \autoref{sec:SnapshotGeometricModeling} for the \sgra data.  For all synthetic data sets we use the same data preparation and snapshot timescale as for the fits to the \sgra data.  We also retain the same model specification, fitting an $m=4$ \model for all \rose analyses and an $m=2$ \model for all \texttt{DPI} analyses.  The \texttt{DPI} analyses are carried out on the low-band data sets only, while the \rose analyses are carried out on both low- and high-band data sets.

\subsection{Full-track geometric modeling specifics}

We carry out full-track geometric modeling of the GRMHD calibration and validation data sets in the same manner as described in \autoref{sec:FullTrackGeometricModeling} for the \sgra data.  For all synthetic data sets we use the same data preparation as for the fits to the \sgra data (see \autoref{sec:fulltrack:dataprep}).  In particular, we derive appropriately individualized priors on the noise model parameters by performing model-agnostic variability quantification (per \autoref{sec:pre-modeling}) on multiday instantiations (corresponding to April 5, 6, 7, and 10; see \autoref{sec:MoDSyntheticData}) of each of the synthetic data sets.  We also retain the same geometric model specification, fitting an $m=4$ \model for all analyses.

\subsection{Validation}

\begin{figure*}
    \centering
    \includegraphics[width=\linewidth]{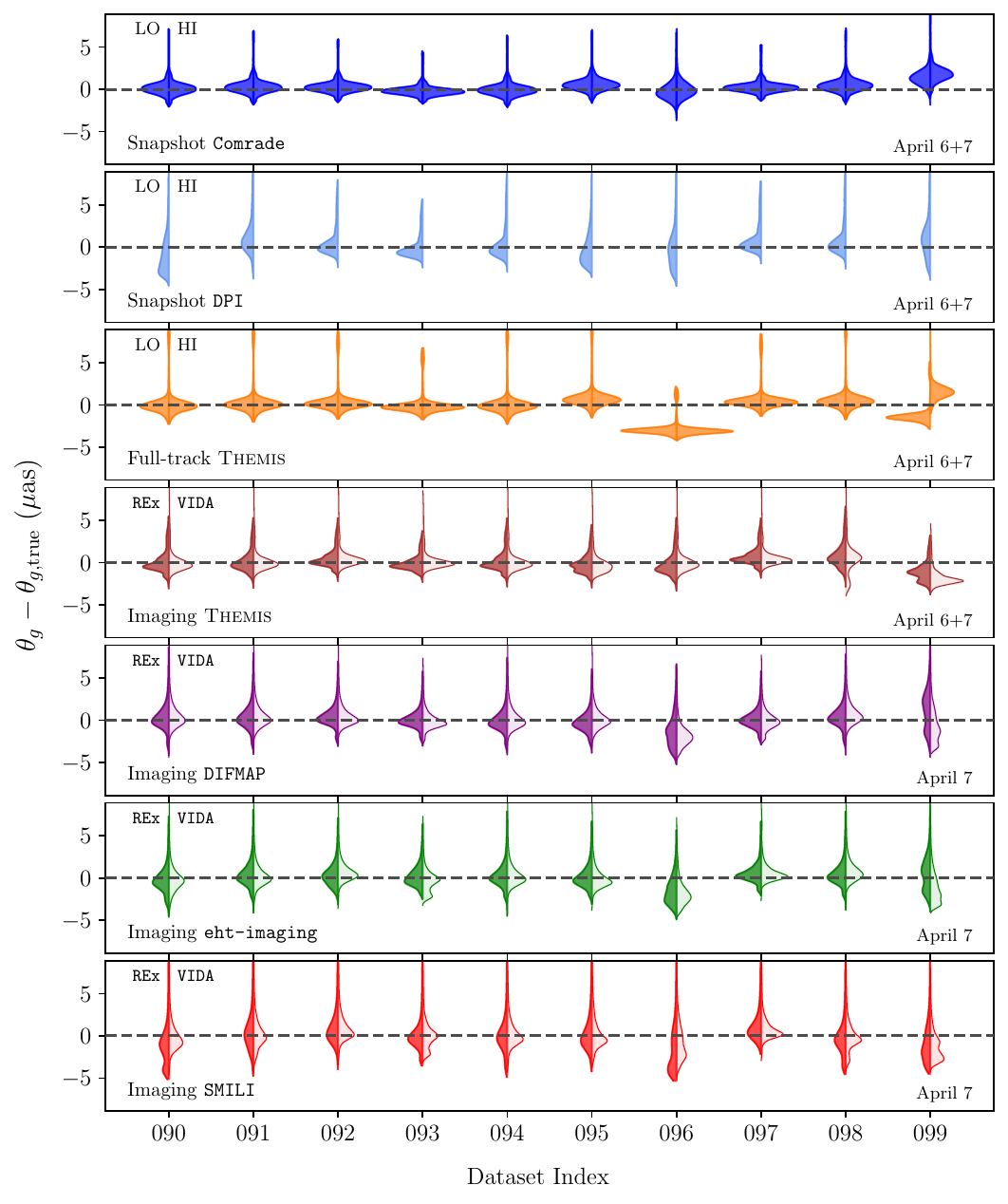}
    \caption{Distributions of recovered $\theta_g$ values relative to the known input value, from the GRMHD validation exercise for each analysis pathway. For geometric modeling results the left- and right-hand distributions show the results from fitting to LO and HI bands, respectively. For the IDFE results, the left- and right-hand distributions show the results from using \rex and \vida, respectively.  No \texttt{metronization}-based culling has been applied to the IDFE results.}
    \label{fig:mod-validation}
\end{figure*}

For each of the analysis pathways, we validate the $\theta_g$ calibration using an additional 10 synthetic GRMHD-based data sets.  \autoref{fig:mod-validation} shows the results of carrying out ring diameter measurements and subsequent $\theta_g$ conversions on these 10 validation suite data sets, for each of the IDFE, snapshot, and full-track analyses.  All analysis pathways are able to successfully recover the correct value of $\theta_g$ to within their determined level of calibration uncertainty.

\subsection{Origin and nature of calibration outliers}

\begin{figure*}[t]
    \centering
    \includegraphics[width=0.46\textwidth]{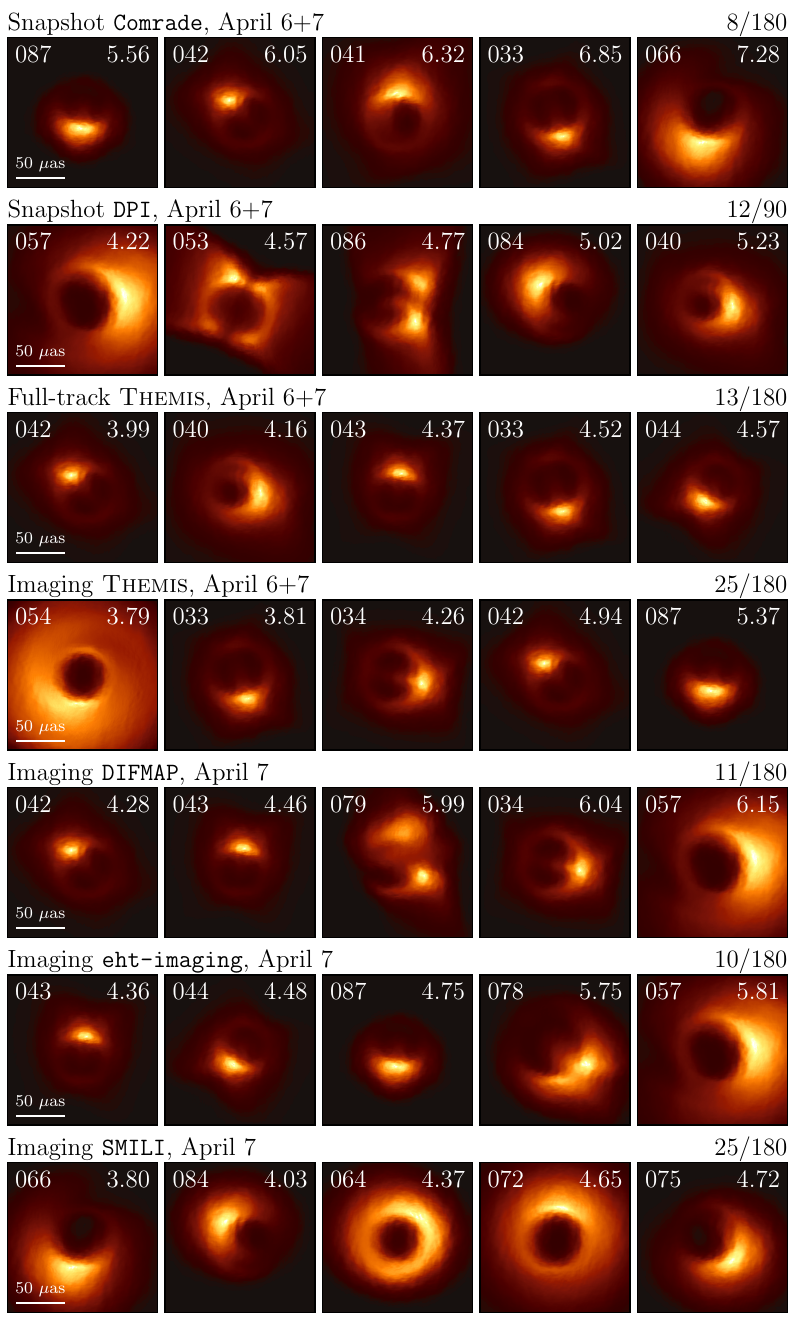}
    \hspace{0.03125in}
    \unskip\ \vrule\
    \hspace{0.03125in}
    \includegraphics[width=0.46\textwidth]{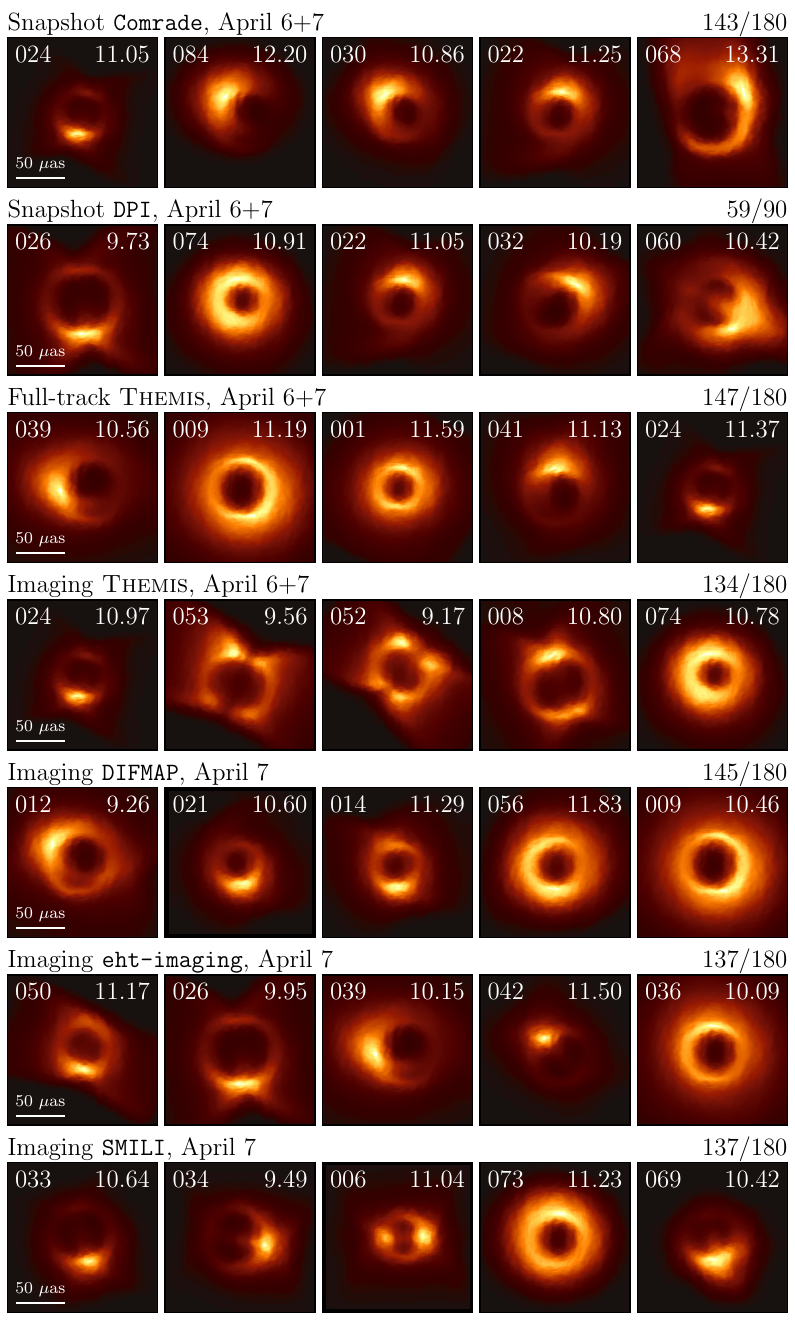}   
    \caption{Time-averaged images for GRMHD simulations that produce anomalously small (left) and typical (right) $\alpha$ calibrations for each geometric modeling and IDFE method used to estimate the mass of \sgra.
    {\em Left:} shown are the GMRHD simulations within the calibration set that result in the five smallest median $\alpha$ values (computed across the applicable posterior or top set).  Above each collection of images, the number of calibration data sets that find $\alpha<7.5$ is listed, in comparison to the total number of calibration experiments (HI/LO band, \rex/\vida ring radius measurements).
    {\em Right:} five GRMHD simulations randomly chosen from within the peak of the distribution of $\alpha$ values.  Above each collection of images, the number of calibration data sets within one standard deviation of the mean $\alpha$ across the calibration set is listed.
    In each panel, the corresponding simulation index in \autoref{tab:GRMHDSynthData} and the median $\alpha$ across the relevant posterior or top set are given in the upper left and right, respectively.}
    \label{fig:mod-pas}
\end{figure*}

The distributions of calibrated $\alpha$ values from each ring diameter measurement technique show heavy tails toward small $\alpha$, which manifest as heavy tails to large $\theta_g$ in \autoref{fig:mod-validation} (see also \autoref{fig:mod_cal}).  This behavior appears to be generic across all classes of geometric modeling and IDFE analyses used in this paper, and it implies that some fraction of the calibration data sets are reconstructed to have systematically smaller rings than would be predicted from the known values of $\theta_g$ in each of the input ground-truth simulations.

The left half of \autoref{fig:mod-pas} shows average images from the input GRMHD calibration suite simulations corresponding to the five smallest $\alpha$ values recovered by each analysis pathway.  The number of simulations for which the reconstructed ring corresponds to a ``small'' value of $\alpha$ depends on the analysis method; for instance, the fraction of reconstructions having median $\alpha < 7.5$ ranges from $\sim$4\% for snapshot geometric modeling with \rose up to $\sim$14\% for imaging with \smili.  We can see in \autoref{fig:mod-pas} that many of these small-$\alpha$ simulations have structures that are not obviously ring-like.  Common morphologies in the small-$\alpha$ simulations include images dominated by compact regions of bright emission (typical of highly edge-on systems), or images containing prominent diffuse emission extending well outside the shadow region of interest.  Such structures cannot necessarily be well measured by, e.g., the \model model or IDFE techniques aimed at extracting signatures of a ring-like emission morphology, and attempts to apply these techniques to such data sets can yield results that are difficult to interpret.  However, we note that not all of the small-$\alpha$ simulations exhibit such morphological difficulties; some of the poor reconstructions are obtained from simulations with readily apparent ring-like structures, indicating that other difficulties (e.g., strong variability) may be playing a more important role in these cases.

The right half of \autoref{fig:mod-pas} shows average images from the input GRMHD calibration suite simulations corresponding to five ``typical'' $\alpha$ values recovered by each analysis pathway; each of these images has been randomly selected from the set whose reconstructed $\alpha$ falls within one standard deviation of the mean.
In contrast to the small-$\alpha$ simulations, these images more commonly exhibit ring-like morphologies of the sort that we would expect to be amenable to \model modeling or ring extraction techniques.  Furthermore, for most analysis methods a large fraction ($\sim$75\%) of the data sets are contained within one standard deviation of the mean; the fact that this fraction is larger than the $\sim$68\% that we would expect for a Gaussian distribution is another manifestation of the heavy tails in the $\alpha$ distributions, and it indicates that the majority of reconstructions are narrowly peaked around the mean (see also \autoref{fig:mod-validation}).  However, even among the well-reconstructed images we still find a small number of less obvious ring structures, including some that are dominated by compact emission regions like many of the small-$\alpha$ simulations.  Again, the presence of such simulations indicates that the ground-truth emission morphology is not the sole driver of whether or not the underlying ring structure can be successfully reconstructed.

The ring measurement analysis techniques developed in this paper are designed to be appropriate for application to the EHT \sgra data.  When applying these techniques to a suite of GRMHD simulations containing very diverse image morphologies, we find that a fraction of the reconstructed rings have unreliable diameter measurements.  These poor reconstructions contribute to the uncertainty in our $\alpha$ calibration, where they manifest as heavy tails in our calibrated $\alpha$ distribution.  The corresponding large uncertainty in $\alpha$ is a consequence of the fact that many of the images in the calibration suite do not resemble \sgra, and thus that analysis techniques designed for the latter do not necessarily function well when applied to the former.
A calibration suite that was more directly tailored to match the properties of the EHT \sgra observations may result in smaller $\alpha$ calibration uncertainties and a correspondingly tighter constraint on $\theta_g$.

\end{appendix}